\documentclass[aps,prd,twocolumn,showpacs,floatfix,preprintnumbers,amsmath,amssymb,nofootinbib, superscriptaddress]{revtex4-1}

\usepackage{graphicx}
\usepackage{color}
\usepackage[font=small]{caption,subfig}
\usepackage{hyperref}
\usepackage{tikz}
\usetikzlibrary{decorations.markings}
\usepackage{lipsum}
\usepackage{url}

\renewcommand{\bm}[1]{\mathbf #1}

\newcommand{\overlrarrow}[1]{\overset{\text{\tiny$\bm\leftrightarrow$}}{#1}}

\newcommand{\be}{\begin{equation}}
\newcommand{\ee}{\end{equation}}
\newcommand{\ba}{\begin{eqnarray}}
\newcommand{\ea}{\end{eqnarray}}
\newcommand{\en}{\nonumber\\}
\newcommand{\pa}{\partial}

\newcommand{\apss}{Astrophysics and Space Science}

\newcommand{\mnras}{Mon. Not. R. Astron. Soc.}

\newcommand{\jcap}{Journal of Cosmology and Astroparticle Physics}
\newcommand{\sovast}{Soviet Astronomy}

\font\FermiSmallfont=cmssq8 scaled 1200

\def\FNALppthead#1#2{
\null
\begin{center}\vskip -1.0truein{\hbox to 7.5truein {
\hfill
\vbox to 1in {\vfill \FermiSmallfont
              \hbox{#1}
              \hbox{#2}
              \vfill}
}}\vskip-0.0truein\end{center}}

\allowdisplaybreaks 

\begin{document}
\title{Sterile neutrino dark matter: A tale of weak interactions in the strong coupling epoch}
\author{Tejaswi Venumadhav}
\affiliation{California Institute of Technology, Mail Code 350-17, Pasadena, California 91125, USA} 
\author{Francis-Yan Cyr-Racine}
\affiliation{California Institute of Technology, Mail Code 350-17, Pasadena, California 91125, USA}
\affiliation{Jet Propulsion Laboratory, California Institute of Technology, Pasadena, California 91109, USA}
\author{Kevork N. Abazajian}
\affiliation{Center for Cosmology, Department of Physics and Astronomy, University of California, Irvine, Irvine, California 92697, USA}
\author{Christopher M. Hirata}
\affiliation{Center for Cosmology and Astroparticle Physics (CCAPP), The Ohio State University, 191 West Woodruff Lane, Columbus, Ohio 43210, USA}

\date{\today}
\preprint{UCI-TR-2015-12, INT-PUB-15-032}

\begin{abstract}
We perform a detailed study of the weak interactions of standard model neutrinos with the primordial plasma and their effect on the resonant production of sterile neutrino dark matter. Motivated by issues in cosmological structure formation on small scales, and reported X-ray signals that could be due to sterile neutrino decay, we consider $7$ keV-scale sterile neutrinos. Oscillation-driven production of such sterile neutrinos occurs at temperatures $T \gtrsim 100$ MeV, where we study two significant effects of weakly charged species in the primordial plasma: (1) the redistribution of an input lepton asymmetry; (2) the opacity for active neutrinos. We calculate the redistribution analytically above and below the quark-hadron transition, and match with lattice QCD calculations through the transition. We estimate opacities due to tree level processes involving leptons and quarks above the quark-hadron transition, and the most important mesons below the transition. We report final sterile neutrino dark matter phase space densities that are significantly influenced by these effects, and yet relatively robust to remaining uncertainties in the nature of the quark-hadron transition. We also provide transfer functions for cosmological density fluctuations with cutoffs at $k \simeq 10 \ h \ {\rm Mpc}^{-1}$, that are relevant to galactic structure formation.

\end{abstract}



\maketitle
\section{Introduction}
\label{sec:introduction}

Deep in the radiation dominated epoch of the Universe, the three neutrinos present in the standard model (SM) of particle physics \cite{Agashe:2014kda} make up a significant population of relativistic species within the primeval cosmic plasma. We have strong evidence of their existence at these early epochs from probes of the primordial Universe such as the Cosmic Microwave Background (CMB) (probing temperature $T \sim 0.25$ eV) \cite{PlanckCosmology}, and the synthesis of light elements during the epoch of Big Bang Nucleosynthesis (BBN) which depends on the neutron-to-proton ratio set at $T_{\rm dec} \sim 1.5$ MeV \cite{BBN2015}, the temperature of weak neutrino decoupling. Above this temperature, SM neutrinos interact with species that carry weak charge, through which they remain coupled to the primordial plasma \cite{McKellar94}. 

There is a long history of speculation about additional neutrino species (see Ref.~\cite{Kusenko:2009up} for a recent review). Owing to the precise measurement of the invisible decay width of the SM $Z$ boson \cite{Agashe:2014kda}, any extra neutrino species must be ``sterile'' (i.e.~electroweak singlets) \cite{Pontecorvo:1967fh}. Furthermore, precise measurements of the CMB \cite{PlanckCosmology,2015JCAP...04..006V} and of the primeval abundance of light elements \cite{2014ApJ...781...31C,2015arXiv150308146A} strongly constrain the presence of extra relativistic species in the early Universe. These constraints indicate that (i) unlike SM neutrinos, light sterile neutrinos never fully thermalize with the rest of the cosmic plasma \cite{2012JCAP...07..025H,Hannestad:2013ana,Dasgupta:2013zpn,Bringmann:2013vra,Ko:2014bka}, or (ii) that sterile neutrinos are massive enough to form the inferred population of dark matter (DM) in the Universe (see e.g. Ref.~\cite{doi:10.1146/annurev.nucl.010909.083654}). Sterile neutrinos with masses in the keV range act as DM in the CMB era, but are relativistic in the BBN era, when they do not significantly impact the expansion rate due to their negligible energy density (compared to the Fermi-Dirac value). 

Early works in this direction studied right-handed sterile neutrinos with masses $m_{\rm s} \approx 0.1-100$ keV, produced by the oscillation of left-handed SM neutrinos \cite{PhysRevD.25.213,Dodelson94,Abazajian01,Dolgov2002339,2007JHEP...01..091A}. The mixing angle between the SM and sterile neutrinos is fixed by the present day DM abundance. In the original Dodelson-Widrow scenario \cite{Dodelson94}, sterile neutrinos are produced with a momentum distribution reflecting that of the active neutrino species, and thus constitute ``warm'' DM \cite{1981ApJ...250..423D, Blumenthal:1982mv,Bode:2000gq,Dalcanton:2000hn}. However, small-scale structure formation \cite{PhysRevD.73.063513,PhysRevLett.97.191303,PhysRevLett.97.071301,1475-7516-2009-05-012,PhysRevD.83.043506,deSouza11072013,PhysRevD.88.043502,PhysRevD.89.025017} and X-ray observations \cite{Abazajian01decay,PhysRevD.74.033009,Boyarsky11072008,1475-7516-2012-03-018,2015arXiv150404027N} appear in significant conflict with the fiducial Dodelson-Widrow scenario, hence prompting the search for alternative sterile neutrino production mechanisms \cite{Shi99,Abazajian01,Asaka200517,Shaposhnikov2006414,1475-7516-2008-06-031,PhysRevLett.97.241301,PhysRevD.77.065014,1475-7516-2008-10-024,1475-7516-2014-03-028,Asaka2006401,PhysRevD.81.085032,1475-7516-2012-07-006,2014PhRvD..90d5030R,2015JHEP...01..006A,2014arXiv1409.6311M,PhysRevD.89.113004,2015PhRvD..91f3502L,2015JCAP...06..011M,Patwardhan:2015kga}. 

In this paper, we examine in detail the resonant production of sterile neutrinos in the presence of a small primordial lepton asymmetry. Originally proposed by Shi and Fuller \cite{Shi99}, this production mechanism makes use of a small lepton asymmetry to modify the plasma's interaction with SM neutrinos in such a manner as to resonantly produce sterile neutrinos at particular momenta \cite{Abazajian01,1475-7516-2008-06-031,Kishimoto08}. This generically results in `colder' DM distribution which improves consistency with models of cosmological structure formation \cite{Boyarsky:2008mt,Boyarsky:2009ix,Maccio:2012uh,Anderhalden:2012jc,Schneider:2013wwa,Kennedy:2013uta,Abazajian:2014gza,2015MNRAS.446.2760M,Bose:2015mga}, while requiring a modest primordial lepton asymmetry, which is relatively poorly constrained \cite{Kneller01,2005PhRvD..72f3003L,2005PhRvD..72f3004A,2006PhRvD..74h5008S,2010JCAP...05..037S}.

Sterile and active neutrino mixing, which is needed for the former's production, also leads to their decay \cite{Shrock:1974nd,PhysRevD.25.766}. For typical values of the sterile neutrino mass this predicts an X-ray flux from the DM distribution in the low redshift Universe \cite{Abazajian01decay, Abazajian01}. This has been the subject of much recent interest, due to hints of an excess flux at $\sim 3.5 ~{\rm keV}$ in stacked X-ray spectra of several galaxy clusters \cite{bulbul:2014sua} and in observations of M31, the Milky Way, and Perseus \cite{boyarsky:2014jta,Boyarsky:2014ska}. There is currently an active debate on the existence, significance and interpretation of this excess \cite{Anderson14,2014arXiv1405.7943R,Jeltema15, Boyarsky14comment, Bulbul14comment, Jeltema14,PhysRevD.90.103506,2015MNRAS.451.2447U}. In the present work, we use this tentative signal as a motivation to study in detail the physics of sterile neutrino production in the early Universe, but the machinery we develop is more generally applicable to the broader parameter space of the Shi-Fuller mechanism.

We present here an updated calculation of resonantly-produced sterile neutrinos and relax several simplifications that had been adopted previously in the literature. Furthermore, we leverage recent advances in our understanding of the quark-hadron transition in order to include a more realistic treatment of the strongly-interacting sector. Our motivation is twofold: a) improve the treatment of lepton asymmetry, which is a crucial beyond-SM ingredient in the mechanism, and b) provide realistic sterile neutrino phase space densities (PSDs) and transfer functions for matter fluctuations, which are starting points for studying cosmological implications on small scales. Our improvements to the sterile neutrino production calculation can broadly be classified in three categories.

Firstly, we study how the cosmic plasma reprocesses a primordial lepton asymmetry. For models that can explain the above X-ray excess, the majority of sterile neutrinos are produced at temperatures above $100$ MeV \cite{Kishimoto08}. At these temperatures, there is a significant population of either quarks or mesons, depending on whether the temperature is above or below the quark-hadron transition. Since these hadronic species are coupled to neutrinos and charged leptons through weak processes, the establishment of chemical equilibrium among the different constituents of the cosmic plasma will automatically transfer a primordial lepton asymmetry to the hadronic sector. An illustrative example is the reaction
\begin{equation}
  \nu_{\mu} + \mu^+ \rightleftharpoons \pi^+ \mbox{,} \label{eq:piondecay}
\end{equation}
which can redistribute an initial neutrino asymmetry into charged lepton and hadronic asymmetries. At lower temperatures, the asymmetry is redistributed to a lesser degree between the leptonic flavors. As we discuss in the body of the paper, this redistribution modifies the dynamics of the resonant sterile neutrino production, resulting in a modified final PSD. 

Secondly, we incorporate several new elements to the calculations of the neutrino opacity (i.e.~the imaginary part of the self-energy) at temperatures $10$ MeV $\leq T \leq 10$ GeV. Accurate neutrino opacities are needed since they basically control the production rate of sterile neutrinos through cosmic epochs. Early works on neutrino interactions in the early Universe \cite{Notzold88, McKellar94, Abazajian01} assumed that neutrinos largely scattered off relativistic particles and thus scaled their cross-sections with the center-of-mass (CM) energy. In addition, these calculations also neglected the effects of particle statistics. Under these two simplifying assumptions, the opacity $\Gamma(E_{\nu_\alpha})$ for an input neutrino of energy $E_{\nu_\alpha}$ is of the form
\begin{equation}
  \Gamma(E_{\nu_\alpha}) = \lambda(T) G_{\rm F}^2 T^4 E_{\nu_\alpha} \mbox{,} \label{eq:pscaling}
\end{equation}
where $G_{\rm F}$ is the Fermi coupling constant, and $\lambda(T)$ is a constant that depends on the number and type of available relativistic species in the cosmic plasma. References \cite{2006JHEP...06..053A,2007JHEP...01..091A} subsequently developed a framework to include particle masses, loop corrections, and particle statistics in the neutrino opacity calculation. In the present work, we add previously-neglected contributions to the opacity such as two- and three-body fusion reactions, and also use chiral perturbation theory to compute the hadronic contribution to the opacity below the quark-hadron transition. We find both quantitative and qualitative modifications to the form of Eq.~\eqref{eq:pscaling}. Wherever we present matrix elements, we use the `--+++' metric signature.

Thirdly, we fold the asymmetry redistribution and opacity calculations into the sterile neutrino production computation, and provide updated PSDs for the range of parameters relevant to the X-ray excess. As part of this process, we carefully review and correct the numerical implementation of the sterile neutrino production used in Ref.~\cite{Abazajian:2014gza}. Our sterile neutrino production code is publicly available at \url{https://github.com/ntveem/sterile-dm}. We finally use the updated sterile neutrino PSDs in a standard cosmological Boltzmann code \cite{Blas11} and provide new dark matter transfer functions.

\begin{figure}[t]
\includegraphics[width=\columnwidth]{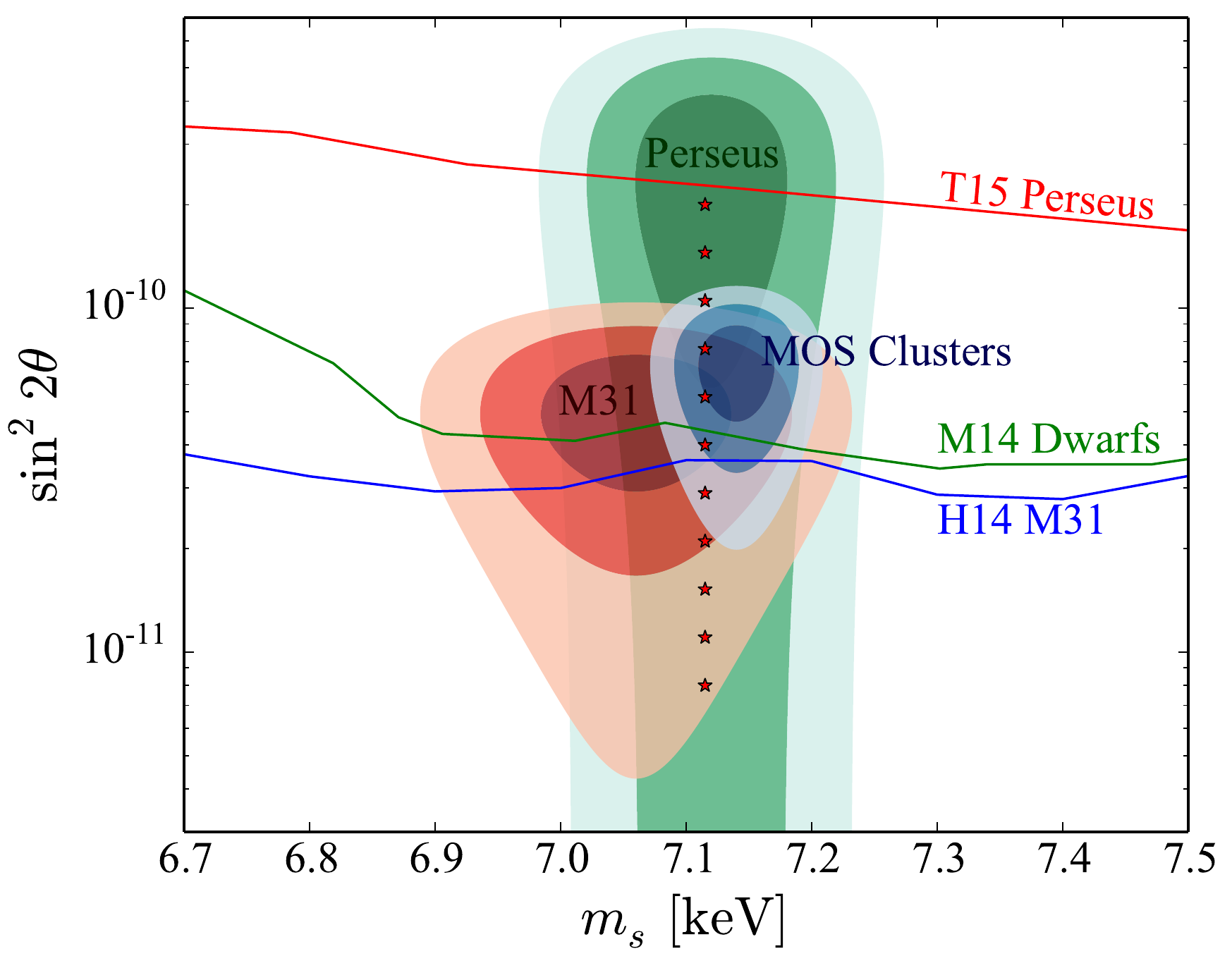}
\caption{Sterile neutrino DM parameter space: shaded regions are consistent with the X-ray signal at $1,2$ and $3$ $\sigma$. The best determined parameters are from the MOS stacked clusters of Ref.~\cite{bulbul:2014sua}. Statistically consistent signals are found in their core-removed Perseus spectrum, and M31 ~\cite{boyarsky:2014jta}. The lines show constraints at the $90\%$ level from {\it Chandra} observations of M31 (H14) \cite{horiuchi:2013noa}, stacked dwarf galaxies (M14) \cite{malyshev:2014xqa}, and Suzaku observations of Perseus (T15) \cite{tamura:2014mta}. Stars mark the models that we study in the body of the paper. \label{fig:parspace}}
\end{figure} 

We organize the paper such that the beginning sections deal with SM physics, while the later ones apply their results to sterile neutrino DM. We introduce the production mechanism in Sec.~\ref{sec:overview}. We then study the asymmetry redistribution in Sec.~\ref{sec:redistribution} and active neutrino opacities in Sec.~\ref{sec:scattering}. Finally, we apply these results to sterile neutrino production in Sec.~\ref{sec:production}, evaluate transfer functions for matter fluctuations in Sec.~\ref{sec:transferfunc}, and finish with a discussion of our assumptions and uncertainties in Sec.~\ref{sec:discussion}. We collect technical details into the appendices.

\section{Overview of resonant sterile neutrino production}
\label{sec:overview}

In this section, we briefly review the resonant production of sterile neutrinos in the early Universe. We first present the specific scenario that we consider in this work, and then discuss the Boltzmann formalism used to compute the out-of-equilibrium production of sterile neutrinos. We finally discuss how the presence of the thermal bath and lepton asymmetry change the neutrino self-energy and govern the sterile neutrino production. We refer the reader to Refs.~\cite{Shi99, Abazajian01, Kishimoto08} for more details.

\subsection{Assumptions}
\label{subsec:assumptions}

In our current study, we focus on the following scenario.
\begin{enumerate}
  \item We consider an extra sterile neutrino species, $\nu_s$, that is massive compared to the SM neutrinos $\nu_{e/\mu/\tau}$, which we take to be effectively massless. The propagation (light/heavy) and interaction (active/sterile) eigenstates are related by a unitary transformation, the most general version of which is a $4 \times 4$ matrix. We assume that the sterile neutrino mixes with only one of the SM ones, which we take to be the muon neutrino, i.e., 
    \begin{align}
      \begin{pmatrix}
	\Psi_{\nu_\mu} \\
	\Psi_{\nu_{\rm s}} 
      \end{pmatrix} & = 
      \begin{pmatrix}
	\cos{\theta} & \sin{\theta} \\
	-\sin{\theta} & \cos{\theta} 
      \end{pmatrix}
      \begin{pmatrix}
	\Psi_0 \\
	\Psi_{m_{\rm s}}
      \end{pmatrix} \mbox{.} \label{eq:mixingdef}
    \end{align}
    The fields on the left- and right-hand sides are interaction and mass ($m_{\rm s}$) eigenstates, respectively, and $\theta$ is the active-sterile mixing angle. The choice of a muon neutrino is arbitrary, and reflects the choice of previous work \cite{Abazajian01,Abazajian:2014gza}.
  \item We assign a non-zero lepton asymmetry to the primordial plasma. In the general case, each SM flavor has its own asymmetry, but we assume a non-zero value only for the mu flavor (i.e. the one that mixes with the sterile neutrino):
    \begin{align}
      \Delta \hat{n}_{\nu_\alpha} + \Delta \hat{n}_{\alpha^-} \equiv \hat{\mathcal{L}}_\alpha = \delta_{\alpha \mu} \hat{\mathcal{L}}_\mu \mbox{,} \label{eq:asymmdef}
    \end{align}
    where the dimensionless asymmetry $\Delta \hat{n}_A$ in species $A$ is the temperature-scaled difference between the particle and anti-particle densities, $\Delta \hat{n}_A \equiv (n_{A} - n_{\bar{A}})/T^3$, and $\delta_{\alpha \mu}$ is the Kronecker delta. In general, entropy-scaled asymmetries are preferable, since they are conserved through epochs of annihilation. However, the definition used in Eq.~\eqref{eq:asymmdef} simplifies comparison with lattice QCD calculations in Sec.~\ref{sec:redistribution}. We fix by hand the mu lepton asymmetry at high temperatures to produce the canonical DM density, $\Omega_{\rm DM} h^2 =0.1188$ in the current epoch \cite{PlanckCosmology}.
\end{enumerate}

In the rest of the paper, we use a hat to indicate temperature scaled quantities. We choose to study the parameter space of interest for resonantly produced sterile neutrino DM consistent with the recent X-ray signal. Figure \ref{fig:parspace} shows a section of the $m_{\rm s}$ and $\sin^2{2\theta}$ plane with contours for the unidentified lines of Refs.~\cite{bulbul:2014sua,boyarsky:2014jta}, along with constraints from {\it Chandra} observations of M31 \cite{horiuchi:2013noa}, stacked dwarf galaxies \cite{malyshev:2014xqa}, and Suzaku observations of Perseus \cite{tamura:2014mta}. The stars show a range of mixing angles at a specific value of $m_s$, and mark models that we study in Sections \ref{sec:production} and \ref{sec:transferfunc}.

For all these models, the bulk of the sterile neutrinos are produced at temperatures well below the masses of the weak gauge bosons ($\sim 80$ GeV), but above weak decoupling at $T \sim 1.5$ MeV \cite{Kishimoto08}. Active-active neutrino oscillations in the primordial plasma are suppressed at these temperatures \cite{Abazajian02}, hence it is consistent to assign individual asymmetries in Eq.~\eqref{eq:asymmdef} and neglect electron and tau neutrino mixing in Eq.~\eqref{eq:mixingdef}.

\subsection{Boltzmann Formalism}
\label{subsec:boltzmann}

In its full generality, out-of-equilibrium sterile neutrino production (via oscillations) is best described by the evolution of the two-state density matrix of the neutrinos in the active--sterile (interaction) basis \cite{Harris82,Stodolsky87,Raffelt92,Raffelt93}.

For the parameter range in Fig.~\ref{fig:parspace}, most sterile neutrinos are produced above temperatures $T \gtrsim 100$ MeV. At these temperatures, the two state system is collisionally dominated, i.e.~the interaction contribution dominates the vacuum oscillations. In this regime, the evolution of the density matrix separates out and yields a quasi-classical Boltzmann transport equation for the diagonal terms, which are the PSDs of the active and sterile components \cite{Bell99,Volkas00,Lee00}. The Boltzmann equation for the sterile neutrino PSD is

\begin{widetext}
\begin{align}
  &\frac{\partial}{\partial{t}}f_{\nu_{\rm s}}(p,t) - H\,p\,
  \frac{\partial}{\partial p }f_{\nu_{\rm s}}(p,t) = \nonumber\\
&\quad \sum_{\nu_x + a + \dots \rightarrow i + \dots} \int{\frac{d^3 p_a}{(2\pi)^3 2 E_a} \dots \frac{d^3 p_i}{(2\pi)^3 2 E_i} \dots (2\pi)^4 \delta^4(p+p_a + \dots - p_i - \dots)}\nonumber\\
&\qquad\qquad\qquad \times \frac12 \left[ \langle P_{\rm m}(\nu_\mu \rightarrow \nu_{\rm s}; p,t) \rangle \left(1-f_{\nu_{\rm s}}\right) \sum |\mathcal{M}|^2_{i + \dots \rightarrow a + \nu_\mu + \dots} f_i \dots \left(1 \mp f_a \right) \left( 1 - f_{\nu_\mu} \right) \dots \right. \nonumber\\
& \qquad\qquad\qquad\qquad ~ \left. - \langle P_{\rm m}(\nu_{\rm s} \rightarrow \nu_\mu; p,t) \rangle f_{\nu_{\rm s}}\left( 1 - f_{\nu_\mu} \right) \sum |\mathcal{M}|^2_{\nu_\mu + a + \dots \rightarrow i + \dots} f_a \dots \left( 1 \mp f_i \right) \dots \right] \mbox{.}
\label{eq:classicalboltz}
\end{align}
\end{widetext}

We can write an analogous equation for the antineutrinos. Here, the $f(p)$ are PSDs for particles with three-momentum $p$ and energy $E$, and $H$ is the Hubble expansion rate. The right-hand side sums over all reactions that consume or produce a muon neutrino. The symbol $\sum |\mathcal M|^2$ denotes the squared and spin-summed matrix element for the reaction, and the multiplicative factors of $(1 \mp f)$ implement Pauli blocking/Bose enhancement respectively. The factor of $1/2$ accounts for the fact that only one (i.e. the muon neutrino) state in the two-state system interacts \cite{Harris82,Stodolsky87,Raffelt92}. The $P_{\rm m}$ are active--sterile oscillation probabilities in matter, which depend on the vacuum mixing angle $\theta$, and are modified by interactions with the medium. The latter are parametrized by the neutrino self energy \cite{Notzold88}, and the `quantum damping' rate for active neutrinos. In terms of these quantities, the oscillation probabilities are \cite{Volkas00,Lee00}
\begin{multline}
  \langle P_{\rm m}(\nu_\mu \leftrightarrow \nu_{\rm s}; p,t) \rangle 
   = \, (1/2) \Delta^2 (p) \sin^2{2\theta} \\
   \times \Bigl\{ \Delta^2 (p) \sin^2{2\theta} + D^2(p) \\
   + \left[\Delta (p) \cos 2\theta - V^{\rm L} - V^{\rm th}(p)\right]^2 \Bigr\}^{-1} \mbox{.} \label{eq:probabilities}
\end{multline}
We have introduced the symbol $\Delta(p)$ for the vacuum oscillation rate, $\Delta(p) \equiv \delta m^2_{\nu_\mu,\nu_{\rm s}}/2p$, and split the neutrino self energy into the lepton asymmetry potential $V^{\rm L}$, and the thermal potential $V^{\rm th}$ (the asymmetry contribution enters with the opposite sign in the version of Eq.~\eqref{eq:probabilities} for antineutrinos). The quantity $D(p)$ is the quantum damping rate, and equals half the net interaction rate of active neutrinos [the factor of half enters for the same reason as it does in Eq.~\eqref{eq:classicalboltz}]. The net interaction rate for a muon neutrino is
\begin{align}
  ~~~ & \!\!\!\!
  \Gamma_{\nu_\mu}(p) \nonumber \\
  & = \!\! \sum_{\nu_x + a + \dots \rightarrow i + \dots} \!\! \int \frac{d^3 p_a}{(2\pi)^3 2 E_a} \dots \frac{d^3 p_i}{(2\pi)^3 2 E_i} \dots \nonumber \\
  & \qquad \qquad \qquad \!\! \times (2\pi)^4 \delta^4(p+p_a + \dots - p_i - \dots) \nonumber \\
  & \qquad \qquad \qquad \!\! \times \sum |\mathcal{M}|^2_{\nu_\mu + a + \dots \rightarrow i + \dots} f_a \dots \left( 1 \mp f_i \right) \dots \label{eq:scatteringrate}
\end{align}
We simplify the phase-space integrals in Eq.~\eqref{eq:classicalboltz} by using detailed balance to equate the forward and backward reaction rates. The resulting Boltzmann equation for quantum-damped and collisionally-driven sterile neutrino production is \cite{Abazajian01}

\begin{align}
  ~ ~ ~ & \!\!\!\!
  \frac{\partial}{\partial{t}}f_{\nu_{\rm s}}(p,t) - H\, p\,
\frac{\partial}{\partial p }f_{\nu_{\rm s}}(p,t) \nonumber \\
& \approx
 \frac{\Gamma_{\nu_\mu}(p)}{2} \langle P_m(\nu_\mu
\leftrightarrow \nu_{\rm s}; p, t) \rangle
\left[f_{\nu_\mu}(p,t) - f_{\nu_{\rm s}}(p,t)\right] \mbox{,}
\label{eq:productionmaster}
\end{align}
with a related equation for antineutrinos. There are subtleties with the effects of quantum-damping in the case of resonance \cite{boyanovsky:2006it}, but tests with the full density matrix formalism find that the quasi-classical treatment is appropriate \cite{Kishimoto08}.

\subsection{Asymmetry and Thermal Potentials}
We now expand on the origins of the asymmetry and thermal potentials appearing in Eq.~\eqref{eq:probabilities}. These potentials encapsulate the self energy of propagating active neutrinos due to interactions with the plasma. Under the conditions we are interested in, there are three contributions to the neutrino self energy: a) an imaginary part proportional to the net neutrino opacity, b) a real part due to finite weak gauge boson masses ($V^{\rm th}$), and c) a real part proportional to asymmetries in weakly interacting particles ($V^{\rm L}$). We follow the treatment in Ref.~\cite{Notzold88}, and recast it in terms of the quantities that we compute later.

Figure \ref{fig:oneloop} shows lowest-order contributions to active neutrinos' self energy. Thick red lines are thermal propagators of weakly charged species in the background plasma. There are two corrections - bubbles and tadpoles, shown in Fig.~\ref{fig:bubble} and \ref{fig:tadpole} respectively. The background fermion is a lepton of the same flavor in the former, and any weakly charged species in the latter. 

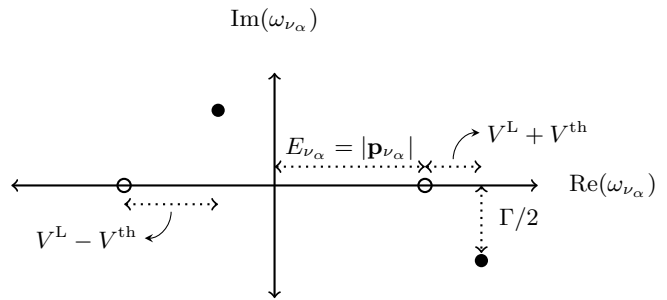
\begin{figure}[t]
  \begin{tikzpicture}
    \draw[black, thick, <->] (-3.5,0) -- (3.5,0);
    \node at (4.5, 0) {${\rm Re}(\omega_{\nu_\alpha})$};
    \draw[black, thick, <->] (0,-1.5) -- (0,1.5);
    \node at (0, 2.2) {${\rm Im}(\omega_{\nu_\alpha})$};

    \draw [black, thick] (2, 0) circle (2.5pt);
    \draw [black, thick] (-2, 0) circle (2.5pt);

    \fill[black] (2.75, -1) circle (2.5pt);
    \fill[black] (-0.75, 1) circle (2.5pt);
    
    \draw[black, thick, dotted, <->] (0,0.25) -- (2,0.25);
    \node at (1, 0.5) {$E_{\nu_\alpha} = \vert \bm p_{\nu_\alpha} \vert$};
    \draw[black, thick, dotted, <->] (2,0.25) -- (2.75,0.25);
    \draw[black, -stealth] (2.375,0.35) arc (180:90:10pt);
    \node at (3.5,0.7) {$V^{\rm L} + V^{\rm th}$};
    
    \draw[black, thick, dotted, <->] (2.75,0) -- (2.75,-0.9);
    \node at (3.25,-0.45) {$\Gamma/2$};

    \draw[black, thick, dotted, <->] (-2,-0.25) -- (-0.75,-0.25);
    \draw[black, -stealth] (-1.375,-0.35) arc (0:-90:10pt);
    \node at (-2.5,-0.7) {$V^{\rm L} - V^{\rm th}$};

  \end{tikzpicture}
  \caption{\label{fig:poleshift} Matter potentials for massless neutrinos in the plasma's rest frame: filled and un-filled circles are poles at finite and zero temperature, respectively. See Sec.~\ref{sec:scattering} for the imaginary shift.}
\end{figure}

A massless active neutrino's `dressed' propagator is
\begin{subequations}
\begin{align}
  G_{\nu_\alpha}^{-1}(p_{\nu_\alpha}) & = {\not}{p_{\nu_\alpha}} - b_{\nu_\alpha}(p_{\nu_\alpha}) {\not}u \left( 1 - \gamma_5 \right)/2 \mbox{,} \label{eq:propcorrect} \\
  b_{\nu_\alpha}(p_{\nu_\alpha}) & = b_{\nu_\alpha}^{(0)} + b_{\nu_\alpha}^{(1)} \omega_{\nu_\alpha} \mbox{,} \qquad \omega_{\nu_\alpha} = - p_{\nu_\alpha} \cdot u \mbox{.} \label{eq:vsplit}
\end{align}
\label{eq:polecorrect}
\end{subequations}
Here, $p_{\nu_\alpha}$ and $u$ are the neutrino and plasma's four-momenta, ${\not}v$ is shorthand for $\gamma^{\mu} v_{\mu}$, and $b_{\nu_\alpha}$ is the left handed neutrino's self energy. Equation \eqref{eq:vsplit} divides this self energy into two contributions that affect the particle and anti-particle poles of Eq.~\eqref{eq:propcorrect} differently. Figure \ref{fig:poleshift} illustrates their association with asymmetry and thermal potentials:
\begin{align}
   V^{\rm L}_{\nu_\alpha} & = b_{\nu_\alpha}^{(0)}, \\
   V^{\rm th}_{\nu_\alpha}(E_{\nu_\alpha}) & = b_{\nu_\alpha}^{(1)} E_{\nu_\alpha}.
\end{align}
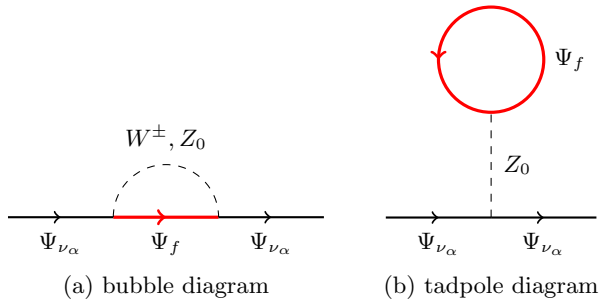
\begin{figure}[t]
  \hspace*{\fill}%
  \subfloat[\label{fig:bubble}bubble diagram]{
    \begin{tikzpicture}[scale=0.7]
      \begin{scope}[decoration={
	markings,
	mark=at position 0.5 with {\arrow{>}} }
      ]
      \draw[black, thick, postaction={decorate}] (-3, 0) -- (-1, 0);
      \node at (-2,-0.5) {$\Psi_{\nu_\alpha}$};
      \draw[red, very thick, postaction={decorate}] (-1, 0) -- (1, 0);
      \draw[black, thick, postaction={decorate}] (1, 0) -- (3, 0);
      \node at (2,-0.5) {$\Psi_{\nu_\alpha}$};
      \draw[black, dashed] (-1, 0) arc (180:0:1cm);
      \node at (0, 1.5) {$W^\pm,Z_0$};
      \node at (0, -0.5) {$\Psi_f$};
      \end{scope}
    \end{tikzpicture}
  }  
  \hfill 
  \subfloat[\label{fig:tadpole}tadpole diagram]{
    \begin{tikzpicture}[scale=0.7]
    \begin{scope}[decoration={
	markings,
	mark=at position 0.5 with {\arrow{>}} }
      ]
      \draw[black, thick, postaction={decorate}] (-2, 0) -- (0, 0);
      \node at (-1,-0.5) {$\Psi_{\nu_\alpha}$};
      \draw[black, thick, postaction={decorate}] (0, 0) -- (2, 0);
      \node at (1,-0.5) {$\Psi_{\nu_\alpha}$};
      \draw[black, dashed] (0, 0) -- (0, 2);
      \node at (0.5, 1) {$Z_0$};
      \draw[red, very thick, postaction={decorate}] (0, 3) circle (1cm);
      \node at (1.5, 3) {$\Psi_f$};
    \end{scope}
    \end{tikzpicture}
  }
  \caption{\label{fig:oneloop}Lowest order contributions to a propagating active neutrino's self energy. Red lines are thermal propagators. In (a), $f$ is any species with weak charge. In (b), $f = \nu_\alpha, \alpha^-$.} 
\end{figure}
Ref.~\cite{Notzold88} computes these terms by summing over all species in Fig.~\ref{fig:oneloop}. Both kinds of diagrams contribute to the asymmetry potential, while only bubble diagrams contribute to the thermal potential. We write the answer in terms of the leptons' asymmetries, and the densities of the strong fluid's conserved quantities:
\begin{subequations}
\label{eq:potentials}
\begin{align}
  V^{\rm L}_{\nu_\alpha} & = \sqrt{2} G_{\rm F} \Bigl[ \sum_{\beta \in \{e,\mu,\tau\}} \!\!\! \bigl( \delta_{\alpha \beta} - \frac{1}{2} + 2\sin^2{\theta_W} \bigr) \Delta n_{\beta^-} \nonumber \\
    & \hspace{45pt} + \!\!\!\!\!\! \sum_{\beta \in \{e,\mu,\tau\}} \!\!\!\!\!\! \left( 1 + \delta_{\alpha \beta} \right) \Delta n_{\nu_\beta} - \frac{1}{2} \Delta n_{\rm B} \nonumber \\
    & \hspace{45pt} + \left( 1 - 2 \sin^2{\theta_W} \right) \Delta n_{\rm Q} \Bigr] \mbox{,} \label{eq:asymmv} \\
    V^{\rm th}_{\nu_\alpha}(E_{\nu_\alpha}) & = - \frac{8\sqrt{2} G_{\rm F}}{3} \left[ \frac{\rho_{\nu_\alpha}}{M_Z^2} + \frac{\rho_{\alpha}}{M_W^2} \right] E_{\nu_\alpha} \mbox{.} \label{eq:thermalv}
\end{align}
\end{subequations}
In the above equations, $\theta_{\rm W}$ is the weak mixing angle, and $M_{Z/W}$ are the masses of the weak gauge bosons. The symbol $\delta_{\alpha \beta}$ is a Kronecker delta, the quantities $\rho_\alpha$ and $\rho_{\nu_\alpha}$ are net energy densities of charged and neutral leptons, respectively, and $\Delta n_{\rm B}$ and $\Delta n_{\rm Q}$ are densities of the baryon number B, and electric charge Q, respectively. The standard model baryon number asymmetry is small compared to the lepton asymmetry of interest \cite{BBN2015}, hence can be set to zero for the purposes of this calculation. 

According to the assumptions in the first part of this section, the plasma starts out with a net lepton asymmetry in the mu flavor. As we showed in Sec.~\ref{sec:introduction}, this asymmetry is redistributed between muons and muon neutrinos. Moreover leptons of other flavors acquire asymmetries that respect Eq.~\eqref{eq:asymmdef}, and the strong fluid acquires a net electric charge density $\Delta n_{\rm Q}$ to maintain overall neutrality. Equation \eqref{eq:asymmv} shows how the asymmetry potential depends on the redistributed asymmetries, which we study in the ensuing section.
%


\section{Redistribution of an input asymmetry}
\label{sec:redistribution}

Weak processes couple leptonic and hadronic degrees of freedom in the primordial plasma. In this section, we study this coupling's effect on lepton asymmetries\footnote{During this preparation of this manuscript, we became aware of Ref.~\cite{2015arXiv150606752G}, which points out the relevance of this effect to sterile neutrino production, and estimates it under the simplifying Stefan-Boltzmann approximation for free quarks.}. We define relevant suceptibilities in Sec.~\ref{subsec:definitions}, and compute them over a range of temperatures in Sec.~\ref{subsec:susceptibilitycalc}. 

\subsection{Definitions and parametrization}
\label{subsec:definitions}

Let us consider the primordial plasma at temperatures above the quark-hadron transition temperature, $T_{\rm QCD}$. The following reactions couple leptons of different flavors, and the quark and lepton sectors:
\begin{subequations}
  \label{eq:redistribution}
  \begin{align}
    \nu_{\alpha} + \beta^{-} & \rightleftharpoons \nu_\beta + \alpha^{-} \mbox{,} \label{eq:invmuondecay} \\
    \nu_{\alpha} + \alpha^{+} & \rightleftharpoons a + \bar{b} \mbox{,} \label{eq:electronpion}
\end{align}
\end{subequations}
where $a$ and $b$ are quarks with charges of $+2/3$ and $-1/3$ respectively. Free quarks no longer exist at temperatures below $T_{\rm QCD}$, and the reactions in Eq.~\eqref{eq:electronpion} transition to ones involving mesons, like Eq.~\eqref{eq:piondecay}. 

In principle, we could study the effect of all these reactions on input asymmetries, but it is a daunting task; one that is further complicated by the quark-hadron transition. The following consideration of the relevant timescales suggests a solution. In the radiation dominated era, the Hubble rate is $H \approx 2 \times 10^5 ~ {\rm s}^{-1} ~ {g_*}^{1/2} (T/{\rm GeV})^2$. At temperatures above the quark-hadron transition, the rates of reactions in \eqref{eq:redistribution} are $\Gamma(T) \simeq G_{\rm F}^2 T^5 \approx 2 \times 10^{14} ~ s^{-1} ~ (T/{\rm GeV})^5$, while the relevant rates below the transition are those of pion decays. The most important channel for the latter is the muonic decay, $\pi^+ \rightarrow \mu^+ + \nu_{\mu}$, which is faster than the Hubble rate ($\Gamma_{\pi^+ \rightarrow \mu^+ + \nu_{\mu}} = 3.8 \times 10^7 ~ {\rm s}^{-1}$). Thus, a significant number of the reactions in Eq.~\eqref{eq:redistribution} are faster than the Hubble rate\footnote{The electronic channel for the pion decay, $\pi^+ \rightarrow e^+ + \nu_{e}$ is helicity-suppressed ($\Gamma_{\pi^+ \rightarrow e^+ + \nu_{e}} = 4.7 \times 10^3 ~ {\rm s}^{-1}$) and of the order of the Hubble rate at temperatures $T \simeq 50$ MeV, hence one might worry that leptons with electronic flavor depart from equilibrium. This is resolved by the observation that they are coupled to muonic species by other non-helicity suppressed, and consequently faster, reactions such as $e^+ + \nu_e \leftrightarrow \mu^+ + \nu_\mu$ and  $\mu^+ \leftrightarrow e^+ + \nu_e + \bar{\nu}_\mu$.}.

This has two primary consequences. Firstly, high reaction rates enforce {\em kinetic} equilibrium, i.e. all active species' PSDs approach the Fermi-Dirac or Bose-Einstein forms. Secondly, forward and backward reactions are in {\em chemical} equilibrium, one effect of which is to equate the chemical potentials for both sides (the Saha equation). However, it has another implication -- the plasma's complicated internal dynamics can be abstracted into a few parameters or susceptibilities that completely specify its response to small external `forces', or in this case, input asymmetries. All that remains is to compute the susceptibilities relevant to our problem.

We now define a few useful quantities and notation. Given any conserved quantity $F$, the symbol $\mu_F$ denotes its chemical potential. The asymmetry $\Delta \hat{n}_A$, in a particle $A$, is a function of its chemical potential $\hat{\mu}_A \equiv \mu_A/T$. The quantities $\Delta \hat{n}_A$ and $\hat{\mu}_A$ are small, and in the linearized limit, related by
\begin{equation}
  \Delta \hat{n}_A = \hat{\chi}_A \hat{\mu}_A \mbox{,} \label{eq:susceptibility}
\end{equation}
where $\hat{\chi}_A \equiv \chi_A/T^2$ is the number-density susceptibility. The lepton asymmetries in the three flavors are
\begin{align}
  \hat{\mathcal{L}}_{\alpha} & = \Delta \hat{n}_{\alpha^-} + \Delta \hat{n}_{\nu_\alpha} \nonumber \\
  & = \hat{\chi}_{\alpha^-} \hat{\mu}_{\alpha^-} + \hat{\chi}_{\nu_\alpha} \hat{\mu}_{\nu_\alpha}, \qquad \alpha \in \{e, \mu, \tau \} \mbox{.} \label{eq:leptasym}
\end{align}
The strong fluid is described by the densities of its conserved quantities: the charge and baryon-number densities $\Delta \hat{n}_{\rm Q}$ and $\Delta \hat{n}_{\rm B}$, respectively\footnote{We do not follow the strangeness, S. since it is not conserved in weak reactions. Above the transitions, it disappears at the Cabbibo--suppressed rate $\Gamma_{\rm S} \simeq \vert V_{\rm u \rm s} \vert^2 G_{\rm F}^2 T^5 \approx 10^{13} s^{-1} (T/{\rm GeV})^5$, while below the transition the relevant rate is the Kaon inverse lifetime, $\Gamma_{K^\pm} = 8.1 \times 10^{7} \ {\rm s}^{-1}$.}. The chemical equilibrium of the reactions in Eq.~\eqref{eq:redistribution} implies
\begin{align}
  \hat{\mu}_{\nu_\alpha} - \hat{\mu}_{\alpha^-} - \hat{\mu}_{\rm Q} = 0, \qquad \alpha \in \{e, \mu, \tau \} \mbox{.} \label{eq:saha}
\end{align}
Here $\hat{\mu}_{\rm Q}$ is the chemical potential for adding a unit of electric charge. The conserved quantities' densities are related to their chemical potentials by their susceptibilities:
\begin{align}
  \begin{pmatrix} 
    \Delta \hat{n}_{\rm Q} \\ 
    \Delta \hat{n}_{\rm B} 
  \end{pmatrix}
  =
  \begin{pmatrix}
    \hat{\chi}^{\rm Q}_2 & \hat{\chi}^{\rm QB}_{11} \\
    \hat{\chi}^{\rm BQ}_{11} & \hat{\chi}^{\rm B}_2
  \end{pmatrix}
  \begin{pmatrix}
    \hat{\mu}_{\rm Q} \\
    \hat{\mu}_{\rm B}
  \end{pmatrix} \mbox{.} 
  \label{eq:chidef}
\end{align}
Equation ~\eqref{eq:chidef}, along with net charge and baryon number conservation, yields the constraint equations
\begin{align}
  \Delta \hat{n}_{\rm B} & = \hat{\chi}^{\rm BQ}_{11} \hat{\mu}_{\rm Q} + \hat{\chi}^{\rm B}_2 \hat{\mu}_{\rm B} \approx 0 \mbox{,} \label{eq:baryonnumber} \\
  0 & = \hat{\chi}^{\rm Q}_2 \hat{\mu}_{\rm Q} + \hat{\chi}^{\rm QB}_{11} \hat{\mu}_{\rm B} - \!\!\!\! \sum_{ \alpha \in \{e, \mu, \tau\} } \!\!\!\! \Delta \hat{n}_{\alpha^-} \nonumber \\
  & = \hat{\chi}^{\rm Q}_2 \hat{\mu}_{\rm Q} + \hat{\chi}^{\rm QB}_{11} \hat{\mu}_{\rm B} - \!\!\!\! \sum_{ \alpha \in \{e, \mu, \tau\} } \!\!\!\! \hat{\chi}_{\alpha^-} \hat{\mu}_{\alpha^-} \mbox{.} \label{eq:charge}
\end{align}
Equations ~\eqref{eq:leptasym}, \eqref{eq:saha}, \eqref{eq:baryonnumber} and \eqref{eq:charge} are eight linear equations for eight unknowns. The resulting asymmetries (obtained via their chemical potentials) are the `redistributed' input lepton asymmetries $\mathcal{L}_\alpha$.

We symbolically represent the solutions as
\begin{align}
  \hat{\mu}_{A} & = \sum_{\alpha \in \{e, \mu, \tau\}} \frac{\partial \hat{\mu}_A}{\partial \hat{\mathcal{L}}_\alpha} \hat{\mathcal{L}}_\alpha \mbox{,} \label{eq:symbolicsoln}
\end{align}
where the coefficients $(\partial \hat{\mu}_A/\partial \hat{\mathcal{L}}_\alpha)$ depend on the susceptibilities of both the leptons and the strong fluid. We also express the redistributed asymmetries as
\begin{align}
  \Delta \hat{n}_A 
  = \sum_{\alpha \in \{e, \mu, \tau\}} \frac{\partial \Delta \hat{n}_A}{\partial \hat{\mathcal{L}}_\alpha} \hat{\mathcal{L}}_\alpha  
  = \sum_{\alpha \in \{e, \mu, \tau\}} \hat{\chi}_A \frac{\partial \hat{\mu}_A}{\partial \hat{\mathcal{L}}_\alpha} \hat{\mathcal{L}}_\alpha \mbox{.}\label{eq:redistcoeff}
\end{align}
At the temperatures of interest, the lepton susceptibilities are essentially given by the free particle, or Stefan-Boltzmann, formula:
\begin{align}
  \hat{\chi}_A (\hat{m}_A) & = -\frac{g_{A}}{\pi^2} \int_0^\infty d\hat{p} \, \hat{p}^2 \hat{n}_{\rm F}^\prime \left(\sqrt{\hat{p}^2 + \hat{m}_A^2}\right) \mbox{.} \label{eq:freeparticlechi}
\end{align}
In this equation, $g_A$ and $\hat{m}_A \equiv m_A/T$ are the spin degeneracy and mass respectively, $\hat{n}_{\rm F}^\prime(x)=(d/dx)\{1/[\exp{(x)}+1]\}$ is the derivative of the Fermi-Dirac distribution. The strongly interacting fluid's susceptibilities are considerably more complicated, especially near the quark-hadron transition. We evaluate them using a number of techniques: perturbative quantum chromodynamics (QCD) at high temperatures, matching to lattice QCD results near the transition, and a hadron resonance gas (HRG) approximation at low temperatures.

\subsection{Susceptibilities of the strongly-interacting plasma}
\label{subsec:susceptibilitycalc}
In this section, we compute the strongly-interacting plasma's susceptibilities to baryon number and electric charge fluctuations. The susceptibilities are the following derivatives of the QCD pressure
\be\label{eq:def_susceptibility2}
\hat{\chi}_2^X = \frac{\pa^2 \hat{p}_{\rm QCD}}{\pa\hat{\mu}^2_X}\bigg|_{\hat{\mu}_X=0},
\ee
\be\label{eq:def_susceptibility11}
\hat{\chi}_{11}^{XY} = \frac{\pa^2 \hat{p}_{\rm QCD}}{\pa\hat{\mu}_X\pa\hat{\mu}_Y}\bigg|_{\hat{\mu}_X,\hat{\mu}_Y=0},
\ee
where $X,\,Y\in \{\rm B,\, \rm Q\}$, $\hat{\mu}_X \equiv \mu_X/T$ is the chemical potential of the conserved charge $X$, and the pressure $\hat{p}_{\rm QCD}$ is given by the logarithm of the QCD partition function $Z_{\rm QCD}$.
\be\label{eq:pQCD_from_partitionZ}
\hat{p}_{\rm QCD} \equiv \frac{p_{\rm QCD}}{T^4} = \frac{1}{VT^3}\ln{ Z_{\rm QCD}(V,T,\mu_{\rm Q},\mu_{\rm B})},
\ee
where $V$ is the volume. In Eq.~(\ref{eq:def_susceptibility11}), the off-diagonal term $\hat{\chi}_{11}^{XY}$ encodes the \emph{correlation} between the fluctuations of conserved charges $X$ and $Y$. Note that the susceptibilities in Eqs.~(\ref{eq:def_susceptibility2}) and (\ref{eq:def_susceptibility11}) are dimensionless. 

The sterile neutrino production calculation carried in the present work requires knowledge of these susceptibilities over a broad range of temperatures, both above and below the quark-hadron transition. At very high temperatures $T\gg T_{\rm QCD}$, the QCD pressure can be computed using a standard perturbative approach, while at intermediate temperatures $T\simeq T_{\rm QCD}$, perturbative techniques become inadequate; we must rely on lattice calculations (see e.g.~\cite{Borsanyi12,Bazavov12}) to compute susceptibilities through the quark-hadron transition. At low temperatures $T < T_{\rm QCD}$, we compute the QCD pressure using the hadron resonance gas (HRG) model \cite{Hagedorn:1984hz,2013NuPhS.234..313M}, which approximates the QCD partition function as a sum over all known hadronic resonances. Our strategy to compute the susceptibility tensor over the whole required range of temperatures is as follows: we first separately calculate it both above and below the quark-hadron transition using either perturbative or HRG techniques, and then smoothly join the results with those from lattice QCD computations in the regions of overlap.
\subsubsection{High-Temperature Limit: Perturbative Approach}\label{sec:highT}
We follow the approach of Ref.~\cite{Laine06} to perturbatively compute the QCD pressure and its derivative up to order $\mathcal{O}(g_{\rm s}^2)$, where $g_{\rm s}$ is the standard QCD gauge coupling constant. The starting point is to write the QCD pressure as 
\be\label{eq:QCD_pressure_highT}
\hat{p}_{\rm QCD} = \alpha_{\rm E1}^{\overline{\rm MS}} + \tilde{g}^2_3\alpha_{\rm E2}^{\overline{\rm MS}}, 
\ee
where $\tilde{g}_3$ is the effective gauge coupling
\be
\tilde{g}_3^2 = g_{\rm s}^2 + \frac{g_{\rm s}^4}{(4\pi)^2}\alpha_{\rm E7}^{\overline{\rm MS}},
\ee
and the functions $\alpha_{{\rm E}n}^{\overline{\rm MS}}$ are given by
\be
\alpha_{\rm E1}^{\overline{\rm MS}} = d_{\rm A}\frac{\pi^2}{45} + 4 C_{\rm A} \sum_{i=1}^{N_{\rm f}}F_1\left(\hat{m}_i^2,\hat{\mu}_i\right),
\ee
\ba
\alpha_{\rm E2}^{\overline{\rm MS}} &=&-d_{\rm A}  \sum_{i=1}^{N_{\rm f}}\Bigg\{\frac{1}{6}F_2\left(\hat{m}_i^2,\hat{\mu}_i\right)\left[1+6F_2\left(\hat{m}_i^2,\hat{\mu}_i\right)\right]\en
&&\qquad+\frac{\hat{m}_i^2}{4\pi^2}\left(3\ln{\frac{\bar{\mu}}{m_i}}+2\right)F_2\left(\hat{m}_i^2,\hat{\mu}_i\right)\en
&&\qquad\qquad-2\hat{m}_iF_4\left(\hat{m}_i^2,\hat{\mu}_i\right)\Bigg\}-\frac{d_{\rm A} C_{\rm A}}{144},
\ea
\ba
\alpha_{\rm E7}^{\overline{\rm MS}} &=&\frac{22 C_{\rm A}}{3}\left[\ln{\left(\frac{\bar{\mu}e^{\gamma_{\rm E}} }{4\pi T}\right)}+\frac{1}{22}\right]\en
&&\qquad-\frac{2}{3} \sum_{i=1}^{N_{\rm f}}\left[2\ln{\frac{\bar{\mu}}{m_i}}+F_3\left(\hat{m}_i^2,\hat{\mu}_i\right)\right].
\ea
Here, $d_{\rm A} \equiv N_{\rm c}^2-1$ and $C_{\rm A} \equiv N_{\rm c}$ stand for the gauge-group constants for the adjoint and fundamental representation of $SU(N_{\rm c})$, respectively. In this work, we adopt the standard value of $N_{\rm c}=3$. In the above, $N_{\rm f}$ is the number of quark flavors, $\bar{\mu}$ is the energy scale at which the masses and the coupling constant are evaluated (not to be confused with the chemical potentials), and $\gamma_{\rm E}$ is the Euler-Mascheroni constant. The functions $F_1,\ldots,F_4$ are given in appendix \ref{app:highT_funcs}.

We also need a prescription for the running of the coupling constant $g_{\rm s}$ and of the quark masses with the energy scale $\bar{\mu}$. As in Ref.~\cite{Laine06}, we adopt a simple 1-loop running which yields
\be
g^2_{\rm s}(\bar{\mu}) = \frac{24\pi^2}{(11 C_{\rm A} -4 T_{\rm F})\ln{(\bar{\mu}/\Lambda_{\overline{\rm MS}})}},
\ee
\be
m_i(\bar{\mu}) = m_i(\bar{\mu}_{\rm ref}) \left(\frac{\ln{(\bar{\mu}_{\rm ref}/\Lambda_{\overline{\rm MS}})}}{\ln{(\bar{\mu}/\Lambda_{\overline{\rm MS}})}}\right)^\frac{9C_{\rm F}}{11C_{\rm A}-4T_{\rm F}},
\ee
where $T_{\rm F} = N_{\rm f}/2$, $C_{\rm F} = (N_{\rm c}^2-1)/(2 N_{\rm c})$, and $\Lambda_{\overline{\rm MS}}$ is the $\overline{\rm MS}$ renormalization scale. Here, we take $\bar{\mu}_{\rm ref} = 2$ GeV. We follow Ref.~\cite{1997NuPhB.503..357K} and use the criterion of minimal sensitivity to set the scale $\bar{\mu}$
\be
\bar{\mu} = 4\pi T e^{-\gamma_{\rm E}} e^{\frac{-N_{\rm c}+4 N_{\rm f}\ln{4}}{22 N_{\rm c} -4 N_{\rm f}} }.
\ee
To compute the susceptibilities, we substitute Eq.~(\ref{eq:QCD_pressure_highT}) into Eqs.~(\ref{eq:def_susceptibility2}) and (\ref{eq:def_susceptibility11}), remembering that the relation between the quark chemical potentials and those of the conserved baryon number and electric charges is
\be
\binom{\hat{\mu}_u}{\hat{\mu}_d} = 
\begin{pmatrix}
\frac{1}{3}  &\frac{2}{3} \vspace{0.05cm}\\ 
\frac{1}{3}  &-\frac{1}{3}
\end{pmatrix}
\binom{\hat{\mu}_{\rm B}}{\hat{\mu}_{\rm Q}},
\ee
where $\hat{\mu}_u$ and $\hat{\mu}_d$ are the chemical potentials for up- and down-type quarks, respectively. We numerically evaluate the integrals in the functions $F_1,\ldots, F_4$. The temperatures relevant to sterile neutrino production are well below the top quark mass; we therefore adopt $N_{\rm f} = 5$. We use quark masses evaluated at the reference scale $\bar{\mu}_{\rm ref}$ from Ref.~\cite{Agashe:2014kda}. The only free parameter in the perturbative calculation is the $\overline{\rm MS}$ renormalization scale, which we set so as to match with the lattice calculation in the regime where both approaches are valid. This entails us setting $\Lambda_{\overline{\rm MS}} = 105$ MeV; the results are only logarithmically dependent on the particular value we choose.
\subsubsection{Intermediate Temperatures: Lattice Calculations}

Susceptibilities at temperatures close to the quark-hadron transition have been previously studied in the context of heavy ion collision experiments \cite{RHICreview94}, where the crossover is signaled by fluctuations in the plasma's quantum numbers \cite{Jeon00, Asakawa00}. At zero chemical potential, these are studied in lattice QCD by mapping to the expectation values of traces. Their auto- and cross-correlations are directly related to the diagonal and off-diagonal elements of the susceptibility matrix of Eq.~\eqref{eq:chidef} \cite{Gottlieb87}.

We directly use the susceptibilities so computed at intermediate temperatures. Specifically, we use the results from the Wuppertal-Budapest (WB) lattice QCD collaboration \cite{Borsanyi12} and the HotQCD collaboration \cite{Bazavov12}. Even though the groups use different staggered fermion actions on the lattice, their results are broadly consistent with one another. They report the susceptibilities $\hat{\chi}_2^{\rm Q}$, $\hat{\chi}_2^{\rm B}$, and $\hat{\chi}_{11}^{\rm QB}$, together with their estimated errors, in (2+1)-flavor QCD extrapolated to the continuum limit\footnote{The WB collaboration does not directly report $\hat{\chi}_{11}^{\rm QB}$, but we infer it from their results via a change-of-basis operation.}. The lattice QCD results are in good agreement with the perturbative calculations described above for the temperature range $250$ MeV $\lesssim T \lesssim300$ MeV, above which they underpredict the primeval plasma's susceptibility owing to the charm quark's influence \cite{Ratti:2013uta}. Therefore, we do not consider the lattice QCD calculations at temperatures above $T \gtrsim 300$ MeV to avoid biasing our results. 

\subsubsection{Low-Temperature Limit: Hadron Resonance Gas}
At temperatures below the quark-hadron transition, we model the strongly-interacting sector as a gas of hadronic resonances. In this HRG model, the pressure entering Eq.~(\ref{eq:pQCD_from_partitionZ}) is given by
\ba\label{eq:HRG_partition_function}
\hat{p}_{\rm HRG} &=& \frac{1}{VT^3}\Bigg( \sum_{i\in\,{\rm mesons}} \ln{Z_i^{\rm M}(V,T,\mu_{\rm Q})}\\
&&\qquad\quad +\sum_{j\in\,{\rm baryons}} \ln{Z_j^{\rm B}(V,T,\mu_{\rm Q},\mu_{\rm B})}\Bigg),\nonumber
\ea
where
\ba
\ln{Z_i^{\rm M}} &=& - \frac{V T^3}{2\pi^2} d_i \int_0^\infty d\hat{p}\, \hat{p}^2 \ln{(1-z_ie^{-\sqrt{\hat{p}^2+\hat{m}_i^2}})}\label{eq:meson_partition},\\
\ln{Z_i^{\rm B}} &=&  \frac{V T^3}{2\pi^2} d_i \int_0^\infty d\hat{p}\, \hat{p}^2 \ln{(1+z_ie^{-\sqrt{\hat{p}^2+\hat{m}_i^2}})}\label{eq:baryon_partition},
\ea
where $d_i$ denotes the degeneracy factor of species $i$, and $z_i$ is the fugacity
\be
z_i = e^{{\rm B}_i \hat{\mu}_{\rm B} + {\rm Q}_i \hat{\mu}_{\rm Q}} \mbox{.}
\ee
Here ${\rm B}_i$ and ${\rm Q}_i$ are the baryon number and electric charge of species $i$, respectively. We construct the partition function given in Eq.~(\ref{eq:HRG_partition_function}) by summing over all hadron resonances with mass below $2$ GeV from the particle data group \cite{Agashe:2014kda}. We then compute the susceptibilities using Eqs.~(\ref{eq:def_susceptibility2}) and (\ref{eq:def_susceptibility11}). We numerically perform the integrals in Eqs.~(\ref{eq:meson_partition}) and (\ref{eq:baryon_partition}).

We find that the HRG results are in good agreement with the lattice QCD calculations for temperatures $125$ MeV$\lesssim T\lesssim150$ MeV, and we smoothly match the HRG-derived susceptibilities to those from the lattice technique in this regime.

\subsubsection{Susceptibilities at all Temperatures} 
\begin{figure*}[t]
  	\hspace*{\fill}%
  	\subfloat[][electric charge subsceptibility]{
		\includegraphics[width=0.48\textwidth]{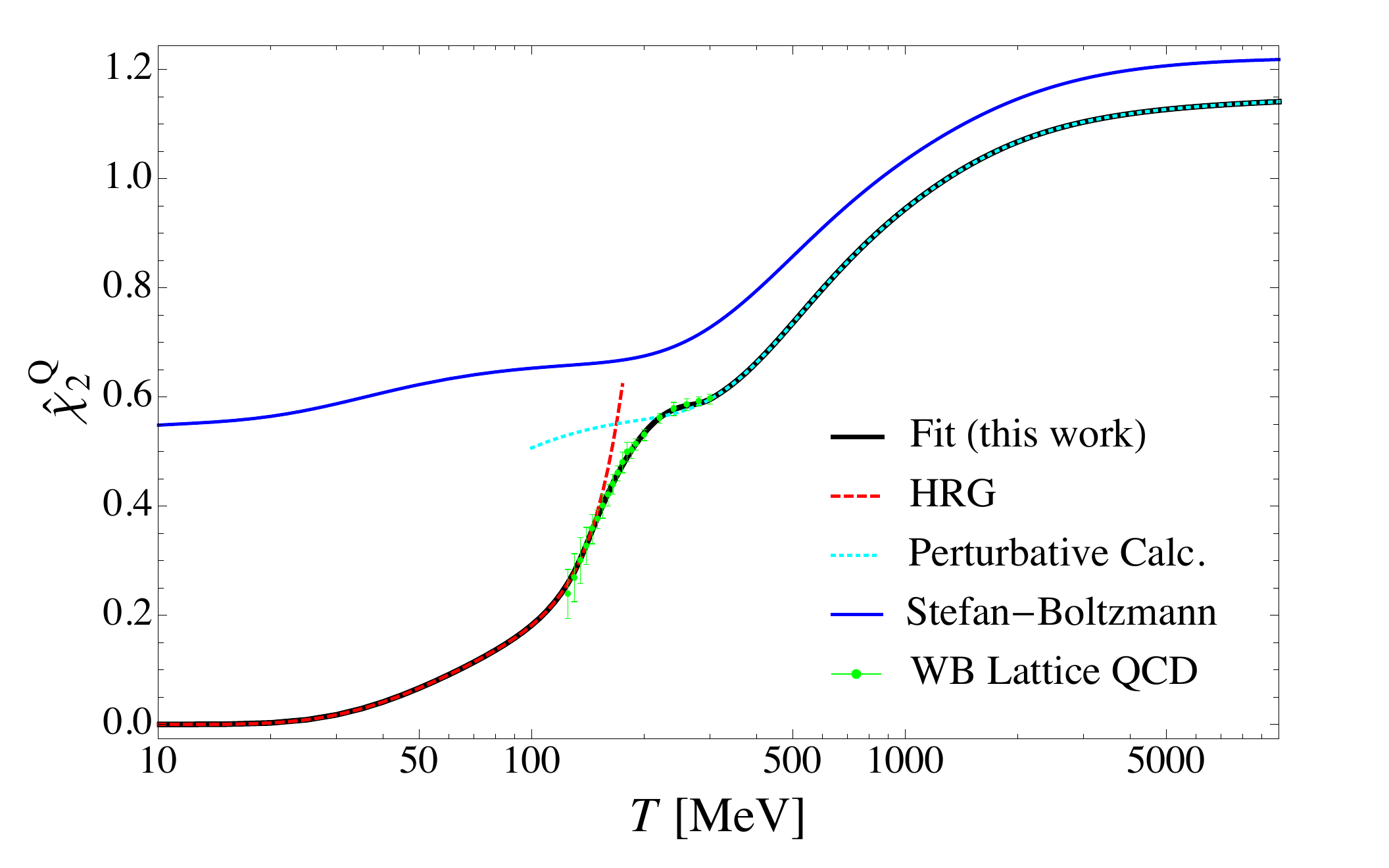}
		\label{fig:susceptibilities_chiQ}
	}
  	\hfill
 	 \subfloat[][baryon number susceptibility]{
  		\includegraphics[width=0.48\textwidth]{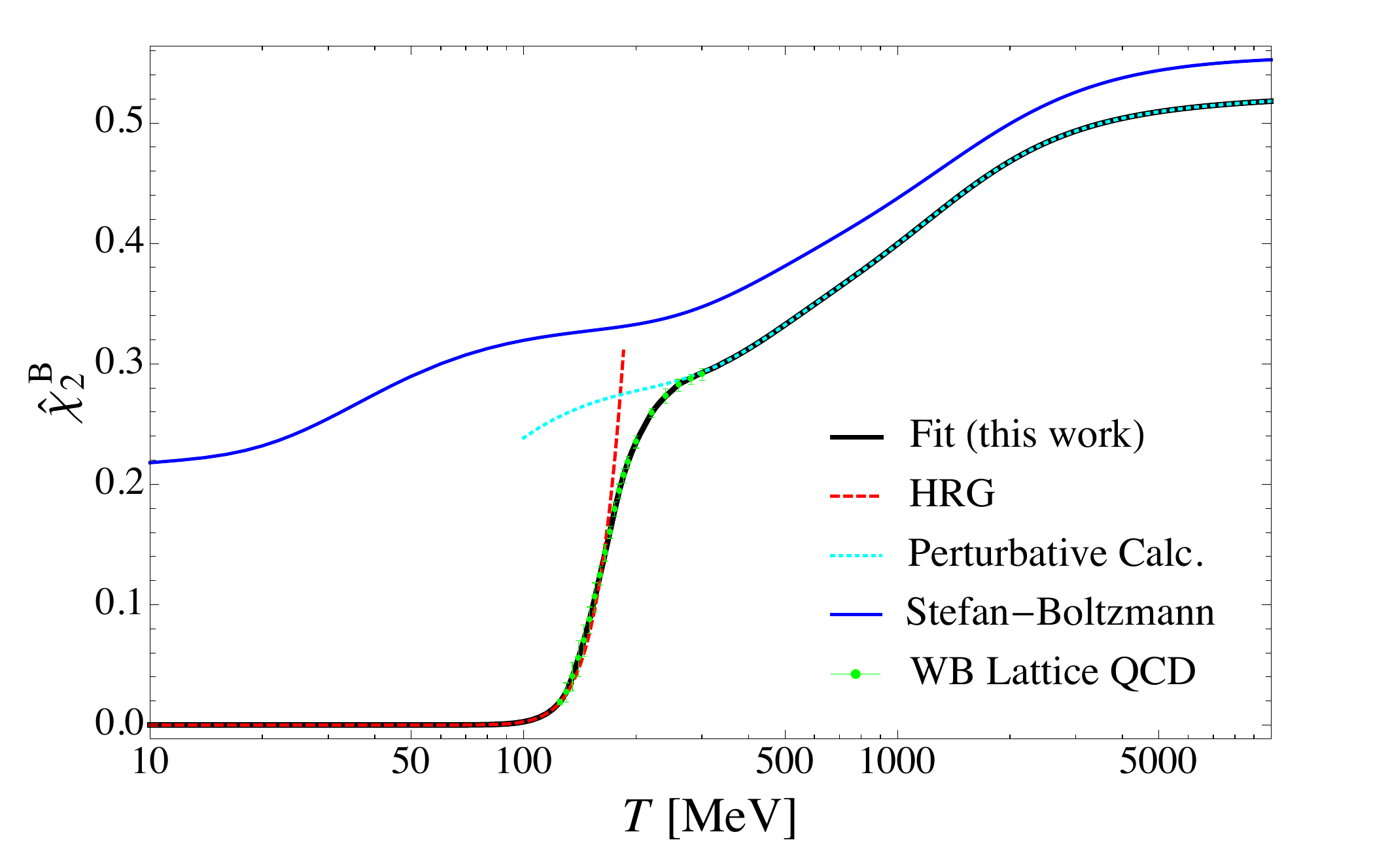}
  		\label{fig:susceptibilities_chiB}
  	}\\
    	\hspace*{\fill}%
    	\subfloat[][charge-baryon number susceptibility]{	
		\includegraphics[width=0.48\textwidth]{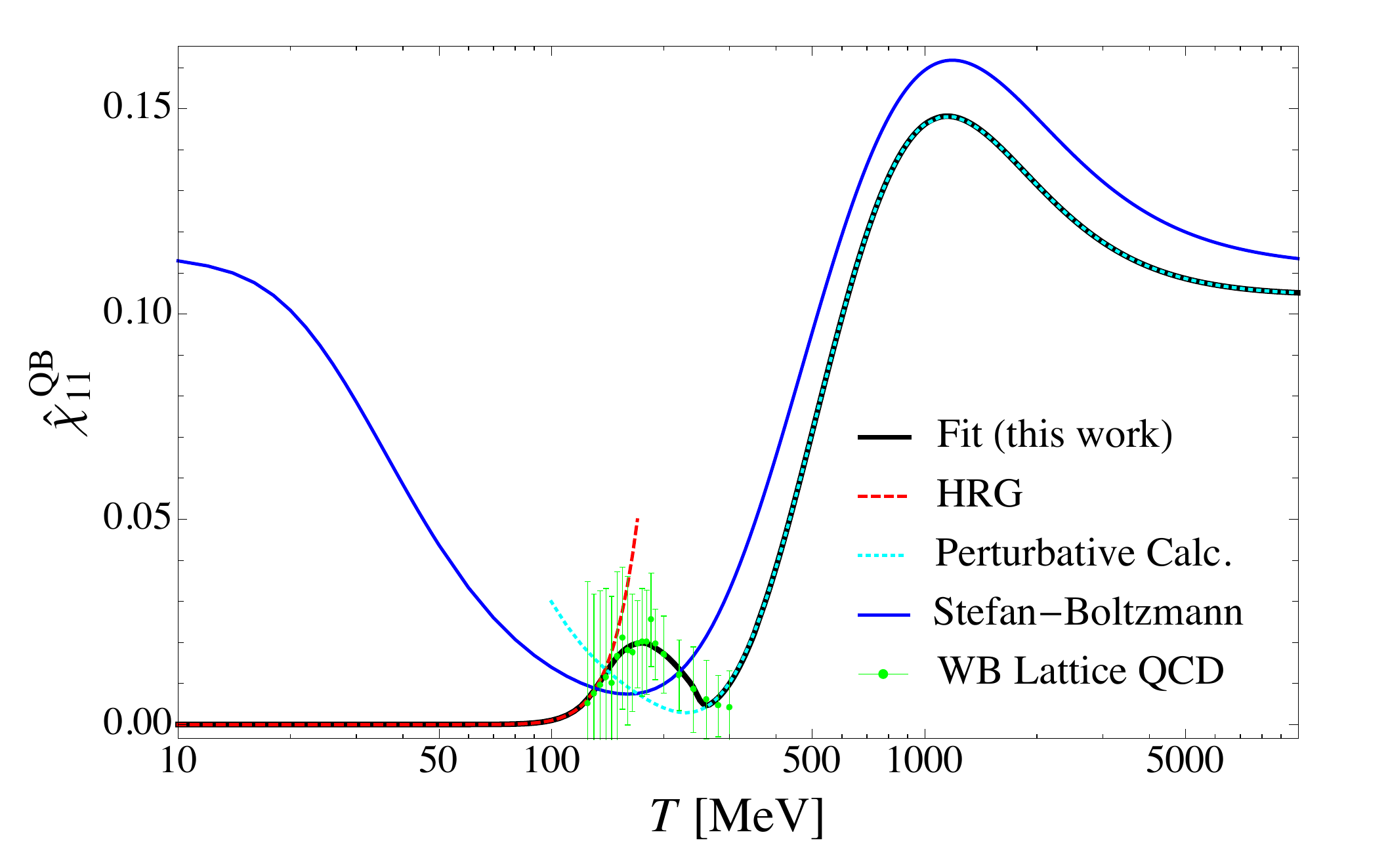}
		\label{fig:susceptibilities_chiQB}
	}
	\hfill
 	 \subfloat[][redistribution functions]{
		 \includegraphics[width=0.48\textwidth]{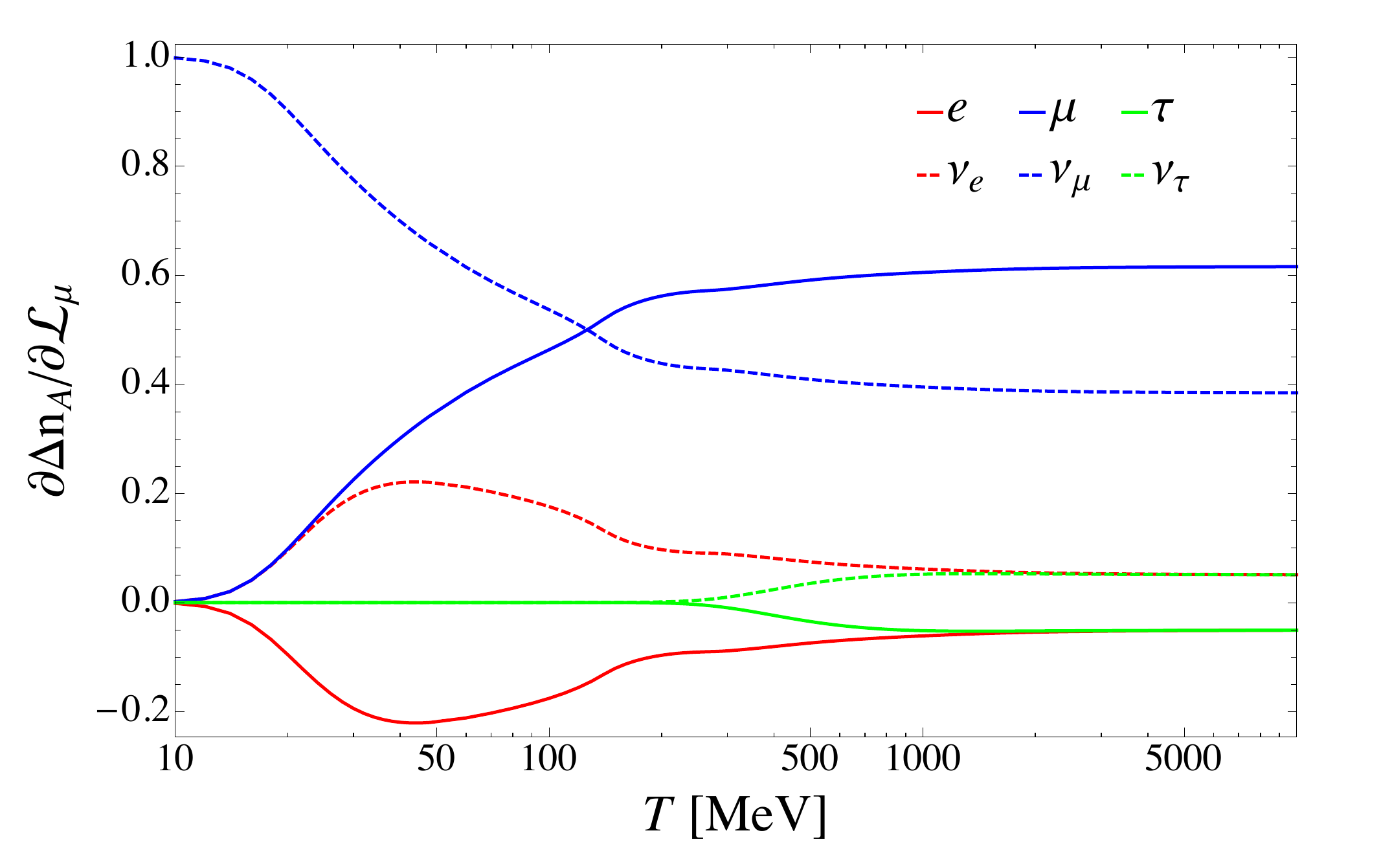}
	 	\label{fig:dndLmu}
	}
	\hspace*{\fill}%
	\caption{\label{fig:susceptibilities} Panels (a)-(c): Components of the quadratic susceptibility tensor for the primordial plasma's electric charge and baryon number. In all panels, the thick solid black line shows our smooth fit used in the computation of sterile neutrino production. At low temperatures, we illustrate the HRG results with dashed red lines, while the high-temperature perturbative results are shown with dotted cyan lines. We also show the results from the WB lattice QCD collaboration\cite{Borsanyi12} with green errorbars. For comparison, we also display the Stefan-Boltzmann approximation to the susceptibilities assuming free quarks at all temperatures. Panel (d): Effective populations of all leptonic degrees of freedom after the redistribution of an infinitesimal mu leptonic asymmetry at all temperatures.}
\end{figure*}

We combine results from the three regimes into a single smooth susceptibility tensor, valid over the range of temperatures relevant to the production of sterile neutrinos with masses of order $\mathcal{O}$(10 keV). Figures ~\ref{fig:susceptibilities_chiQ}, \ref{fig:susceptibilities_chiB}, and \ref{fig:susceptibilities_chiQB} display the susceptibilities $\hat{\chi}_2^{\rm Q}$, $\hat{\chi}_2^{\rm B}$, and $\hat{\chi}_{11}^{\rm QB}$ for temperatures satisfying $10$ MeV $<T<10$ GeV. The thick solid black lines are our smooth fits to the three regimes, while the dashed red and cyan dotted lines are the HRG and perturbative results, respectively. We also show the results from the WB lattice QCD collaboration in the neighborhood of the quark-hadron transition. For comparison, we also show the susceptibilities computed in the Stefan-Boltzmann limit, i.e., assuming free quarks throughout and using Eq.~\eqref{eq:freeparticlechi}.

We observe that the HRG calculation agrees well with the lattice QCD result for $T\lesssim150$ MeV, but systematically overpredicts the susceptibilities at higher temperatures. The perturbative approach is consistent with the available lattice QCD data at $T\gtrsim225$ MeV, but again systematically overpredicts the susceptibilities at lower temperatures. Generally, the Stefan-Boltzmann approximation overestimates the susceptibilities by a factor of order unity, except near the quark-hadron transition. Interestingly, we observe an accidental cancellation in the off-diagonal susceptibility, $\hat{\chi}_{11}^{\rm QB}$, in the (2+1)-flavor model which does not appear in the $N_{\rm f}=5$ theory. This arises because the sum of the electric charges of the up, down, and strange quarks exactly vanishes. Hence, we expect $\hat{\chi}_{11}^{\rm QB}\rightarrow0$ for temperatures above the strange quark mass in the (2+1)-flavor model. In the $N_{\rm f}=5$ model however, the charm quark becomes rapidly important at $T\gtrsim300$ MeV, leading to a sharp turnover in $\hat{\chi}_{11}^{\rm QB}$ near this temperature.

Given a set of infinitesimal lepton asymmetries, we solve for the chemical potentials using the above susceptibilities in Eqs.~\eqref{eq:leptasym}, \eqref{eq:saha}, \eqref{eq:baryonnumber} and \eqref{eq:charge}, We obtain the redistributed asymmetries in all the constituent species by using these chemical potentials, along with the appropriate susceptibilities in Eq.~\eqref{eq:redistcoeff}. Figure \ref{fig:dndLmu} plots the redistributed asymmetries for an infinitesimal input mu leptonic asymmetry. We note the following features: 
\begin{enumerate}
   \item At temperatures $T > 2$ GeV, the redistribution is efficient and $\simeq 60\%$ of the mu leptonic asymmetry ends up in the muons. All the charged leptons are effectively massless at this epoch, hence the populations of the electron and tau flavors are identical. 
   \item The rise in the mu and tau lepton populations above temperatures of $\simeq 25$ MeV and $300$ MeV reflects, in part, the rise in their particle number susceptibilities as the temperature becomes comparable to their masses [see Eq.~\eqref{eq:freeparticlechi}]. However, the largest contribution to the former is from the disappearance of the hadronic degrees of freedom below the quark-hadron transition, and the associated drop in the strongly interacting fluid's susceptibilities.
   \item The `kink' in all the redistributed asymmetries close to temperatures $T \simeq 170$ MeV is a signature of the sharp change in the strongly interacting fluid's susceptibilities at the quark-hadron transition [see Figs.~\ref{fig:susceptibilities_chiQ}, \ref{fig:susceptibilities_chiB} and \ref{fig:susceptibilities_chiQB}]. 
   \item At lower temperatures, $T \lesssim 30$ MeV, the redistribution is inefficient and most of the asymmetry ends up in the muon neutrinos. Moreover, the electron neutrino and the muon have identical (small) populations. This is characteristic of inelastic neutrino scattering, $\nu_\mu + e^- \rightarrow \nu_e + \mu^-$, which is the most important channel at these temperatures (the hadronic susceptibilities are negligible at this epoch).
\end{enumerate}

These redistributed asymmetries impact sterile neutrino production via the asymmetry potential, $V^{\rm L}_{\nu_\mu}$. Equation \eqref{eq:asymmv} expresses this potential in terms of the asymmetries in the populations of the individual charged and neutral leptons, along with those in the charge and baryon number of the strongly interacting fluid. As earlier, for an infinitesimal input mu leptonic asymmetry, the individual asymmetries are formally represented by the functions in Eq.~\eqref{eq:redistcoeff}; the solutions for the charged and neutral leptons are as plotted in Fig.~\ref{fig:dndLmu}. We obtain the electric charge density of the strongly interacting fluid, $\Delta n_{\rm Q}$, using net electric charge neutrality, i.e. Eqs.~\eqref{eq:chidef} and \eqref{eq:charge}. Tables of susceptibilities, along with the functions in Eq.~\eqref{eq:redistcoeff} at a number of temperatures from $10$ GeV down to $10$ MeV can be found at \url{https://github.com/ntveem/sterile-dm/tree/master/data/tables}.

\begin{figure}[t]
   \begin{center}
      \includegraphics[height=6cm]{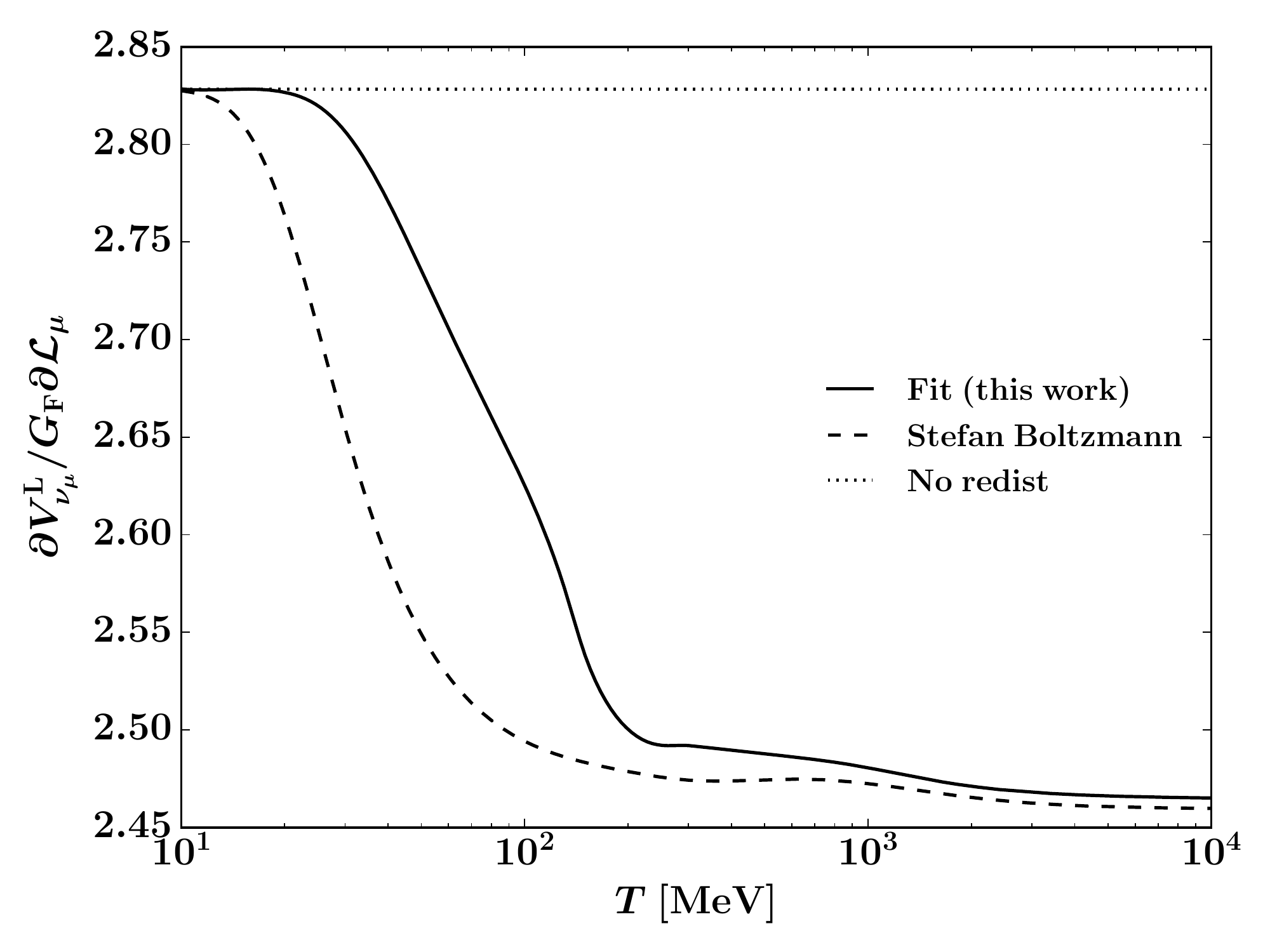}
      \caption{Asymmetry potential $V^{\rm L}$ per unit {\em unscaled} mu lepton asymmetry: Solid line shows the effect of redistribution using a combination of perturbative QCD, lattice calculations and the HRG approximation. Dashed line shows the result using Stefan-Boltzmann approximation for free quarks. The value is constant ($=2\sqrt{2}=2.83$) in the absence of redistribution [see Eq.~\eqref{eq:asymmv}]. The redistribution is a $\simeq 10 \%$ correction at the most relevant temperatures for sterile neutrino production ($T \gtrsim 100$ MeV).}
      \label{fig:asymmv}
   \end{center}
 \end{figure}

Figure \ref{fig:asymmv} shows the potential per unit {\em physical} $\mu$ lepton asymmetry using these solutions; this quantity is constant and equals $2\sqrt{2} = 2.83$ in the absence of redistribution. As shown in the figure, asymmetry redistribution corrects the potential at the ten-percent level above temperatures $T \gtrsim 100$ MeV, which is where the bulk of the sterile neutrinos are produced. This contribution changes the resonant momenta, and the resultant sterile neutrino dark matter's phase-space densities; we explore this further in Sec.~\ref{sec:production}.

\section{Neutrino opacity}
\label{sec:scattering}

In this section, we outline our calculation of muon neutrino opacities in the early universe. Initial work in this area focused on reactions involving leptons, in the context of neutrino decoupling, active--active neutrino oscillations and supernova calculations \cite{McKellar94, Hannestad95, Yueh76}. In particular, Ref.~\cite{Hannestad95} lists a number of relevant matrix elements. Our calculations apply to earlier epochs, with a larger number of reactions due to the population of hadronic species above the quark-hadron transition. 

Early work on sterile neutrino production used simple prescriptions for the resultant increase in reaction rates \cite{Kishimoto08, Abazajian01}. Recent work in Refs.~\cite{2006JHEP...06..053A,2007JHEP...01..091A} provides a theoretical framework to include particle masses and statistics in the neutrino opacity calculation, and formalism for loop corrections. We include a number of additional contributions to the neutrino opacities that are significant at the temperatures relevant to sterile neutrino production. We adopt the following simplifying assumptions:
\begin{enumerate}
  \item We neglect small asymmetries in the participating species' populations (as for the thermal potential). This is justified since the scattering rates are non-zero even in a CP symmetric plasma. Moreover, we assume thermal and kinetic equilibrium, due to which the populations of all active species are Fermi-Dirac/Bose-Einstein distributions. 
  \item We integrate out the massive gauge bosons, $Z$ and $W^{\pm}$ and approximate the weak interaction by a four-fermion contact term. Consequently, the reactions separate into leptonic and hadronic processes, depending on the species involved. Moreover, we neglect the thermal populations of $Z^0$ and $W^\pm$. These steps are valid at low temperatures and momentum transfers, i.e., $T, s/t/u \ll M_{W^{\pm}/Z^{0}} \approx 80 \ {\rm GeV}$. We operate in the temperature and energy ranges
    \begin{align}
      10 \ {\rm MeV} &< T < 10 \ {\rm GeV} \mbox{,} \\
      10^{-4} &< E_{\nu_{\mu}}/T < 20 \mbox{.}
    \end{align}
    The approximation fails at the higher energies at the upper end of the temperature range. However, as we see in Sec.~\ref{sec:production}, the bulk of the sterile neutrinos are produced at lower temperatures.  
  \item We assume incoming and outgoing particles to be non-interacting within two limits -- below and above the quark-hadron transition (see \S 3.3 of Ref.~\cite{Kolb90}). Below the transition, we include hadronic channels with pseudoscalar and vector mesons\footnote{We also include quark production in s-channel reactions at high CM energies. See Sec.~\ref{subsubsec:matrixelts} for details.}, and neglect the small population of baryons. Above the transition, we include reactions with free quarks, i.e. we neglect the strong coupling constant. 
    
    This approximation fails at temperatures $T \simeq T_{\rm QCD}$ \cite{Laine06}. We show opacities interpolated through the transition using a few prescriptions, whose consequences for sterile neutrino production we explore in Sec.~\ref{sec:production}.
\end{enumerate}

The collision integral for a massless muon neutrino is 
\begin{align}
  C[f_{\nu_\mu}(E_{\nu_\mu})] & = -\Gamma(E_{\nu_\mu}) f_{\nu_\mu}(E_{\nu_\mu}) \nonumber \\
  & ~ ~ ~ + \Gamma(E_{\nu_\mu}) e^{-E_{\nu_\mu}/T} ( 1 - f_{\nu_\mu}(E_{\nu_\mu}) ) \mbox{,} \label{eq:collisionintegral}
\end{align}
where $\Gamma$ and $f_{\nu_\mu}$ are the interaction rate (opacity) and PSD, respectively (This expression satisfies detailed balance, see assumption \# 1). The interaction rate is given by a sum over all reactions that consume the muon neutrino [see Eq.~\eqref{eq:scatteringrate}].

It is useful to define the scaled interaction rate
\begin{align}
  \widetilde{\Gamma}(E_{\nu_\mu}) & = \frac{\Gamma(E_{\nu_\mu})}{G_{\rm F}^2 T^4 E_{\nu_\mu}} \mbox{.}
\end{align}
In the limit where all the particles involved are relativistic, weak cross-sections are proportional to the squared energy in the CM reference frame. If we ignore particle statistics, reaction rates follow the scaling of Eq.~\eqref{eq:pscaling}, hence the scaled rate is proportional to the number of relativistic degrees of freedom involved \cite{Notzold88}. We present the scaled rates in the rest of this section, in order to contrast our results with this intuition.

In the rest of the section, we enumerate reactions contributing to the opacity and present final rates under the above approximations. We elaborate the calculation of matrix elements in Appendix \ref{sec:matrixelements}. 

\begin{table}[t]
  \caption{\label{tab:leptonreactions} Two-particle to two-particle leptonic reactions contributing to the muon neutrino opacity: antineutrinos are similar. Symbol $\alpha$ runs over the other leptonic flavors, i.e. $\alpha \in \{ e, \tau \}$, and $\overset{(-)}{\nu}$ stands for $\nu/\bar{\nu}$.}
   \begin{ruledtabular}
     \begingroup\makeatletter\def\f@size{7}\check@mathfonts
     \begin{tabular}{ccc}
        s-channel & t-channel & mixed \\
        \hline
	$\nu_\mu + \mu^+ \rightarrow \nu_\alpha + \alpha^+$ & $\nu_\mu + \alpha^{\pm} \rightarrow \nu_\mu + \alpha^{\pm}$ & $\nu_\mu + \overset{(-)}{\nu}_\mu \rightarrow \nu_\mu + \overset{(-)}{\nu}_\mu$ \\
	$\nu_\mu + \bar{\nu}_\mu \rightarrow \nu_\alpha + \bar{\nu}_\alpha$ & $\nu_\mu + \overset{(-)}{\nu}_\alpha \rightarrow \nu_\mu + \overset{(-)}{\nu}_\alpha$ & $\nu_\mu + \mu^{\pm} \rightarrow \nu_\mu + \mu^{\pm}$ \\
	$\nu_\mu + \bar{\nu}_\mu \rightarrow \alpha^- + \alpha^+$ & $\nu_\mu + \alpha^- \rightarrow \mu^- + \nu_\alpha$ & $\nu_\mu + \bar{\nu}_\mu \rightarrow \mu^- + \mu^+$ \\
	& $\nu_\mu + \bar{\nu}_\alpha \rightarrow \mu^- + \alpha^+$ & \\
     \end{tabular}
   \endgroup
   \end{ruledtabular}
\end{table}

Table \ref{tab:leptonreactions} lists the purely leptonic two-particle to two-particle reactions contributing to the muon neutrino opacity. It is similar to the list in Ref.~\cite{Hannestad95}, albeit with tau leptons, which are important at temperatures $T \gtrsim 400$ MeV. We also include `three-body fusions', since they arise at the same level of approximation. These are generated by omitting in turn the products in the reactions of Table~\ref{tab:leptonreactions}, adding their charge conjugates to the reactants, subject to the constraint that the product's rest mass is strictly greater than the sum of the reactants'. An  example is tau lepton production via $\nu_\mu + \mu^+ + \bar{\nu}_\tau \rightarrow \tau^+$.

\begin{figure}[t]
  \centering
    \includegraphics[width=\columnwidth]{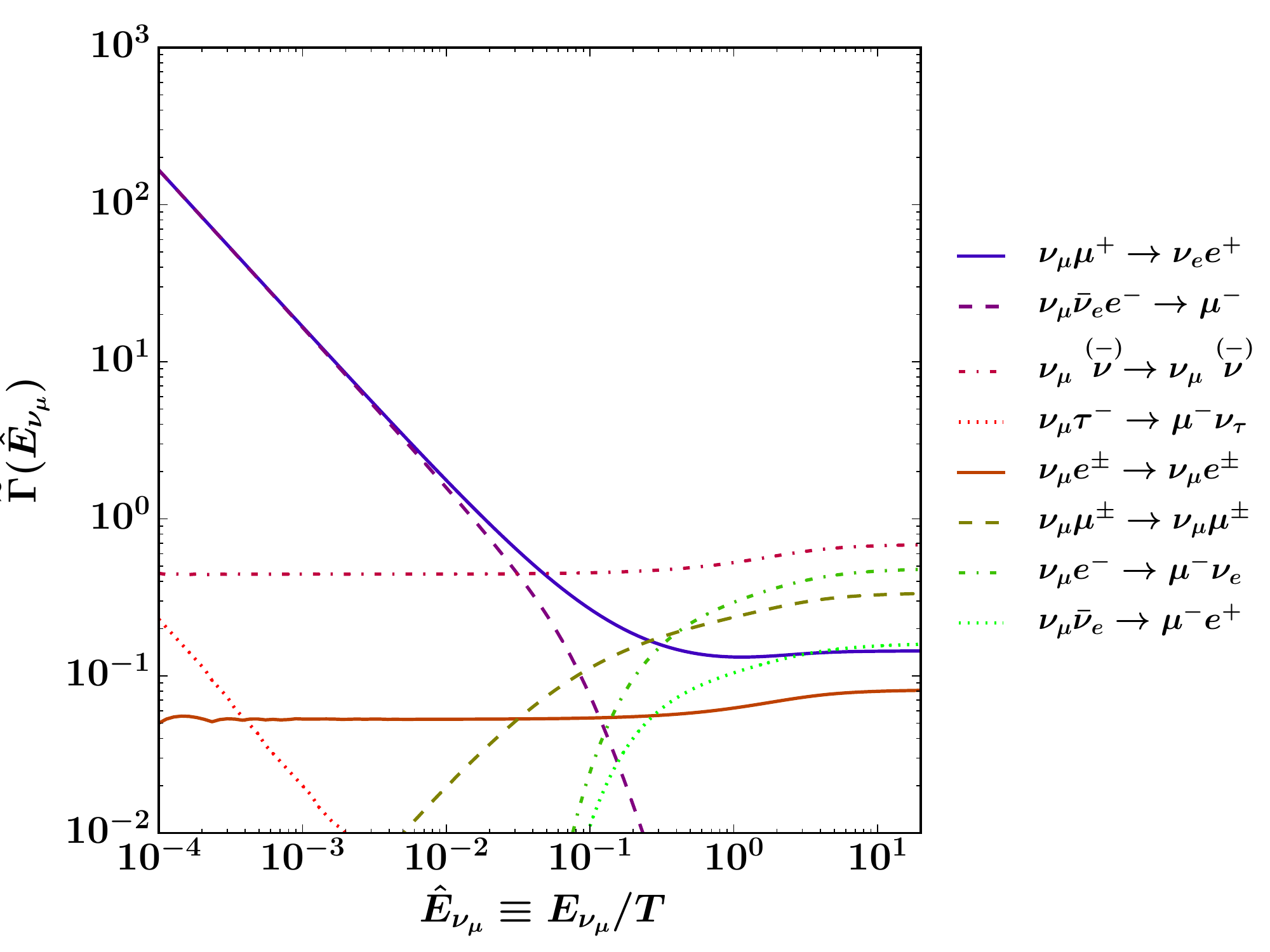}
    \caption{\label{fig:leptonicrates} Scaled muon neutrino opacities through leptonic reactions, vs energy at $T = 100$ MeV. Only reactions in the top five at any particular momentum bin are shown. The symbol $\overset{(-)}{\nu}$ stands for $\nu/\bar{\nu}$.}
\end{figure}

Figure \ref{fig:leptonicrates} shows the leptonic contribution to the muon neutrino opacity at a temperature $T = 100$ MeV, using the matrix elements for reactions in Table~\ref{tab:leptonreactions}, and related three-body fusions. For convenience, we only show reactions in the top five at any particular momentum bin. In the numerical implementation, we evaluate the dimensionless quantity $\Gamma(E_{\nu_\mu})/G_{\rm F}^2 T^5$ (proportional to the {\em unscaled} rates) to an accuracy of $10^{-6}$ after simplifying the collision integrals in Eq.~\eqref{eq:scatteringintegral} and \eqref{eq:3bodyscatteringintegral}.

The quark-hadron transition considerably complicates the hadronic reactions. We appeal to assumption \# 3 and evaluate their rates in two limits: at low and high temperatures, i.e. $T<T_{\rm QCD}$ and $T>T_{\rm QCD}$ respectively. Table \ref{tab:hadronreactions} lists the hadronic two-particle to two-particle reactions contributing to the muon neutrino opacity. At high temperatures, we list all reactions involving free quarks, while at temperatures $T \lesssim T_{\rm QCD}$, we assume that all  incoming hadronic degrees of freedom belong to the pseudoscalar meson nonet (the heaviest member of which is the eta meson, with a mass of $m_\eta = 547.8$ MeV). As in the leptonic case, we also include three-body fusions involving pseudoscalar mesons or quarks. Examples are $K^0$ and charm quark production via $\nu_\mu + \mu^+ + \pi^- \rightarrow K^0$ and $\nu_\mu + \mu^+ + s \rightarrow c$, respectively. 

A complication is that at low temperatures, free quark and parton currents contribute to the initial and final states for large momentum transfer in the t- and s-channel respectively (see Appendix \ref{subsubsec:matrixelts}). For s-channel reactions, we thus treat individual meson resonances for center of mass energies $<1$ GeV, and use the free quark model for inclusive cross sections at $>1$ GeV. Also important are `two-body fusions', i.e. reactions with two particles in the initial state and one in the final state, with the latter being a pseudoscalar or vector meson. Table \ref{tab:2.body.rate} in Appendix \ref{subsec:oneparticle} lists all such reactions included in our opacities.

\begin{table}[t]
  \caption{\label{tab:hadronreactions} Two-particle to two-particle hadronic reactions contributing to the $\mu$ neutrino opacity: antineutrinos are similar. Symbols $a$ and $b$ run over quarks with charge $+(2/3)e$, and $-(1/3)e$, respectively, and $\overset{(-)}{a}$ stands for $a/\bar{a}$.}
   \begin{ruledtabular}
     \begin{tabular}{cc}
        s-channel & t-channel \\
        \hline
	\multicolumn{2}{c}{$T > T_{\rm QCD}$} \\
	$\nu_\mu + \mu^+ \rightarrow a + \bar{b}$ & $\nu_\mu + \overset{(-)}{a} \rightarrow \nu_\mu + \overset{(-)}{a}$ \\
	$\nu_\mu + \bar{\nu}_\mu \rightarrow a + \bar{a}$ & $\nu_\mu + \overset{(-)}{b} \rightarrow \nu_\mu + \overset{(-)}{b}$ \\
	$\nu_\mu + \bar{\nu}_\mu \rightarrow b + \bar{b}$ & $\nu_\mu + b \rightarrow \mu^- + a$ \\
	& $\nu_\mu + \bar{a} \rightarrow \mu^- + \bar{b}$ \\
	\hline
	\multicolumn{2}{c}{$T < T_{\rm QCD}$\footnotemark[1]} \\
	$\nu_\mu + \mu^+ \rightarrow \pi^+ + \pi^0$ & $\nu_\mu + \pi^{\pm} \rightarrow \nu_\mu + \pi^{\pm}$ \\
	$\nu_\mu + \mu^+ \rightarrow K^+ + \overline{K^0}$ & $\nu_\mu + K^{\pm} \rightarrow \nu_\mu + K^{\pm}$ \\
	$\nu_\mu + \mu^+ \rightarrow \pi^+ + K^0$ & $\nu_\mu + \pi^- \rightarrow \mu^- + \pi^0$ \\
	$\nu_\mu + \mu^+ \rightarrow K^+ + \pi^0$ & $\nu_\mu + K^- \rightarrow \mu^- + \overline{K^0}$ \\
	$\nu_\mu + \mu^+ \rightarrow K^+ + \eta$ & $\nu_\mu + \pi^- \rightarrow \mu^- + K^0$ \\
	$\nu_\mu + \bar{\nu}_\mu \rightarrow \pi^+ + \pi^-$ & $\nu_\mu + K^- \rightarrow \mu^- + \pi^0$ \\
	$\nu_\mu + \bar{\nu}_\mu \rightarrow K^+ + K^-$ & $\nu_\mu + K^- \rightarrow \mu^- + \eta$ \\
	& $\nu_\mu + \pi^0 \rightarrow \mu^- + \pi^+$ \\
	& $\nu_\mu + K^0 \rightarrow \mu^- + K^+$ \\
	& $\nu_\mu + \overline{K^0} \rightarrow \mu^- + \pi^+$ \\
	& $\nu_\mu + \pi^0 \rightarrow \mu^- + K^+$ \\
	& $\nu_\mu + \eta \rightarrow \mu^- + K^+$ \\
     \end{tabular}
     \footnotetext{Input neutrinos can produce quarks (s-channel) or probe mesons' quark content (t-channel) at $T<T_{\rm QCD}$ for large momentum transfer in $Z^0, W^\pm$. See Appendix \ref{subsubsec:matrixelts} for details.}
   \end{ruledtabular}
\end{table}

\begin{figure}[t]
   \centering
   \subfloat[][$T=100$ MeV]{
      \includegraphics[width=\columnwidth]{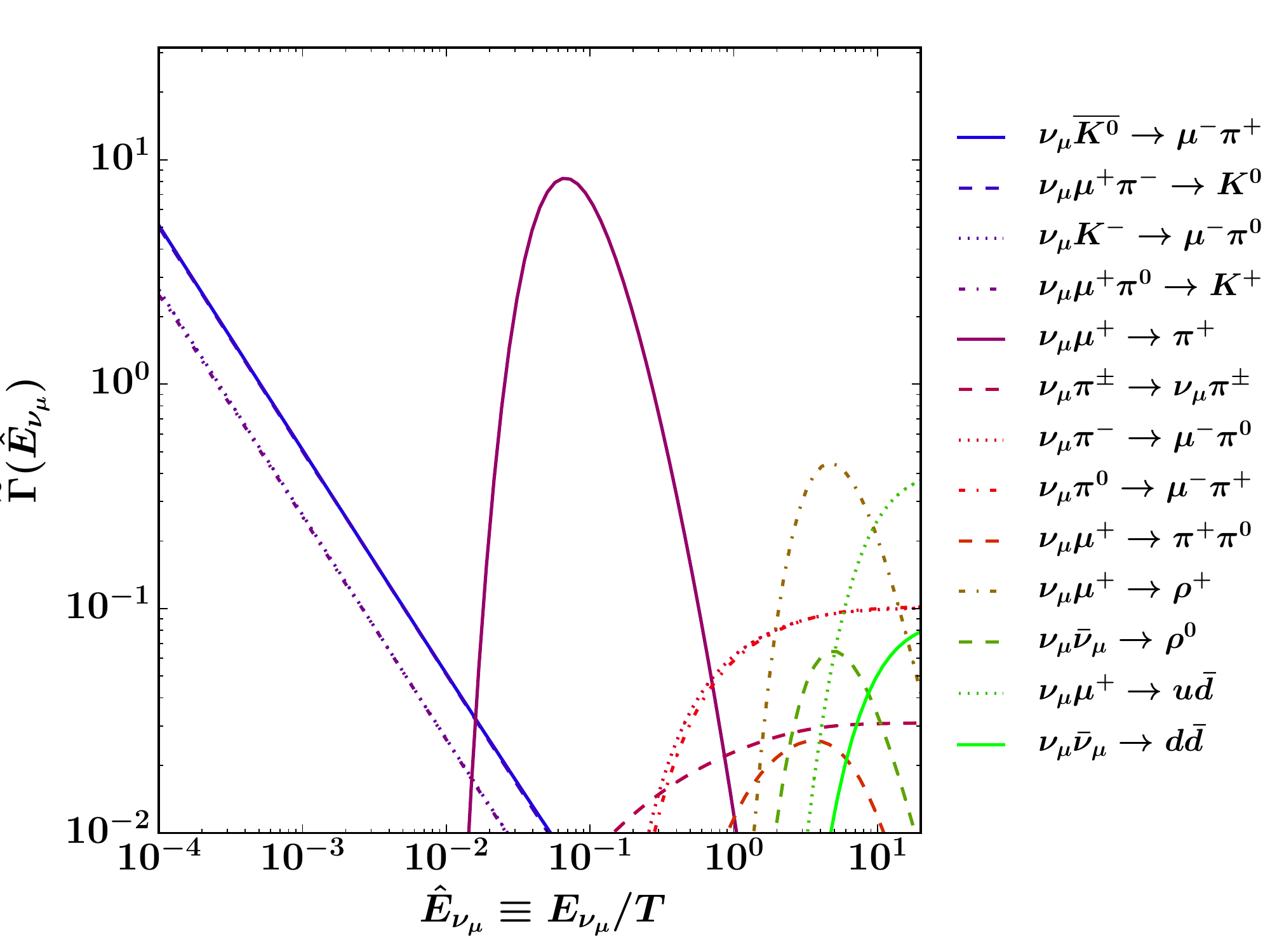}
      \label{fig:hadronic100mev}
   }\\
   \subfloat[][$T=2$ GeV]{
      \includegraphics[width=\columnwidth]{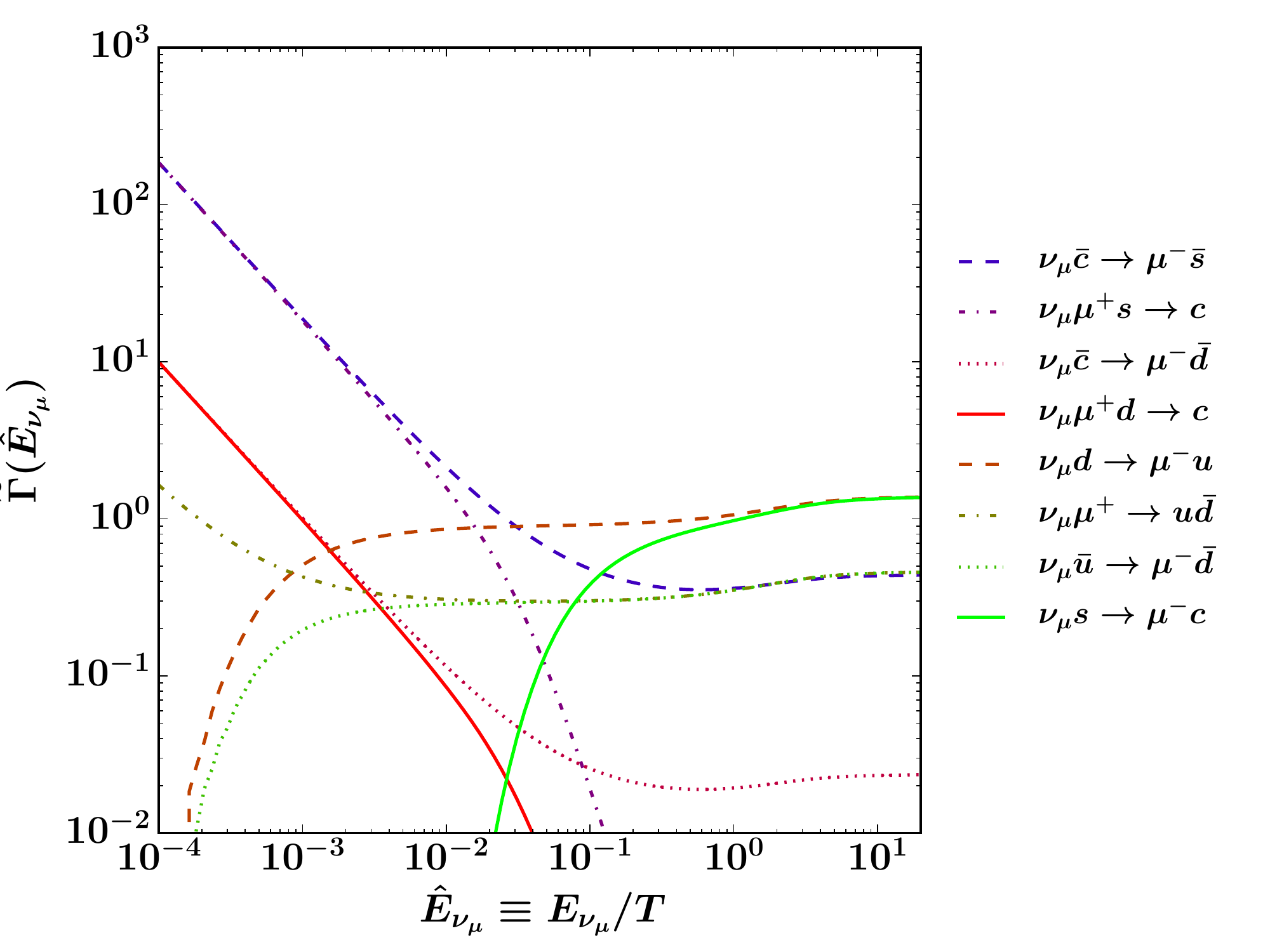}
      \label{fig:hadronic2gev}
   }
   \caption{\label{fig:hadronicrates}Scaled muon neutrino opacities through hadronic reactions, vs energy: panels (a) and (b) show rates at $T=100$ MeV and $2$ GeV, respectively. Only reactions in the top five at any particular momentum bin are shown.}
\end{figure}

Figure \ref{fig:hadronicrates} shows the hadronic contribution to the muon neutrino opacity at low and high temperatures, using the matrix elements for two-particle to two-particle reactions in Table~\ref{tab:hadronreactions}, the associated three-body fusions, and two-body fusions in Table~\ref{tab:2.body.rate} of Appendix \ref{subsec:oneparticle}. As earlier, we only show reactions in the top five at any momentum bin; the numerical implementation of the first two classes is unchanged.

\begin{figure*}
\hspace{-0.57cm}%
  \subfloat[][rates vs energy]{
    \includegraphics[height=6cm]{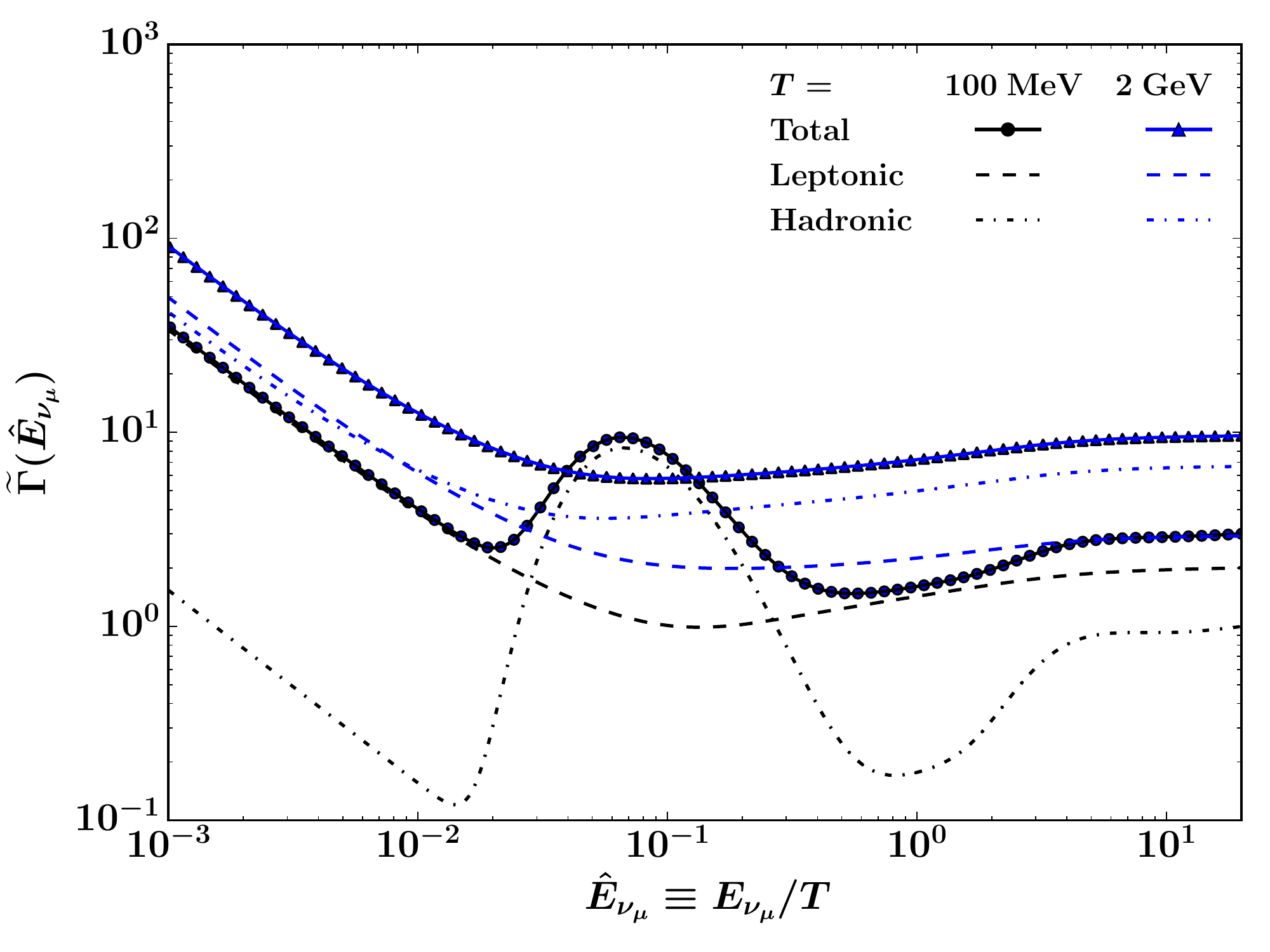}
    \label{fig:T_scattering_rates}
  }
  \hspace{0.13cm}%
  \subfloat[][rates vs temperature]{
    \includegraphics[height=6cm]{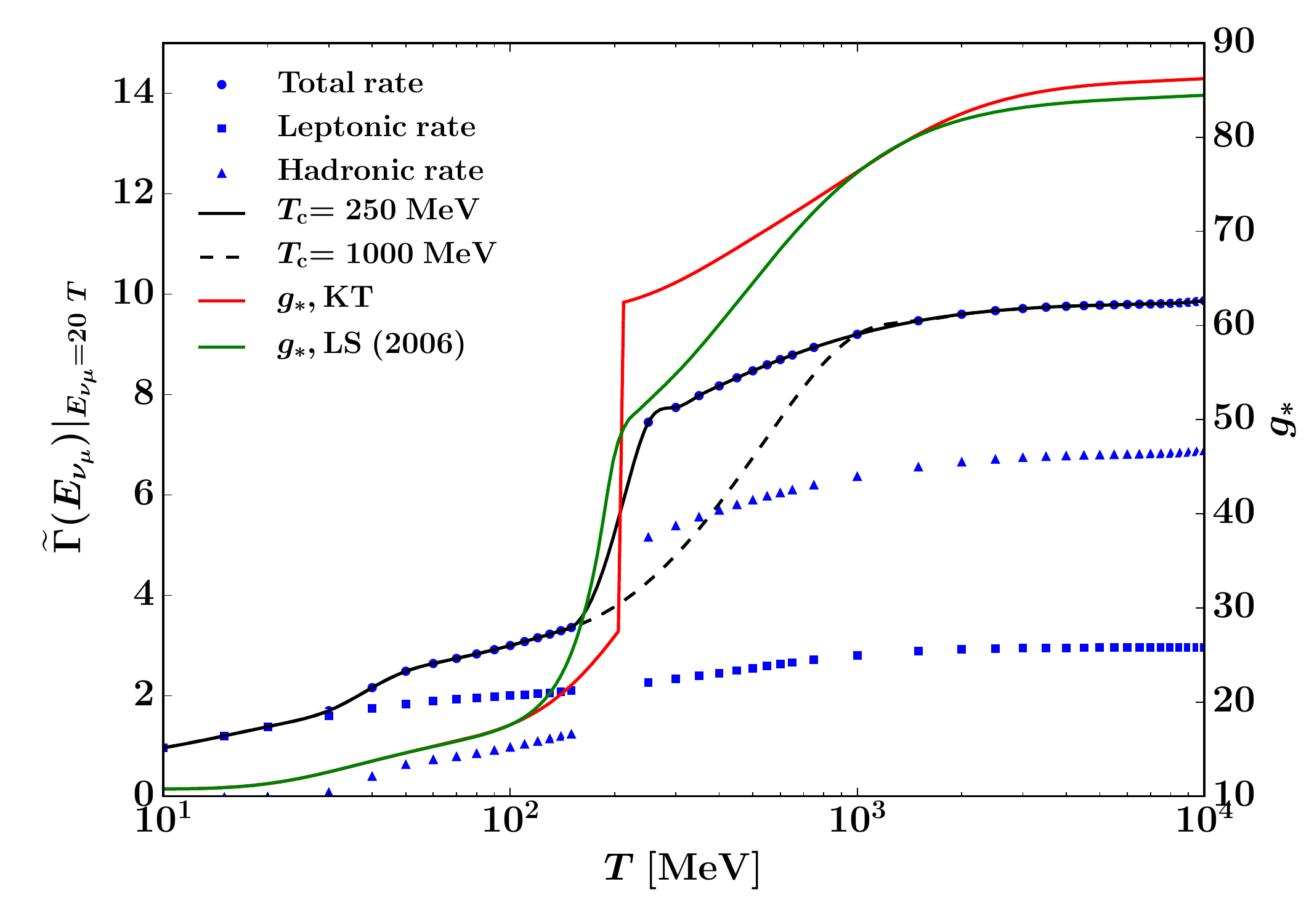}
    \label{fig:highp_scattering_rates}
  }\\
  \hspace*{\fill}%
   \subfloat[][$T_{\rm c} = 250 \ {\rm MeV}$]{
    \includegraphics[height=7.87cm]{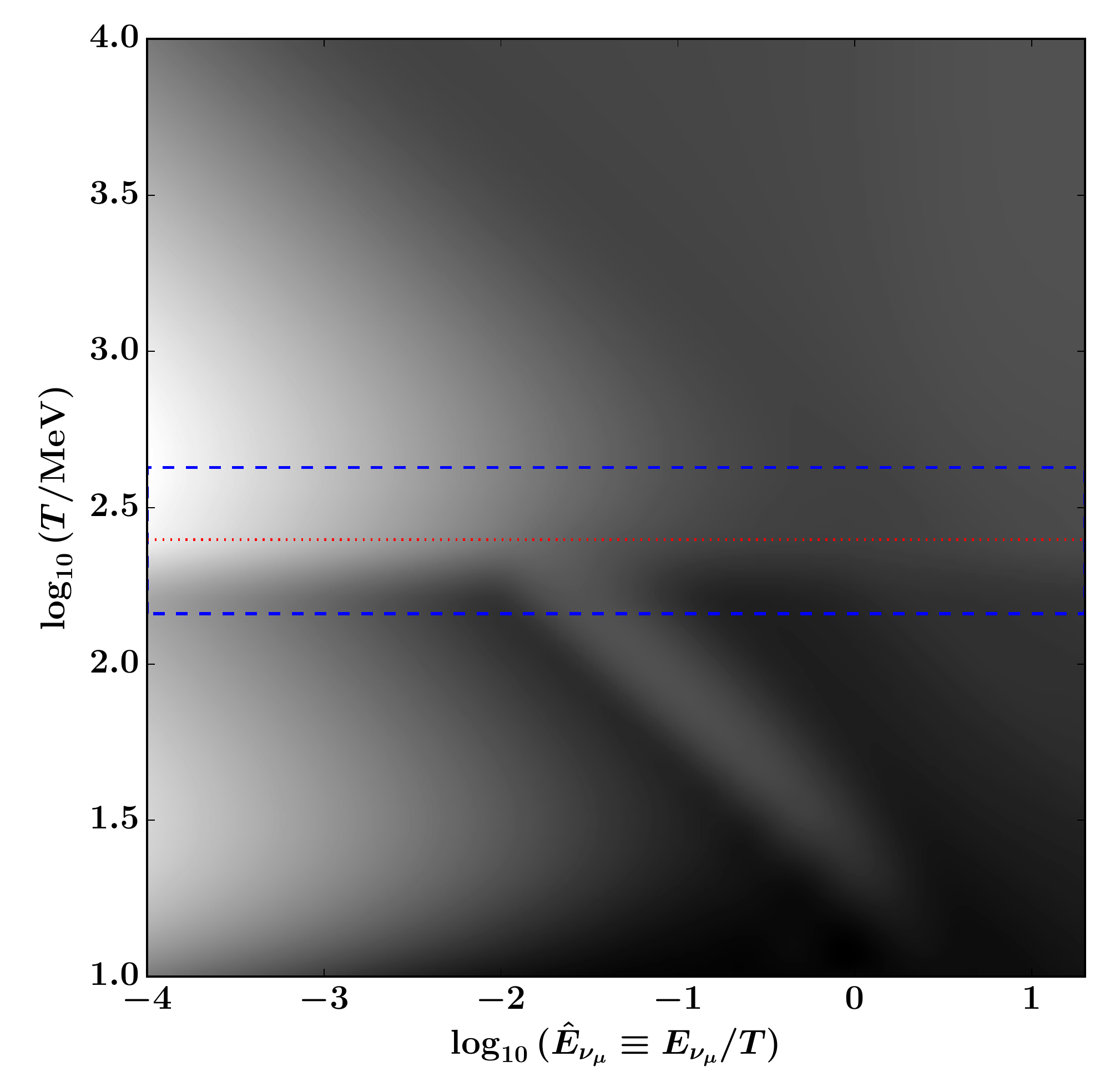}
    \label{fig:T_c_250MeV}
  }
  \hfill%
   \subfloat[][$T_{\rm c} = 1000 \ {\rm MeV}$]{
    \includegraphics[height=7.87cm]{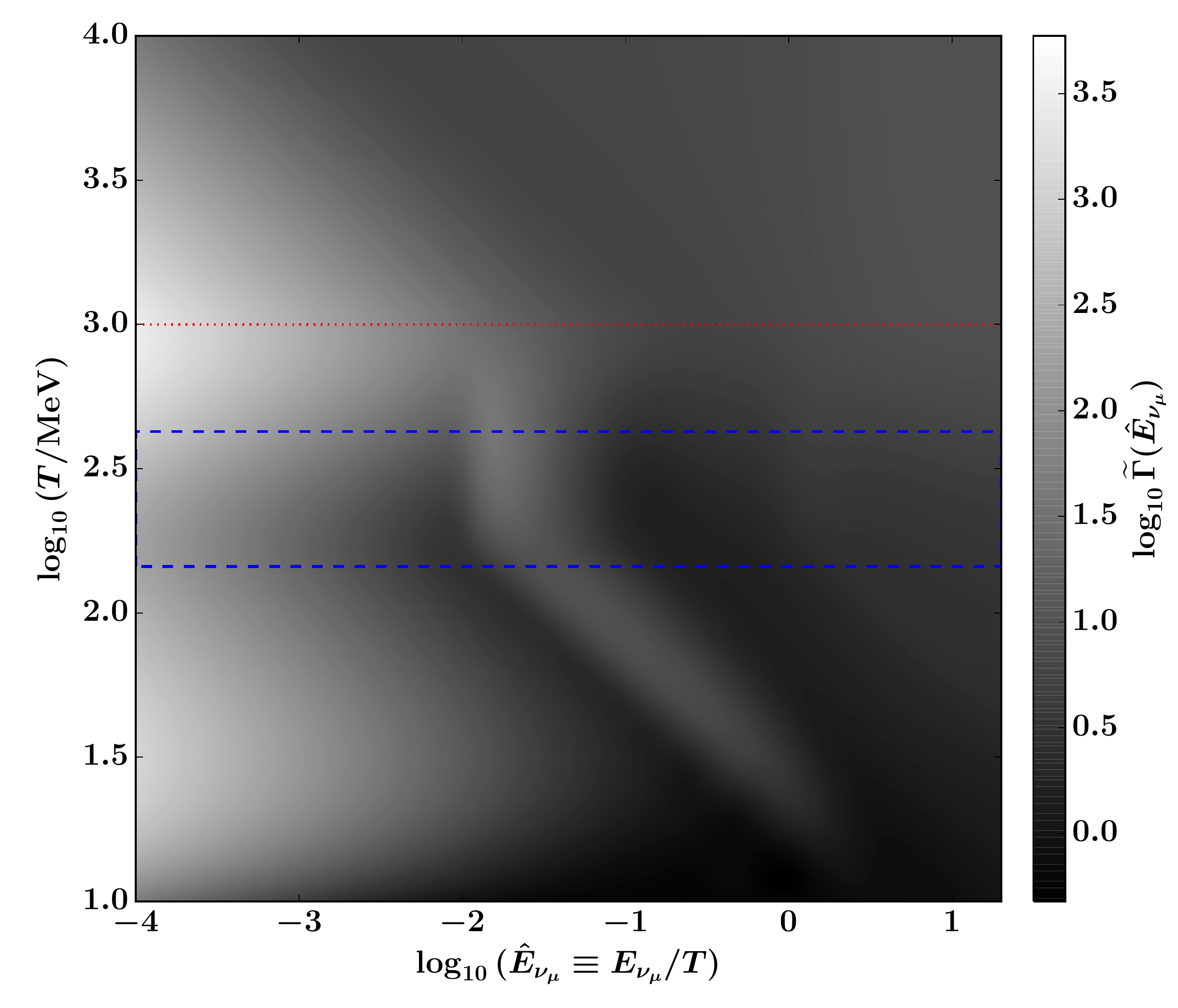}
    \label{fig:T_c_1000MeV}
  }
  \caption{\label{fig:scattering_rates} Scaled muon neutrino opacities for a range of energies and temperatures: panel (a) shows total, leptonic, and hadronic opacities vs energy at $T = 100 \ {\rm MeV}$ and $2 \ {\rm GeV}$. Panel (b) shows opacities at high energies ($E_{\nu_\mu} = 20 T$) vs temperature: black lines are two interpolations through $T_{\rm QCD}$. They are cubic splines labeled by their cutoff temperature, $T_{\rm c}$, as defined in the text. Colored lines are numbers of relativistic degrees of freedom: $g_*,{\rm KT}$ under assumption \# 3, i.e., that of Kolb and Turner \cite{Kolb90} with $T_{\rm QCD} = 210$ MeV, chosen to match Ref.~\cite{Laine06} whose results are $g_*,{\rm LS}(2006)$. Panels (c) and (d) show interpolated opacities vs energy and temperature. Blue dashed lines mark ranges where the two values of $g_*$ differ by more than $10 \%$, and are a rough guide to where these rates can be trusted. Red dotted lines mark the cutoff temperature.}
\end{figure*}

Figure \ref{fig:T_scattering_rates} shows the {\em total} opacities with muon neutrino energy at temperatures of $100 \ {\rm MeV}$ and $2 \ {\rm GeV}$. We note a few salient features of these rates.

Firstly, we note that the leptonic and hadronic two-particle to two-particle reaction rates approach the scaling of Eq.~\eqref{eq:pscaling} at large energies; the downturn at lower energies is due to Pauli blocking.

Secondly, both sets of rates exhibit a rise at low energies, which reflects non-zero limiting values of the {\em unscaled} rates. This is due to the behavior of cross sections for inelastic collisions involving massive particles, such as the three body collision $\nu_\mu + e^- + \bar{\nu}_e \rightarrow \mu^-$ or the scattering process $\nu_\mu + \mu^+ \rightarrow \nu_e + e^+$. We illustrate this by calculating the cross-section for the latter, while neglecting the positron's rest mass and Pauli blocking for simplicity. The squared and spin-summed/averaged matrix element for this process is
\begin{align}
  \left\langle \vert \mathcal{M} \vert^2 \right\rangle & = 128 G_{\rm F}^2 \left( p_{\nu_\mu} \cdot p_{e^+} \right) \left( p_{\mu^+} \cdot p_{\nu_e} \right) \mbox{.}
\end{align}
In the limit of zero neutrino energy
\begin{align}
  \left( p_{\mu^+} \cdot p_{\nu_e} \right) & = -m_{\mu}^2/2 + \mathcal{O}\left(E_{\nu_\mu}\right) \mbox{,}
\end{align}
which implies that modulus squared is
\begin{align}
  \left\langle \vert \mathcal{M} \vert^2 \right\rangle & = - 64 G_{\rm F}^2 m_{\mu}^2  \left( p_{\nu_\mu} \cdot p_{e^+} \right) + \mathcal{O}\left( E_{\nu_\mu} \right) \mbox{.}
\end{align}
Hence the cross section for the $\mu-$neutrino, integrated over outgoing particles' directions, is
\begin{align}
  \sigma_{\nu_\mu} & = \frac{G_{\rm F}^2 m_{\mu}^2}{\pi} + \mathcal{O}\left( E_{\nu_\mu} \right) \mbox{.}
\end{align}
Such non-zero limiting values are responsible for the rise in the {\em scaled} rates for soft neutrinos in Figs.~\ref{fig:leptonicrates} and \ref{fig:hadronicrates}

Thirdly, the hadronic opacities at low temperatures, i.e. $T<T_{\rm QCD}$, exhibit a series of peaks. These are signatures of two-body fusions, which are broad resonances in the propagators of the weak gauge bosons. These include the production of pseudoscalar mesons (e.g. pion production via $\nu_\mu + \mu^+ \rightarrow \pi^+$) and vector mesons (e.g. $\rho^0$ production via $\nu_\mu + \bar{\nu}_\mu \rightarrow \rho^0$). In the total opacities of Fig.~\ref{fig:T_scattering_rates}, the former is visible as a peak at intermediate momenta, while the latter are smeared out at large momenta.

Finally, we observe from Fig.~\ref{fig:T_scattering_rates} that the total opacities at high energies exhibit a jump as the temperature passes through $T_{\rm QCD}$. This is due to the increase in the number of hadronic degrees of freedom, as evidenced by the sizes of the jumps in hadronic- and leptonic contributions (the latter due to the tau lepton turning on).

This is shown clearly in Fig.~\ref{fig:highp_scattering_rates}, which shows the scaled muon neutrino opacities at high energies for a range of temperatures. Note that these rates assume that the hadronic species above and below the transition are free quarks and mesons respectively (assumption \# 3\ in our list above). For comparison, the figure shows the number of relativistic degrees of freedom, $g_*$, both under this assumption and from Ref.~\cite{Laine06}, which implements the running of the strong coupling constant. We note the significant deviation close to the quark-hadron transition ($T_{\rm QCD} = 210$ MeV in the lattice calculations underlying Ref.~\cite{Laine06}). 

Motivated by this, we explore two methods of interpolating opacities through the quark-hadron transition. In each of them, we choose a cutoff temperature, $T_{\rm c}$, above which we use the free quark results, and use a cubic spline interpolation in between. We emphasize that this is not physically motivated; the actual rates, and their matrix elements, need to incorporate the strong coupling constant and its running. The figure shows interpolations with $T_{\rm c} = 250$ MeV and $1000$ MeV, which we expect to bracket the range of rates.

With this caveat, Figs.~\ref{fig:T_c_250MeV} and \ref{fig:T_c_1000MeV} show interpolated $\mu$ neutrino opacities for a range of energies and temperatures, with $T_{\rm c} = 250$ MeV and $1000$ MeV, respectively. In the rest of the paper, we use these scattering rates and the potentials defined in Sec.~\ref{sec:overview} to study sterile neutrino production via oscillation in the early universe. We use both interpolations through the quark-hadron transition in order to illustrate the results' sensitivity to the scattering rates.

\section{Sterile neutrino production}
\label{sec:production}

In this section, we incorporate the standard model calculations of Sections \ref{sec:redistribution} and \ref{sec:scattering} into the sterile neutrino production mechanism, whose broad outline we provided in Sec.~\ref{sec:overview}.

We evolve the sterile neutrino and anti-neutrino PSDs, $f_{\nu_{\rm s}}(p)$ and $f_{\bar{\nu}_{\rm s}}(p)$, using the Boltzmann equation of Eq.~\eqref{eq:productionmaster}. We use the primordial plasma's temperature $T$ as a clock, and numerically integrate a thousand logarithmically spaced Lagrangian momentum bins from a temperature of $10$ GeV down to $10$ MeV. For the models illustrated by stars in Fig.~\ref{fig:parspace}, the vast majority of the production happens between these temperatures. We use the muon neutrino opacities derived in Sec.~\ref{sec:scattering} and provide results using the two interpolation scheme presented in Fig.~\ref{fig:scattering_rates}, which bracket the range of uncertainties due to the quark-hadron transition. We use Eq.~\eqref{eq:thermalv} for the thermal potential $V^{\rm th}_{\nu_\mu}$, and the results presented in Fig.~\ref{fig:asymmv} for the asymmetry potential $V^{\rm L}_{\nu_\mu}$ incorporating the redistribution of Sec.~\ref{sec:redistribution}. 

In order to close the system of equations, we also need the evolution of the plasma temperature $T$ and mu leptonic asymmetry $\hat{\mathcal{L}}_{\mu}$ with coordinate time $t$. Before discussing the details of the sterile neutrino production, we briefly review these two relations.
\subsection{Time-temperature relation}
In this subsection, we derive the time-temperature relationship prior to the epoch of weak decoupling. The Hubble rate, $H$, is
  \begin{equation}
    \frac{d}{d t} \ln a = H = \sqrt{ \frac{8\pi}{3 m_{\rm P}^2} (\rho_{\rm SM} + \rho_{\nu_{\rm s}})} \mbox{,} \label{eq:hubble}
  \end{equation}
  where $a$ is the scale factor, $m_{\rm P} = 1.2 \times 10^{19}$ GeV is the Planck mass, and $\rho_{\rm SM}$ and $\rho_{\nu_{\rm s}}$ are energy densities in standard model particles and sterile neutrinos, respectively. The latter is given by an integral over the PSDs, $\rho_{\nu_{\rm s}} = (1/2\pi^2) \int p^2 dp \sqrt{p^2 + m_{\rm s}^2} [ f_{\nu_{\rm s}}(p) + f_{\bar{\nu}_{\rm s}}(p) ]$. During Hubble expansion from $a$ to $a + \delta a$: a) the sterile neutrino PSDs evolve to $f_{\nu_{\rm s}/\bar{\nu}_{\rm s}}(p) + \delta f_{\nu_{\rm s}/\bar{\nu}_{\rm s}}(p)$ due to a combination of mixing with muon neutrinos, and their momentum redshifting as $p a \equiv {\rm const.}$ b) due to large neutrino opacities, all active species maintain equilibrium PSDs with a common temperature, whose evolution is affected by the production of sterile neutrinos.

The continuity equation for the {\em total} stress-energy tensor is
\begin{align}
  3\frac{d}{d T} \ln a & = - \frac{d}{d T} \left( \rho_{\rm SM} + \rho_{\nu_{\rm s}} \right) \left( \rho_{\rm SM} + P_{\rm SM} + \rho_{\nu_{\rm s}} + P_{\nu_{\rm s}} \right)^{-1} \mbox{,} \label{eq:energyconv}
\end{align}
where $P_{\rm SM/\nu_s}$ are SM and sterile neutrino pressures, respectively. The sterile energy density evolves according to
\begin{align}
  \frac{d \rho_{\nu_{\rm s}} }{d T} & = \frac{\partial \rho_{\nu_{\rm s}} }{\partial \ln a} \frac{d \ln a}{d T} + \frac{\partial \rho_{\nu_{\rm s}} }{\partial t} \frac{d t}{d T} \mbox{.} \label{eq:decomposition}
\end{align}
The two terms on the right-hand side are the free-streaming and oscillation contributions, respectively.
\begin{align}
  \frac{\partial \rho_{\nu_{\rm s}} }{\partial \ln a} & = - 3( \rho_{\nu_{\rm s}} + P_{\nu_{\rm s}} ) \label{eq:freestreaming} \mbox{,} \\
  \frac{\partial \rho_{\nu_{\rm s}} }{\partial t} & = \! \int \frac{dp \ p^2}{2\pi^2} \sqrt{p^2 + m_{\rm s}^2} \frac{\partial}{\partial t} \left[ f_{\nu_{\rm s}}(p) + f_{\bar{\nu}_{\rm s}}(p) \right] \label{eq:oscillation_cont}\mbox{.}
\end{align}
We substitute Eqs.~\eqref{eq:decomposition} and \eqref{eq:freestreaming} into Eq.~\eqref{eq:energyconv} and solve for the relation between the scale factor and temperature
\begin{align}
  3\frac{d \ln a}{d T}  & = - \left( \frac{d \rho_{\rm SM}}{d T} +  \frac{\partial \rho_{\nu_{\rm s}} }{\partial t} \frac{d t}{d T} \right) \left( \rho_{\rm SM} + P_{\rm SM} \right)^{-1} \mbox{.}
\end{align}
Substituting Eq.~\eqref{eq:hubble}, we obtain the time-temperature relation\footnote{We note that we correct here an error introduced in Ref.~\cite{Abazajian01}.}
\begin{align}
  \frac{d T}{d t} & = - \frac{ 3 H [\rho_{\rm SM} + P_{\rm SM} ] + (\partial \rho_{\nu_{\rm s}}/\partial t) }{ d \rho_{\rm SM}/d T } \mbox{.}
\end{align}
Defining the number of SM relativistic degrees of freedom for the energy and entropy densities via
\begin{align}
    \rho_{\rm SM} & = \frac{\pi^2}{30} g_* T^4 \mbox{,} \\
    s_{\rm SM} = \frac{\rho_{\rm SM} + P_{\rm SM}}{T} & = \frac{2\pi^2}{45} g_{*,s} T^3 \mbox{,}
\end{align}
we have the final form of the time-temperature relation
\begin{align}
  \frac{d T}{d t} & = - \frac{ 4 H g_{*,s} T^4 + (30/\pi^2) (\partial \rho_{\nu_{\rm s}}/\partial t) }{ d [g_* T^4]/d T } \mbox{.} \label{eq:timetemp}
\end{align}
We use numbers of relativistic degrees of freedom $g_*$ and $g_{*,s}$ from Ref.~\cite{Laine06} in our numerical implementation. 

\subsection{Time-evolution of asymmetry}
The temperature-scaled muon asymmetry, $\hat{\mathcal{L}}_\mu$, evolves both from the depletion of relativistic degrees of freedom due to annihilations and from the production of sterile neutrinos. There are subtleties in dealing with the latter in the case of resonant production \cite{2006PhRvL..97n1301K}, but for the semi-classical approach outlined in Sec.~\ref{subsec:boltzmann}, we can write down the contribution in terms of the evolution of sterile neutrino PSD. Keeping in mind the definition of the lepton asymmetry in Eq.~\eqref{eq:asymmdef}, the asymmetry evolution due to both contributions together is
\begin{align}
~~~ & \!\!\!\!
  \frac{d \hat{\mathcal{L}}_\mu }{d t} \nonumber \\
  & = \frac{d}{d t} \int \frac{d\hat{p} \, \hat{p}^2}{2 \pi^2} \left[ f_{\nu_\mu}(p) - f_{\bar{\nu}_\mu}(p) + 2 f_{\mu^-}(p) - 2 f_{\mu^+}(p) \right] \nonumber \\
  & = -3 \left[ H + \frac{d \ln T}{dt} \right] \hat{\mathcal{L}}_\mu  - \int \frac{d\hat{p} \, \hat{p}^2}{2 \pi^2} \frac{\partial}{\partial t} \left[ f_{\nu_{\rm s}}(p) - f_{\bar{\nu}_{\rm s}}(p) \right] \mbox{,}
    \label{eq:asymmevol}
\end{align}
where the symbol $\hat{p}$ is the temperature-scaled momentum, $\hat{p} \equiv p/T$. The first term in the square bracket in the last line above can be evaluated with the help of Eqs.~\eqref{eq:hubble} and \eqref{eq:timetemp}, while the second term can be evaluated using Eq.~\eqref{eq:productionmaster}. Our large number of momentum bins $(1000)$ allows us to use spline integration at every time step in order to perform the momentum integrals in Eqs.~\eqref{eq:asymmevol} and \eqref{eq:oscillation_cont}.  We set up our Lagrangian momentum bins such that $5\times10^{-3}\leq p/T\leq 20$ at temperature $T=10$ GeV. We have checked that this range is more than sufficient to accurately capture the most relevant range of the sterile neutrino PSDs. 

\begin{figure}[t]
\begin{center}
   \includegraphics[height=6cm]{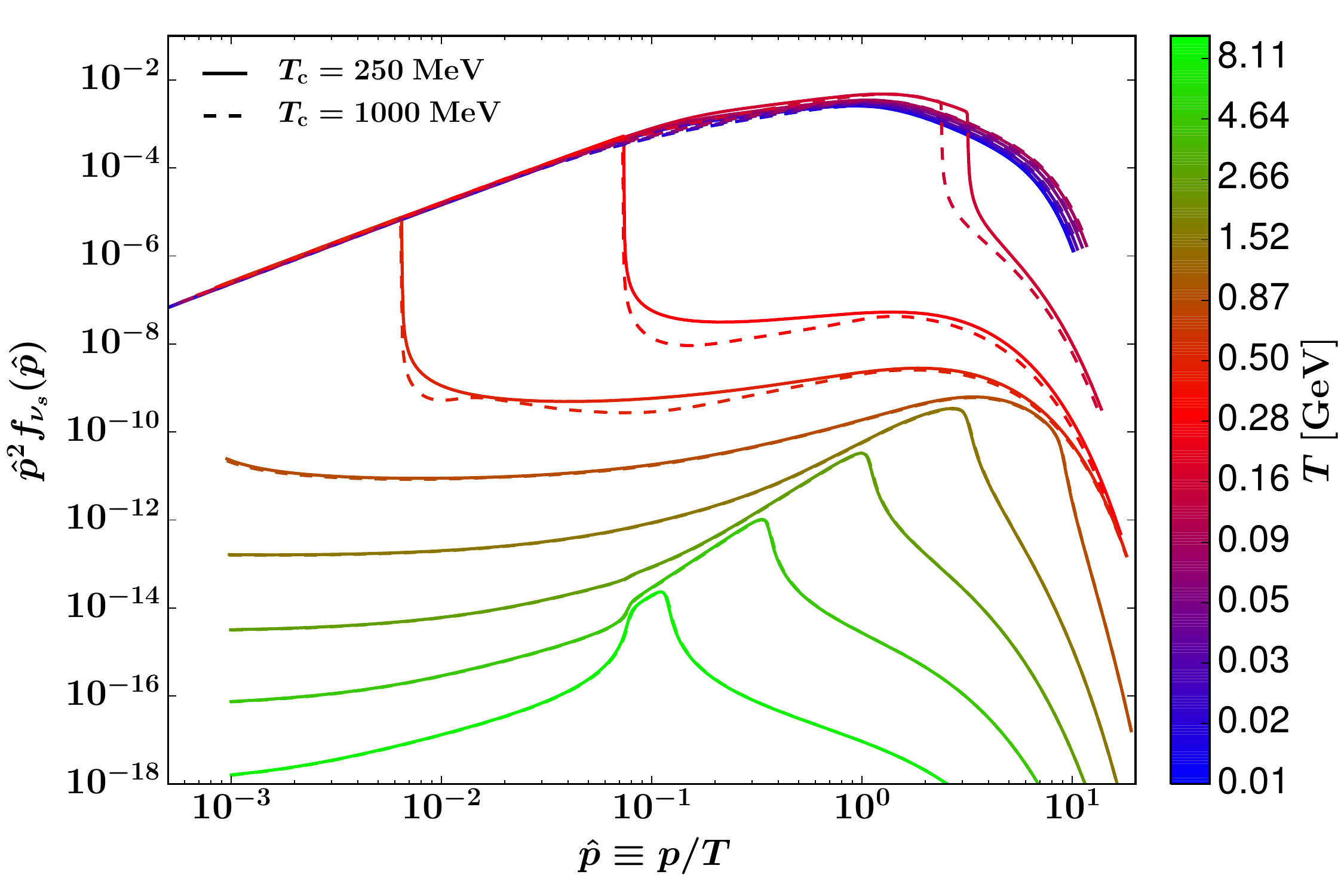}
\caption{We illustrate the temperature-evolution of the sterile neutrino's PSD  for the central model of Fig.~\ref{fig:parspace} with $(m_{\rm s}, \sin^2{2\theta}) = (7.1 \ {\rm keV}, 4 \times 10^{-11})$. Solid and dashed lines distinguish results with neutrino opacities from Fig.~\ref{fig:T_c_250MeV} and \ref{fig:T_c_1000MeV}, respectively.}
\label{fig:loge2fe_temp}
\end{center}
\end{figure}
\subsection{Resonant Production}
\begin{figure*}[t]
  \hspace*{\fill}%
  \subfloat[][mu lepton asymmetry]{
    \includegraphics[height=6cm]{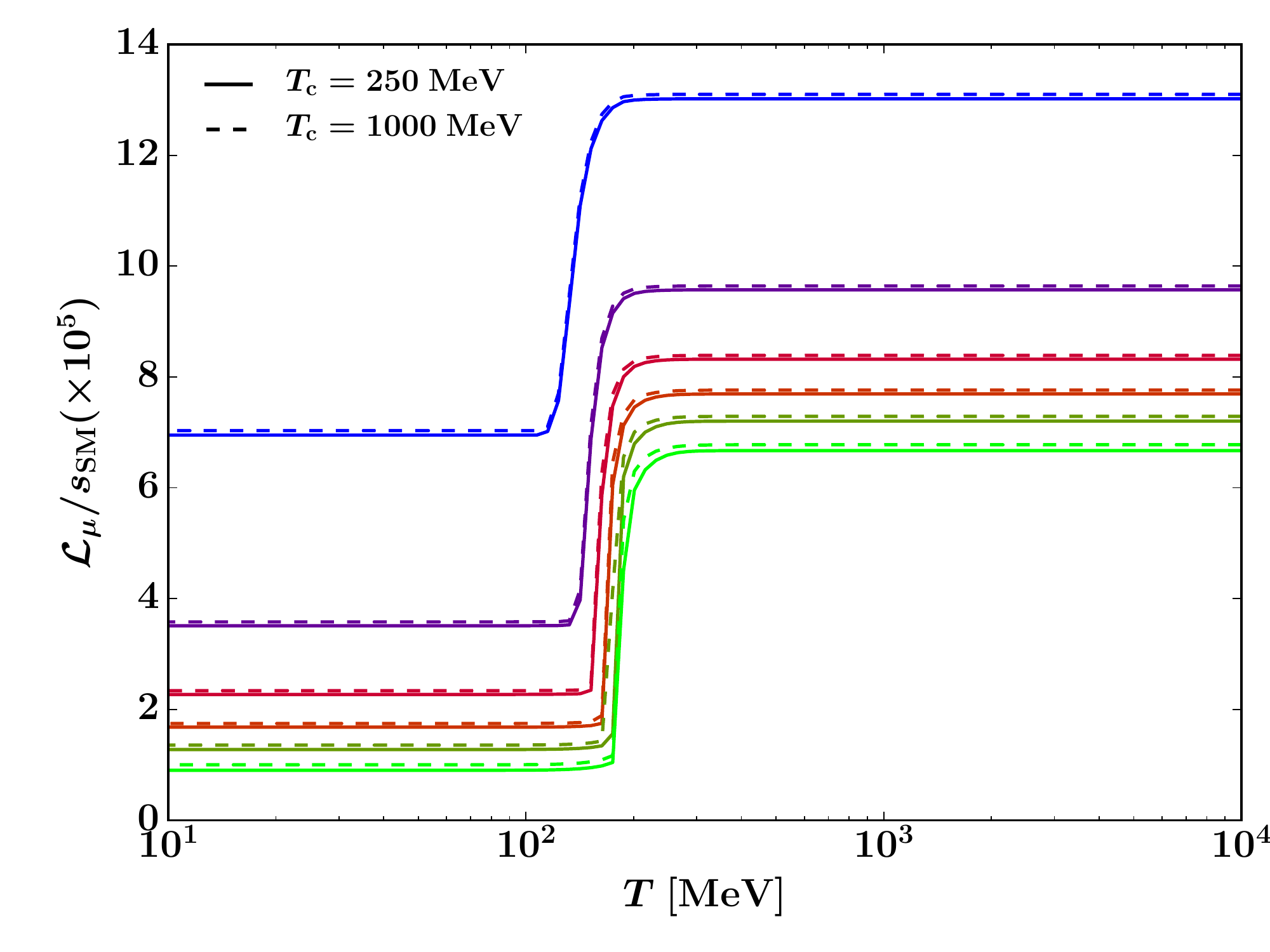}
    \label{fig:L_s_all}
  }
  \hfill
  \subfloat[][sterile neutrino PSDs at $T=10$ MeV]{
    \includegraphics[height=6cm]{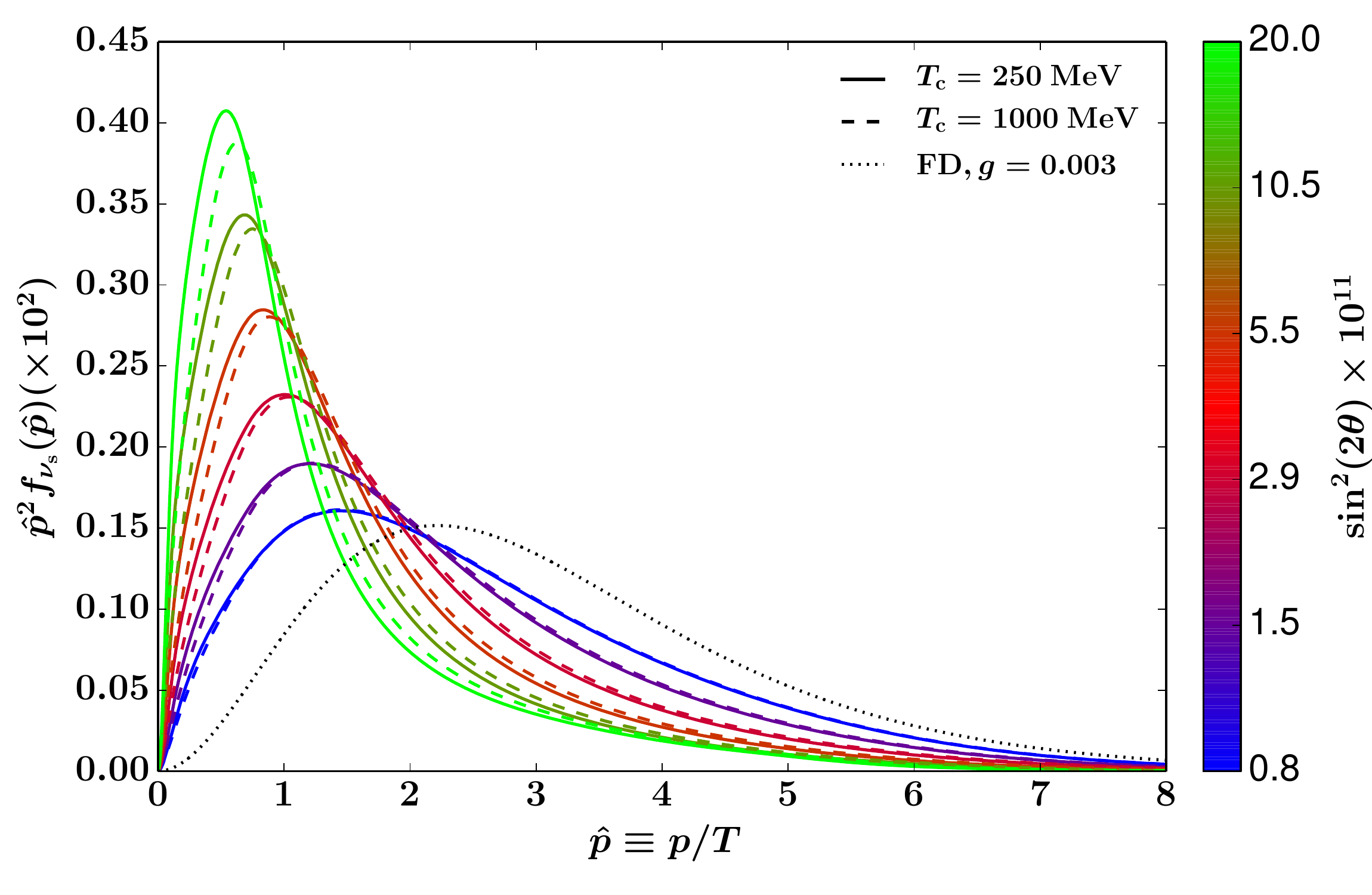}
  \label{fig:e2fe_all}
  }\\
  \hspace*{\fill}%
  \subfloat[][net sterile density]{
    \includegraphics[height=6cm]{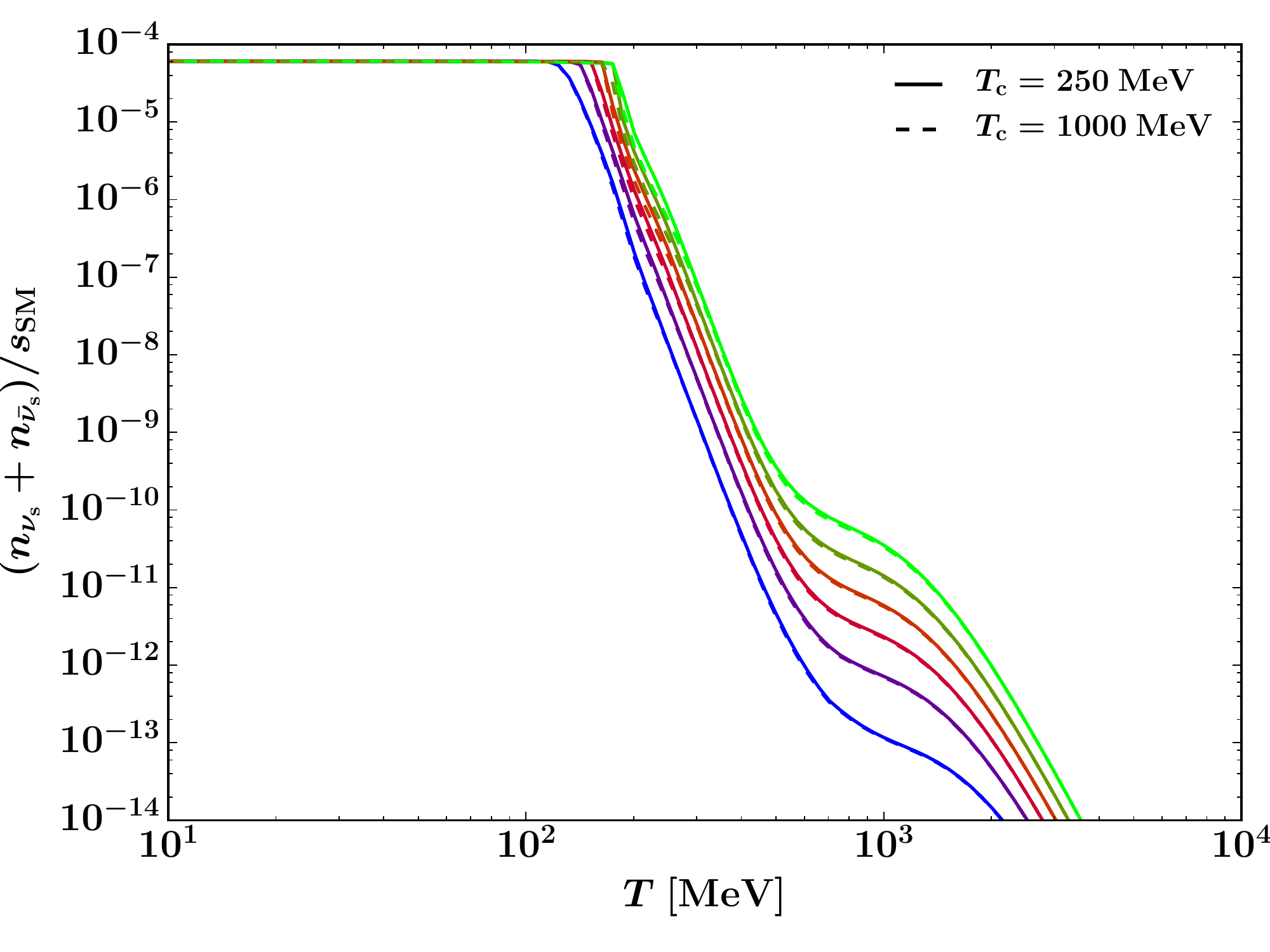}
    \label{fig:n_s_all}
  }
  \hfill
  \subfloat[][sterile antineutrino PSDs at $T=10$ MeV]{
    \includegraphics[height=6cm]{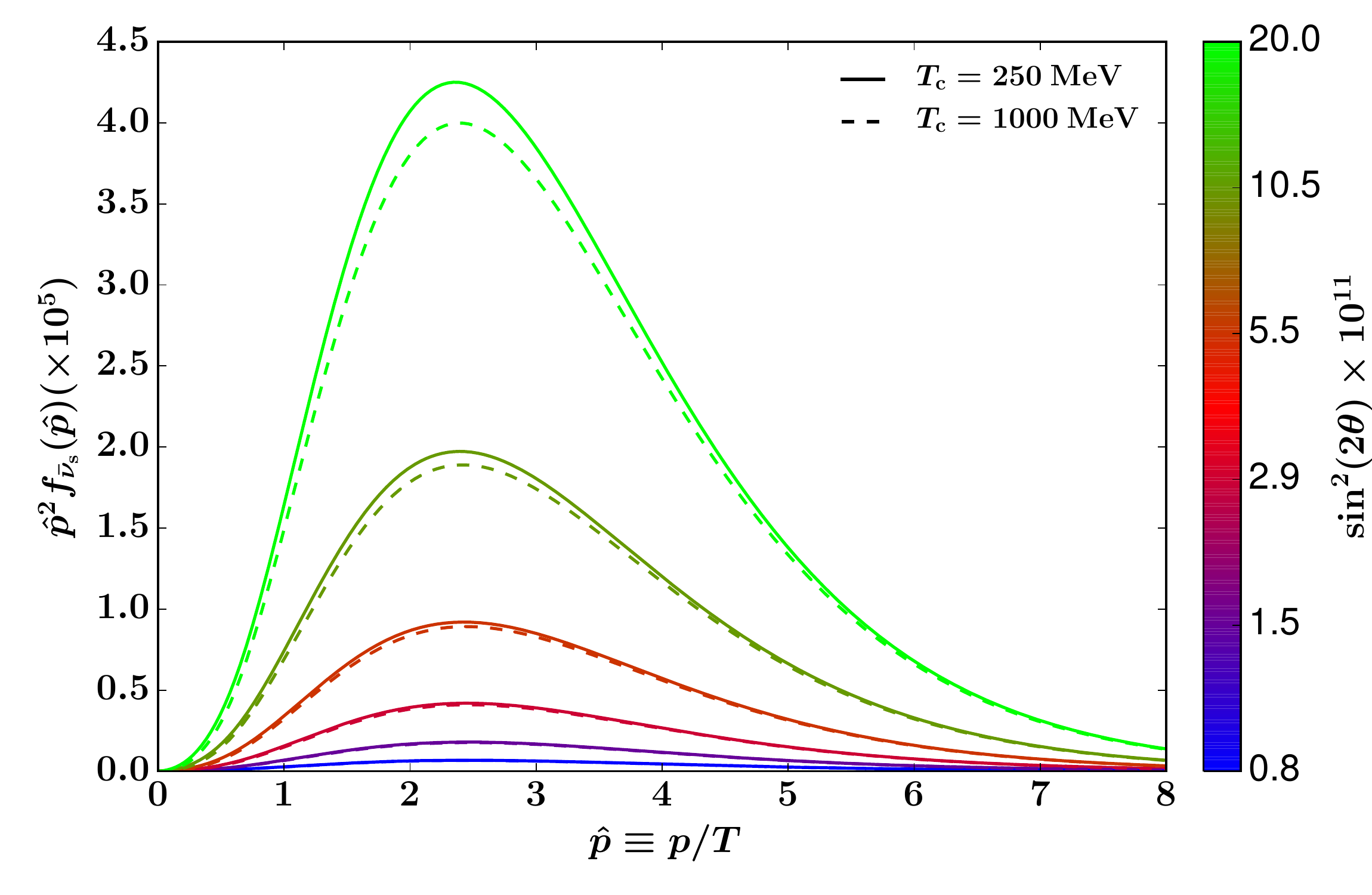}
    \label{fig:e2fe_nusbar_all}
  }
  \caption{\label{fig:production} Sterile neutrino production mechanism: Panels (a) and (c) show the entropy scaled mu lepton asymmetry and the net sterile number density with temperature. For each model with a given mass and mixing angle, the mu lepton asymmetry at high temperatures is tuned by hand to produce the right relic abundance. Panels (b) and (d) show sterile neutrino and antineutrino PSDs, respectively, at $T=10$ MeV. Colors differentiate models in Fig.~\ref{fig:parspace}, and solid and dashed lines distinguish results with neutrino opacities from Fig.~\ref{fig:T_c_250MeV} and \ref{fig:T_c_1000MeV} respectively. Note the different numerical factors multiplying the y-axis of panels (b) and (d). The dotted line in panel (b) is a massless Fermi-Dirac distribution with degeneracy $g = 0.003$.}
\end{figure*}

As described in Sec.~\ref{sec:overview}, the presence of a lepton asymmetry leads to a resonant production of sterile neutrinos with specific momenta. Through Eq.~\eqref{eq:productionmaster}, the resonant momenta at a particular temperature satisfy
\begin{align}
  \Delta(p) \cos{2\theta} - V^{\rm L} - V^{\rm th}(p) & = 0 \mbox{.}
\end{align}
Substituting the definition of $\Delta(p)$ and the potentials from Eq.~\eqref{eq:potentials}, we obtain
\begin{align}
  \frac{m_{\rm s}^2}{2 p} - \frac{d V^{\rm L}}{d \mathcal{L}_\mu} \mathcal{L}_\mu - \frac{d V^{\rm th}(p)}{d p} p & = 0 \mbox{.}
\end{align}
There are two roots, i.e. two momenta resonant at any temperature \cite{Abazajian01}. Consideration of the terms' approximate temperature scaling shows that each scaled root ($\hat{p}\equiv p/T$) sweeps to larger values at lower temperatures (ignoring changes in the numbers of relativistic degrees of freedom). This is reflected in Fig.~\ref{fig:loge2fe_temp}, which shows the sterile neutrino PSD's evolution with temperature for the central model in Fig.~\ref{fig:parspace} with $m_{\rm s} = 7.1$ keV and $\sin^2{2\theta} = 4 \times 10^{-11}$. We observe that most of the neutrinos are produced at the lower resonance and at temperatures close to $T_{\rm QCD}$. This is illustrated by Figs.~\ref{fig:L_s_all} and \ref{fig:n_s_all}, which show the evolution of the entropy-scaled\footnote{We show this scaling rather than the one with temperature, since it is conserved through epochs of annihilation.} $\mu$ lepton asymmetry and the net sterile neutrino and antineutrino density for the range of models marked by stars in Fig.~\ref{fig:parspace}. The latter is also sensitive to thermal (nonresonant) production, which operates at all temperatures, but is subdominant for the mixing angles of interest. 

Figures.~\ref{fig:e2fe_all} and \ref{fig:e2fe_nusbar_all} show the sterile neutrino and antineutrino PSDs at $T=10$ MeV for these models. We note that the sterile antineutrinos are produced off-resonance for the positive lepton asymmetries we consider here, and their abundance is thus significantly suppressed compared to that of the sterile neutrinos. Solid and dashed lines in Fig.~\ref{fig:production} show results for the two interpolations of the $\mu$ neutrino opacities through $T_{\rm QCD}$ presented in Fig.~\ref{fig:scattering_rates}, which differ in the temperature range $150 \ {\rm MeV} < T < 1 \ {\rm GeV}$. For small values of the mixing angle, we observe that there is little difference between the PSDs computed using our two different interpolation schemes for the neutrino opacity. For these models, most of the production happens at temperature below the quark-hadron transition where our two opacity approximation schemes are essentially the same, hence leading to similar PSDs. As the mixing angle is increased, the production is pushed toward higher temperatures (see Fig.~\ref{fig:n_s_all}) where the difference between our two interpolation schemes is greater, leading to a larger uncertainties in the final PSDs. 

Table \ref{tab:production_data} lists parameters describing the production and final sterile neutrino DM PSDs for the models marked in Fig.~\ref{fig:parspace}. Also provided are the ranges for different interpolated $\mu$ neutrino opacities through the quark-hadron transition as in Fig.~\ref{fig:scattering_rates}. Note that the sterile PSDs in Figs.~\ref{fig:e2fe_all} and \ref{fig:e2fe_nusbar_all} are non-thermal; we show the mean momentum $\langle p/T \rangle$ relative to the active neutrino temperature scale.   

A key element to take away from Table \ref{tab:production_data} and Figs.~\ref{fig:e2fe_all} and \ref{fig:e2fe_nusbar_all} is that the `warmer' models with larger values of $\langle p/T \rangle$ are less sensitive to the uncertainty in the quark-hadron transition. This is important since these warmer models can be most easily constrained by small-scale structure formation. Therefore, uncertainties in the strong plasma near $T_{\rm QCD}$ are unlikely to affect the robustness of the these constraints. 

\begin{table}[t]
   \caption{\label{tab:production_data} Parameters for the models marked in Fig.~\ref{fig:parspace}, with $m_{\rm s} = 7.1$ keV and $\Omega_{\rm DM} h^2 = 0.119$ \cite{PlanckCosmology}. The ranges displayed in the three last columns account for the uncertainties in the neutrino opacities near the quark-hadron transition. }
   \begin{ruledtabular}
     \begin{tabular}{cccc}
        $\sin^2{2\theta}$ & $\left(\mathcal{L}_{\mu}/s_{\rm SM}\right)_{\rm i}$ & $\left(\mathcal{L}_{\mu}/s_{\rm SM}\right)_{\rm f}$ & $\langle p/T \rangle$ \footnotemark[1] \\
         & at $T = 10$ GeV & at $T = 10$ MeV & \\
         $\times 10^{-11}$ & $\times 10^{-5}$ & $\times 10^{-5}$ & \\
        \hline
        0.800 & 13.0 -- 13.1 & 6.95 -- 7.03 & 2.60 -- 2.61 \\
        1.104 & 10.80 -- 10.88 & 4.74 -- 4.81 & 2.45 -- 2.47 \\
        1.523 & 9.57 -- 9.64 & 3.51 -- 3.58 & 2.28 -- 2.32 \\
        2.101 & 8.81 -- 8.88 & 2.76 -- 2.83 & 2.12 -- 2.16 \\
        2.899 & 8.32 -- 8.39 & 2.27 -- 2.34 & 1.95 -- 2.01 \\
        4.000 & 7.96 -- 8.03 & 1.93 -- 2.00 & 1.80 -- 1.87 \\
        5.519 & 7.69 -- 7.76 & 1.68 -- 1.74 & 1.66 -- 1.74 \\
        7.615 & 7.45 -- 7.53 & 1.47 -- 1.54 & 1.53 -- 1.62 \\
        10.506 & 7.20 -- 7.29 & 1.28 -- 1.36 & 1.43 -- 1.52 \\
        14.496 & 6.95 -- 7.05 & 1.09 -- 1.18 & 1.35 -- 1.44 \\
        20.000 & 6.7 -- 6.8 & 0.9 -- 1.0 & 1.29 -- 1.38 \\
     \end{tabular}
     \footnotetext[1]{The sterile DM distributions are non-thermal; we compute $\langle p/T \rangle$ using the active neutrino temperature. Below the epoch of $e^{\pm}$ annihilation, the latter is related to the CMB temperature by the factor $(4/11)^{1/3} = 0.714$. We note that for a Fermi-Dirac distribution  $\langle p/T \rangle\simeq3.15$ }
   \end{ruledtabular}
\end{table}


\section{Transfer functions for matter fluctuations}
\label{sec:transferfunc}

In this section, we study the effect of sterile neutrinos on the growth of density fluctuations in the early universe. We focus on the lepton asymmetry-driven mechanism outlined in Sec.~\ref{sec:overview}, and on modes of the matter distribution with co-moving wavenumbers $k \in [1, 100] \ h\, {\rm Mpc}^{-1}$. These scales are probed by the Lyman-$\alpha$ forest in quasar spectra (see \cite{Viel13} and references therein), and populations of dwarf galaxies in the Local Group (see \cite{Horiuchi14,Polisensky14} and references therein). All these scales enter the horizon after the redshift $z_{\rm H} \simeq 4 \times 10^{7}$, when the temperature of the photon-baryon plasma is $T \simeq 10$ keV. The sterile neutrino models shown in Fig.~\ref{fig:parspace} cease to be produced below temperatures $T \sim 100$ MeV; hence we can assume they are essentially collisionless in this section.

The main effect of such a collisionless component on matter fluctuations is suppression due to free-streaming in the epochs where it is relativistic \cite{Gilbert66,Bisnovaty-Kogan71}. Previous works extensively studied this in the context of warm and/or neutrino DM models (see Refs.~\cite{Bond80,Bond83} and references therein), and identified the characteristic scales at which the suppression set in as a function of the neutrinos' mass and mean momentum \cite{Abazajian01}.

In order to obtain the suppression's detailed form, we need to incorporate the PSDs of the sterile neutrinos and antineutrinos into the Boltzmann equation for the DM component. This entails solving a perturbed form of Eq.~\eqref{eq:productionmaster}, with additional terms due to inhomogeneities, but without the source (production) terms. The scales of interest are non-linear in the current epoch, but we only provide the linear transfer functions at $z=0$, which can be used as initial conditions for cosmological N-body simulations.

We use the publicly available CLASS solver \cite{Blas11} to integrate the perturbed linear Boltzmann equation\footnote{Our choice was motivated by the availability of well-documented modules to deal with non-cold relics. We have checked our results against those obtained from a modified version of the publicly available CAMB solver \cite{Lewis99}.}. We initiate the solver with the Planck background parameters \cite{BBN2015}, except with the CDM component replaced by collisionless components with PSDs as shown in Figs.~\ref{fig:e2fe_all} and \ref{fig:e2fe_nusbar_all}. Since we are interested in the detailed shape of the transfer function, we turn off the default fluid approximation for non-cold relics \cite{Lesgourges11}. 

\begin{figure}[t]
  \begin{center}
    \includegraphics[height=6cm]{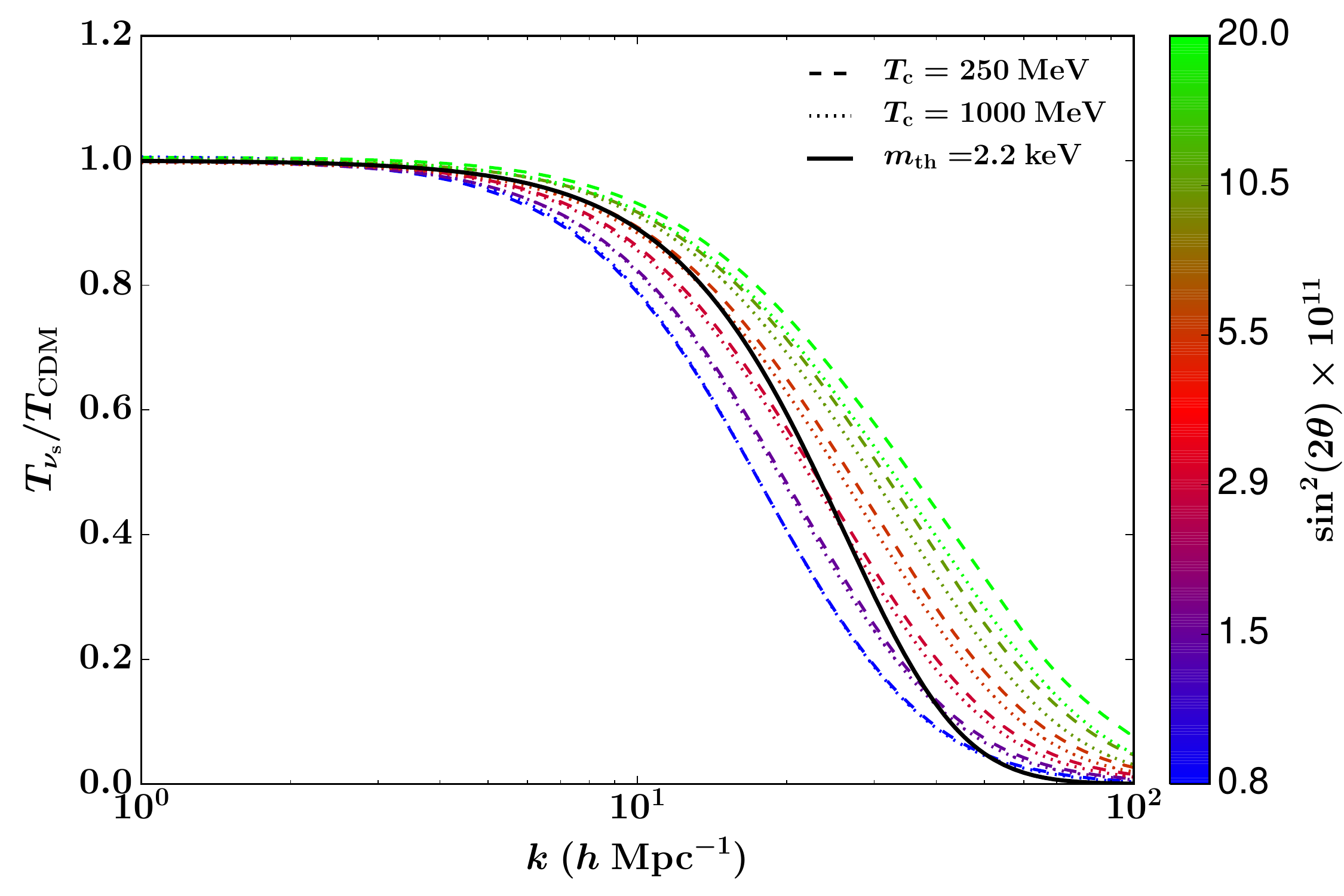}
  \end{center}
  \caption{Suppression of the transfer functions of overall density fluctuations relative to the $\Lambda$CDM ones for sterile neutrino models in Fig.~\ref{fig:parspace}, as a function of wavenumber. Dashed and dotted lines show results for the interpolated $\mu$ neutrino scattering rates of Figs.~\ref{fig:T_c_250MeV} and \ref{fig:T_c_1000MeV}, respectively. The solid black line is the numerical fit for a thermal warm DM transfer function as given in Ref.~\cite{Viel05}. }
  \label{fig:transferfuncs}
\end{figure}

Figure \ref{fig:transferfuncs} shows the resulting suppression as a function of the comoving wavenumber. We illustrate the suppression in the fluctuations' transfer functions relative to their values in $\Lambda$CDM. Also shown is the commonly-used fit to a thermal warm DM transfer function given in Refs.~\cite{Bode:2000gq,Viel05} with an `equivalent thermal mass' of $m_{\rm th} = 2.2$ keV; fits for models marked with stars in Fig.~\ref{fig:parspace} have a range of $1.6$ to $3.2$ keV. 

However, the strong difference in shape with the thermal WDM transfer function warrants use of the exact sterile neutrino dark matter transfer functions. The thermal warm DM PSDs relevant to the fit are rescaled versions of the Fermi-Dirac distribution; as can be seen from Fig.~\ref{fig:e2fe_all}, the resonantly-produced DM's PSD has an excess at low momenta that cannot be reproduced by such a rescaling. Hence, our DM transfer functions do not exhibit the fits' steep $\sim k^{-10}$ dependence at large wavenumbers and the resultant severe suppression of power on small scales. This indicates that the models considered in the present work are more likely to be in agreement with small-scale structure formation constraints, as recently pointed out in Refs.~\cite{Boyarsky:2008mt,Boyarsky:2009ix,Maccio:2012uh,Anderhalden:2012jc,Schneider:2013wwa,Kennedy:2013uta,Bose:2015mga}.  


\section{Discussion and conclusions}
\label{sec:discussion}

Sterile neutrinos are a well-motivated extension of the standard model of particle physics, and offer a promising candidate for the inferred DM population of the Universe. In this paper, we performed a detailed study of the resonant production of sterile neutrinos with masses and mixing angles relevant to the recent X-ray excess. In doing so, we explored the rich phenomenology associated with the active neutrinos' weak interaction with the primordial plasma. These interactions efficiently redistribute primordial lepton asymmetries among all the available degrees of freedom, and impact the temperature and momentum dependence of neutrino opacities. We incorporated these effects into the sterile production calculation, corrected and extended the existing numerical implementation, and obtained revised DM phase space densities. We finally computed transfer functions for fluctuations in the matter density, which can be used as starting points for $N$-body simulations of cosmological structure formation.

For the parameters relevant to the X-ray excess, resonant sterile neutrino production coincidentally occurs in the vicinity of the quark-hadron transition (see Fig.~\ref{fig:n_s_all}). Strongly interacting degrees of freedom affect the production in two ways: a) they influence both asymmetry redistribution and neutrino opacities through their interaction with the weak gauge bosons ($Z$ and $W^\pm$), and b) the transition from free quarks to hadrons at $T_{\rm QCD}$ influences the time-temperature relation [Eq.~\eqref{eq:timetemp}]. We now consider the robustness of each of these elements to the remaining uncertainties in the quark-hadron transition.

The asymmetry redistribution among the strongly-interacting degrees of freedom depends on the susceptibility of the quark-hadron plasma to baryon number and electric charge fluctuations. At high temperatures, we use tree-level perturbative QCD to compute the susceptibilities. There are uncertainties concerning the exact values of the quark masses, loop corrections, and the exact implementation of the $\overline{\rm MS}$ renormalization scheme. We expect these to have little effect on the final sterile neutrino PSDs since the bulk of the production occurs at lower temperatures, where the lattice QCD- and HRG-derived susceptibilities are most relevant. Thus, uncertainties in the asymmetry redistribution are likely dominated by systematic errors in the lattice calculations \cite{Borsanyi12}, measurement errors in the hadronic resonances' masses, and inaccuracies inherent in the HRG approach near the quark-hadron transition. Our confidence in the fit we use in this work is bolstered by the facts that a) an independent lattice QCD calculation \cite{Bazavov12} find very similar susceptibilities to those we used, and b) the HRG approach -- without any free parameter -- is in very good agreement with the lattice calculation for $T\lesssim150$ MeV. It is therefore unlikely that uncertainties in the susceptibilities will lead to dramatic changes in the sterile neutrino PSDs.
 
The validity of our neutrino opacities is much less clear -- we have attempted to calculate them in as much detail as possible, but the hadronic parts still retain significant uncertainties due to the quark-hadron transition. We expect that opacities at high and low temperatures are well described by the rates of reactions involving free-quarks, and the lightest pseudoscalar and vector mesons, which are shown in Sec.~\ref{sec:scattering}. For temperatures near $T_{\rm QCD}$, we have considered two interpolation schemes (shown  in Fig.~\ref{fig:scattering_rates}) that we expect might bracket the range of possibilities. We have computed the sterile neutrino PSDs for both cases and shown that they are fairly robust to the choice used, especially for models with larger values of the average momentum $\langle p/T \rangle$. We leave the calculation of self-consistent opacities through the transition to future work. Yet another approximation we have made is that of equilibrium distributions for all active species, which has been studied in a different context in Ref.~\cite{Hannestad:2015tea}. We expect this to be valid at the temperatures relevant to the models we study.

To compute the Hubble expansion rate and time-temperature relation, we have used the plasma's equations of state provided in Ref.~\cite{Laine06}, which are obtained by matching to the lattice QCD results of \cite{Boyd96}. As the former's authors point out, this result is still uncertain at temperatures close to the quark-hadron transition. It would interesting to update their result with the latest lattice QCD computations, which suggest a lower transition temperature \cite{Bhattacharya:2014ara}. We expect the uncertainties associated with the plasma's equation of state to be at most similar in magnitude to those coming from the neutrino opacity \cite{2007JHEP...01..091A}.

Another simplification we adopted is the semi-classical Boltzmann equation, which greatly facilitates our study of the oscillation-driven production. As mentioned in Sec.~\ref{subsec:boltzmann}, the most general analysis considers the evolution of a two-state density matrix, rather than phase-space densities. The validity of the semi-classical approach rests on the assumption that collisions dominate the off-diagonal element of the Hamiltonian that is responsible for vacuum oscillations \cite{Bell99,Volkas00,Lee00}. For typical momenta at the temperatures of interest, the ratio of these terms is $\Delta(T) \sin^2{2\theta}/D(T) \simeq 0.6 \times (T/100 \ {\rm MeV})^{-6} (m_{\rm s}/7 {\rm keV})^2 (\sin^2{2\theta}/10^{-11})^{1/2}$. The production of sterile neutrinos happens at temperatures above, but close to where these terms become comparable (note the ratio's steep temperature dependence). Thus, we expect that the results in this paper are relatively unaffected by this approximation, but further work in this direction can settle this question.

Finally, we examine the assumptions underlying the model itself, which were enumerated in Section \ref{subsec:assumptions}. If there is indeed an extra neutrino that is an electroweak singlet, it is not restricted to mix with only one flavor. However, the general case where the sterile neutrino mixes with all flavors introduces extra mixing angles, which cannot be constrained as easily from observations. The same can be said about the assumption of a lepton asymmetry in a single flavor. We briefly remark on the possibility of the sterile neutrinos mixing with electron or tau flavors instead. The redistribution of Sec.~\ref{sec:redistribution} is almost identical for the cases with electronic and muonic lepton asymmetries, but is different in the tauonic case. This is due to the significantly larger mass of the corresponding charged lepton ($m_\tau = 1.77$ GeV \cite{Agashe:2014kda}), which annihilates away at higher temperatures. Thus most of an input tau asymmetry ends up in the tau neutrino below $T \lesssim 400$ MeV, and the quark hadron transition does not impact the redistribution. The electron and tau neutrino opacities are different from the muonic case, and so is the balance between the thermal and asymmetry potentials, which affects the resonant momenta and ultimately the final dark matter PSDs -- we leave for future work the possibility of sterile neutrinos mixing with those flavors.

Also worth considering is active--active neutrino mixing, which does not conserve asymmetries in the individual flavors. This was studied in Ref.~\cite{Abazajian02}, which showed that such asymmetries are frozen in at the temperatures of interest. An interesting possibility is to revisit this study and use the redistributed asymmetries of Sec.~\ref{sec:redistribution} to calculate the active neutrino self energies at this epoch.

In conclusion, we find remarkable that sterile neutrino models that are in agreement with the X-ray excess have transfer function shapes that can significantly impact structure formation on subgalactic scales. Fixing the leptonic asymmetry to produce the right DM relic density, the resonantly-produced sterile neutrino transfer function goes from `warm' to `cold' as the mixing angle is increased from small to large values. This indicates that upcoming X-ray observations \cite{2014SPIE.9144E..25T,2015arXiv150605519F} and ongoing efforts to study small-scale structure can together cover all of the allowed mixing angle parameter space, and consequently confirm or disfavor the model.


\begin{acknowledgments}
We thank Olivier Dor\'{e} and Roland de Putter for fruitful discussions. We are grateful to Mikko Laine for providing us the data for the plasma's equation of state. T.V. and C.M.H. are supported by the David and Lucile Packard Foundation, the Simons Foundation, and the U.S. Department of Energy. The work of F.-Y. C.-R. was performed in part at the California Institute of Technology for the Keck Institute for Space Studies, which is funded by the W. M. Keck Foundation. Part of the research described in this paper was carried out at the Jet Propulsion Laboratory, California Institute of Technology, under a contract with the National Aeronautics and Space Administration (NASA). K.N.A. is partially supported by NSF CAREER Grant No. PHY-1159224 and NSF Grant No. PHY-1316792. K.N.A. acknowledges support from the Institute for Nuclear Theory program ``Neutrino Astrophysics and Fundamental Properties'' 15-2a where part of this work was done.
\end{acknowledgments}

\onecolumngrid
\appendix

\section{Functions for high-temperature QCD pressure}\label{app:highT_funcs}
The functions defined in section \ref{sec:highT} are \cite{Laine06}
\begin{align}
F_1(y,\hat{\mu}) &\equiv \frac{1}{24\pi^2}\int_0^\infty dx\,x\,\left[\frac{x}{x+y}\right]^{\frac{1}{2}}\left[\hat{n}_{\rm F}(\sqrt{x+y}-\hat{\mu})+\hat{n}_{\rm F}(\sqrt{x+y}+\hat{\mu})\right],\\
F_2(y,\hat{\mu}) &\equiv \frac{1}{8\pi^2}\int_0^\infty dx \left[\frac{x}{x+y}\right]^{\frac{1}{2}}\left[\hat{n}_{\rm F}(\sqrt{x+y}-\hat{\mu})+\hat{n}_{\rm F}(\sqrt{x+y}+\hat{\mu})\right],\\
F_3(y,\hat{\mu}) &\equiv -\int_0^\infty dx \frac{1}{x}\left[\frac{x}{x+y}\right]^{\frac{1}{2}}\left[\hat{n}_{\rm F}(\sqrt{x+y}-\hat{\mu})+\hat{n}_{\rm F}(\sqrt{x+y}+\hat{\mu})\right],\\
F_4(y,\hat{\mu}) &\equiv \frac{1}{(4\pi)^4}\int_0^\infty dx_1 \int_0^\infty dx_2  \frac{1}{\sqrt{x_1+y}\sqrt{x_2+y}}\en
& \qquad \times \Bigg\{ \left[\hat{n}_{\rm F}(\sqrt{x_1+y}-\hat{\mu})\hat{n}_{\rm F}(\sqrt{x_2+y}+\hat{\mu})+\hat{n}_{\rm F}(\sqrt{x_1+y}+\hat{\mu})\hat{n}_{\rm F}(\sqrt{x_2+y}-\hat{\mu})\right]\en
&\qquad \qquad \times\ln{\left[\frac{\sqrt{x_1+y}\sqrt{x_2+y}+y-\sqrt{x_1x_2}}{\sqrt{x_1+y}\sqrt{x_2+y}+y+\sqrt{x_1x_2}}\right]}\en
&\qquad \quad+ \left[\hat{n}_{\rm F}(\sqrt{x_1+y}-\hat{\mu})\hat{n}_{\rm F}(\sqrt{x_2+y}-\hat{\mu})+\hat{n}_{\rm F}(\sqrt{x_1+y}+\hat{\mu})\hat{n}_{\rm F}(\sqrt{x_2+y}+\hat{\mu})\right]\en
&\qquad \qquad\times\ln{\left[\frac{\sqrt{x_1+y}\sqrt{x_2+y}-y+\sqrt{x_1x_2}}{\sqrt{x_1+y}\sqrt{x_2+y}-y-\sqrt{x_1x_2}}\right]}\Bigg\} \mbox{,}
\end{align}
where $\hat{n}_{\rm F}(x) = 1/[\exp{x} + 1]$ is the Fermi-Dirac distribution.
%

\section{Neutrino opacities: matrix elements and collision integrals}
\label{sec:matrixelements}

In this appendix, we expand on the details underlying the neutrino opacities that were presented in Sec.~\ref{sec:scattering}. Under the set of assumptions presented therein, we add contributions from a large number of reaction rates, which we present in an organized manner in the rest of this section. 

\subsection{Rates for neutrinos to go to two-particle final states}
\label{subsec:twoparticle}

We compute reaction rates for momenta and temperatures where we can integrate the weak gauge bosons out and approximate the weak interaction by a four-particle vertex. For tree level processes under this approximation, if one of the ingoing particles is a neutrino, one of the other particles is either a neutrino or a charged lepton belonging to the same generation. We classify reactions as leptonic or hadronic based on the nature of the remaining two particles. 

We now describe our calculation of these reactions' matrix elements, and the associated contributions to the neutrino opacity.

\subsubsection{Matrix elements for two-particle to two-particle reactions}
\label{subsubsec:matrixelts}

It is a lengthy, but straightforward, task to enumerate all leptonic reactions that contribute to the neutrino opacity. Ref.~\cite{Hannestad95} lists a complete set of reactions at temperatures of a few MeV. Our calculations extend to higher temperatures, hence we also include reactions involving tau leptons. The reactions are enumerated in Table~\ref{tab:leptonreactions}. It is harder to study hadronic reactions in a consistent manner through the quark-hadron transition temperature, $T_{\rm QCD}$. We adopt assumption \# 3 of Sec.~\ref{sec:scattering}: we neglect the strong coupling constant and its running at temperatures $T>T_{\rm QCD}$, and hence calculate opacities with free quarks. We enumerate all reactions involving quarks in the same manner as the leptonic ones, but with standard model quark currents that couple to $Z^0$ and $W^\pm$. These reactions are listed in the upper half of Table~\ref{tab:hadronreactions}. Table \ref{tab:twoparticle} shows squared and spin-summed matrix elements for a representative reaction involving only leptons, and for one involving quarks. 

\begin{table}[t]
   \caption{\label{tab:twoparticle} Representative examples of two-particle to two-particle reactions that contribute to the neutrino opacity, along with their matrix elements written using a four-fermion vertex for the weak interaction (for reactions involving leptons and quarks) and three quark chiral perturbation theory (for reactions involving the pseudoscalar meson octet). For a reaction $\nu_\mu + A \rightarrow B + C$, the momenta $p_i: i =1,2,3,4$ within matrix elements are mapped to $\nu_\mu, A, B$ and $C$ respectively.}
   \begin{ruledtabular}
     \begin{tabular}{ccc}
        Reaction & S $\sum \vert \mathcal{M} \vert^2$ & Remarks \\
        \hline
        \multicolumn{3}{c}{Reactions involving leptons} \\
        $\nu_\mu + \tau^- \rightarrow \mu^- + \nu_\tau$ & $128 G_{\rm F}^2 \left( p_1 \cdot p_2 \right) \left( p_3 \cdot p_4 \right) $ & \\
        \hline
        \multicolumn{3}{c}{Reactions involving quarks} \\
        $\nu_\mu + \mu^+ \rightarrow u + \bar{d}$ & $384 \vert V_{ud} \vert^2 G_{\rm F}^2 \left( p_1 \cdot p_4 \right) \left( p_2 \cdot p_3 \right)$ & $T>T_{\rm QCD}$ \\
        \hline
        \multicolumn{3}{c}{Reactions involving the pseudoscalar meson octet} \\
        $\nu_\mu + \mu^+ \rightarrow \pi^+ + \pi^0$ & $8 \vert V_{ud} \vert^2 G_{\rm F}^2 \left[ 2 p_2 \cdot \left( p_4 - p_3 \right) p_1 \cdot \left( p_4 - p_3 \right) - \left( p_1 \cdot p_2 \right) \left( p_4 - p_3 \right)^2 \right]$ & $T<T_{\rm QCD}$, s-channel \footnotemark[1] \\
        $\nu_\mu + \pi^+ \rightarrow \nu_\mu + \pi^+$ & $4 \left( 1 - 2 \sin^2{\theta_{\rm W}} \right)^2 G_{\rm F}^2 \left[ 2 p_3 \cdot \left( p_4 + p_2 \right) p_1 \cdot \left( p_4 + p_2 \right) - \left( p_1 \cdot p_3 \right) \left( p_4 + p_2 \right)^2 \right] $ & $T<T_{\rm QCD}$, t-channel \footnotemark[2] \\
     \end{tabular}
     \footnotetext[1]{Input neutrinos can produce quarks at low temperatures, $T<T_{\rm QCD}$, for large CM energies in the s-channel.}
     \footnotetext[2]{Input neutrinos can probe the quark content of mesons at low temperatures, $T<T_{\rm QCD}$, for large momentum transfers in the t-channel.}
   \end{ruledtabular}
\end{table}

The physical rates for hadronic reactions diverge from our calculated ones close to the transition, since the strong coupling constant is non-zero. Treating this self-consistently is beyond the scope of this paper. In the main body, we present results for a few unphysical interpolations through the transition. At even lower temperatures, $T < T_{\rm QCD}$, we cannot use the free quark approximation. The most important hadronic degrees of freedom are the pseudoscalar meson octet, which are pseudo-Goldstone bosons associated with the spontaneous breaking of the axial part of an approximate $SU(3)_L \times SU(3)_R$ flavor symmetry \cite{Srednicki07}. We use three quark chiral perturbation theory ($3\chi$PT) to write down the currents that couple to $Z^0$ and $W^\pm$, and through them evaluate the mesonic contribution to the neutrino opacity.

Consider a $3 \times 3$ unitary matrix, $U(x)$, which represents low-lying hadronic excitations at temperatures $T < T_{\rm QCD}$. We express $U(x)$ in terms of the pion fields $\pi^a(x)$ as follows:
\begin{align}
  U(x) & = \exp{\Bigl[ 2 i \frac{ \pi^a(x) T^a }{ f_\pi } \Bigr]} \mbox{,} \qquad a \in [1, 8] \mbox{,} \\
  \frac{ 2 \pi^a(x) T^a }{ f_\pi } & = 
     \frac{1}{f_\pi} 
     \begin{pmatrix} 
        \pi^0 + \frac{1}{\sqrt{3}} \eta & \sqrt{2} \pi^+ & \sqrt{2} K^+ \\
        \sqrt{2} \pi^- & - \pi^0 + \frac{1}{\sqrt{3}} \eta & \sqrt{2} K^0 \\
        \sqrt{2} K^- & \sqrt{2} \overline{K^0} & -\frac{2}{\sqrt{3}} \eta
     \end{pmatrix} \mbox{,}
\end{align}
where $f_\pi$ is an energy-scale associated with the breaking of the $SU(3)_A$ symmetry, and $T^a$ are generators of $SU(3)$. The most massive member of this octet, the $\eta$ meson, has a mass of $m_\eta = 547.8$ MeV \cite{Agashe:2014kda}. We only use this prescription at $T \leq 150$ MeV, so these low lying excitations are sufficient to describe all relevant {\em incoming} hadronic degrees of freedom. 

In the framework of $3\chi$PT, the dynamics of the pion fields are described by an effective Lagrangian for $U(x)$ coupled to matrix valued $SU(3)_L$ and $SU(3)_R$ gauge fields $l_\mu$ and $r_\mu$, respectively. The first approximation to the Lagrangian is the lowest term in a derivative expansion:
\begin{align}
   \mathcal{L} = -\frac14 f_\pi^2 {\rm Tr}[ D^\mu U^\dagger D_\mu U ] \mbox{,} \qquad {\rm with} \qquad D_\mu U = \partial_\mu U - i l_\mu U + i U r_\mu \mbox{.} \label{eq:covariantd}
\end{align}
The gauge fields $l_\mu$ and $r_\mu$ are Hermitian matrices, which we decompose as
\begin{align}
     (l/r)_\mu = (l/r)_{\mu}^a T^a + ( \mathcal{V}_{\mu} \mp \mathcal{A}_{\mu} ) I \mbox{,} \qquad a \in [1, 8] \mbox{.} 
\end{align}
The fields $\mathcal{V}_{\mu}$ and $\mathcal{A}_{\mu}$ are vector and axial-vector parts of $l_\mu$ and $r_\mu$. We identify the electroweak gauge bosons of the standard model, $Z_\mu^0, W_\mu^\pm$ and $A_\mu$ [or equivalently, the underlying $SU(2) \times U(1)$ gauge fields $A_\mu^a$ and $B_\mu$], with elements of $(l^a/r^a/\mathcal{V}/\mathcal{A})_\mu$ by equating their action on the pion fields $\pi^a(x)$ or the excitation $U(x)$ via the right-hand side of Eq.~\eqref{eq:covariantd}. The results of this procedure are
\begin{subequations}
\label{eq:smgbosons}
\begin{align}
  g_2 A_\mu^{1} & = l_\mu^1 \mbox{,} \\
  g_2 A_\mu^{2} & = l_\mu^{2} \mbox{,} \\
  g_2 A_\mu^{3} & = l_\mu^{3} + \frac{1}{\sqrt{3}} l_\mu^8 - \frac{1}{12} \mathcal{V}_\mu + \frac{1}{12} \mathcal{A}_\mu \mbox{,} \\
  e A_\mu & = l_\mu^3 + r_\mu^3 + \frac{1}{\sqrt{3}} l_\mu^8 + \frac{1}{\sqrt{3}} r_\mu^8 \mbox{.}
\end{align}
\end{subequations}
Here $g_2$ is the $SU(2)$ coupling constant, and $e = g_2 \sin{\theta_{\rm W}}$. The overlap of $A_\mu^3$ with the anomalous $\mathcal{A}_\mu$ is an artifact of the omission of the charm quark from our analysis. If the charm quark were included (as it has to be to make the SM non-anomalous) then there is no coupling to an anomalous current. As long as there are no charm quarks (always true at the temperatures where $3\chi$PT is applied) this current looks like that associated with the anomalous $U(1)_A$.

We read off the currents that couple to the gauge fields using the definition $J_{l/r/\mathcal{V}/\mathcal{A}}^{(a)\mu} = \partial \mathcal{L}/\partial (l^a/r^a/\mathcal{V}/\mathcal{A})_\mu$, and the Lagrangian of Eq.~\eqref{eq:covariantd}. We transform the resultings to obtain the currents that couple to the SM electroweak gauge fields, $Z^0_\mu, W_\mu^\pm$ and $A_\mu$: 
\begin{subequations}
\label{eq:weakcurrents}
\begin{align}
   J^{+\mu} & = \frac{1}{\sqrt{2}} \Bigl[ V_{ud}^\ast \Bigl( f_\pi \partial^\mu \pi^+ + i \pi^0 \overlrarrow{\partial^\mu} \pi^+ - \frac{i}{\sqrt{2}} \overline{K^0} \overlrarrow{\partial^\mu} K^+ \Bigr) + V_{us}^\ast \Bigl( - \frac{i}{\sqrt{2}} K^0 \overlrarrow{\partial^\mu} \pi^+ + \frac{i}{2} \pi^0 \overlrarrow{\partial^\mu} K^+ + \frac{\sqrt{3}i}{2} \eta \overlrarrow{\partial^\mu} K^+ \Bigr)  \Bigr] \mbox{,} \\
   J^{-\mu} & = \frac{1}{\sqrt{2}} \Bigl[ V_{ud} \Bigl( f_\pi \partial^\mu \pi^- - i \pi^0 \overlrarrow{\partial^\mu} \pi^- + \frac{i}{\sqrt{2}} K^0 \overlrarrow{\partial^\mu} K^- \Bigr) + V_{us} \Bigl( \frac{i}{\sqrt{2}} \overline{K^0} \overlrarrow{\partial^\mu} \pi^- - \frac{i}{2} \pi^0 \overlrarrow{\partial^\mu} K^- - \frac{\sqrt{3}i}{2} \eta \overlrarrow{\partial^\mu} K^- \Bigr)  \Bigr] \mbox{,} \\
   J^{\mu}_z & = J^{\mu}_3 - \sin^2{\theta_W} J^{\mu}_{\rm EM} \mbox{,} \\
   J^{\mu}_3 & = \frac12 \Bigl[ f_\pi \Bigl( \partial^\mu \pi^0 + \frac{1}{\sqrt{3}} \partial^\mu \eta \Bigr) + i \pi^+ \overlrarrow{\partial^\mu} \pi^- + i K^+ \overlrarrow{\partial^\mu} K^- \Bigr] \mbox{,} \\
   J^{\mu}_{\rm EM} & = i \pi^+ \overlrarrow{\partial^\mu} \pi^- + i K^+ \overlrarrow{\partial^\mu} K^- \mbox{.}   
\end{align}
\end{subequations}
These currents agree with the leading order parts of the functionals computed in Ref.~\cite{Unterdorfer05}. The lower half of Table \ref{tab:hadronreactions} enumerates the allowed reactions involving pseudoscalar mesons. Table \ref{tab:twoparticle} shows the squared and spin-summed matrix elements for two such reactions, computed using the currents listed in Eq.~\eqref{eq:weakcurrents}

A final complication is that $3\chi$PT, and the currents derived from it, are valid only when the momentum in the intermediate weak gauge bosons is low compared to the energy scale $4\pi f_\pi \sim 1$ GeV \cite{Srednicki07}). The physical currents that couple to the SM electroweak gauge fields are continuous functions of this momentum; they approach the SM free quark currents for large momentum values. This manifests as the production of quarks in the large CM energy limit in s-channel reactions, and as `deep-inelastic scattering' off the mesons' quark content in the large momentum-transfer limit in t-channel reactions. These limits are important to consider at the higher energies for which we calculate neutrino opacities using $3\chi$PT (the total energy range is shown in Figures \ref{fig:T_scattering_rates} and \ref{fig:scattering_rates}). 

We do not self-consistently compute these corrections to the currents, as it is beyond the scope of this paper. Instead, we modify the s-channel reaction rates in a phenomenological manner: we apply a cutoff in the CM energy at $1$ GeV with a width of $50$ MeV, below which we use the $3\chi$PT currents, and above which we use the SM free quark currents. We do not incorporate any corrections to t-channel reactions; this would involve some knowledge of the parton distribution functions for the mesons involved.

We observe that the squared and spin-summed matrix elements for tree level processes, be it for processes involving leptons and free quarks (computed using the SM currents), or for those involving pseudoscalar mesons (computed using $3\chi$PT), are at-most quadratic functions of the Mandelstam variables. This greatly facilitates a semi-automated computation of the two-particle to two-particle reactions' contribution to the neutrino opacity, which we very briefly describe next.

\subsubsection{Rates for two-particle to two-particle reactions}
\label{subsubsec:twopartrates}

Consider a general two-particle to two-particle reaction, $\nu_\alpha + A \rightarrow B + C$, that consumes a massless input neutrino, $\nu_\alpha$. The particles $A, B$ and $C$ can all be fermions (leptons or quarks), or contain a pair of bosons (pseudoscalar mesons). We expand Eq.~\eqref{eq:scatteringrate} to write down the following expression for the scattering rate as a collision integral:
\begin{align}
  \Gamma(E_{\nu_\alpha}) & = \frac{1}{2 E_{\nu_\alpha}} \int d^3 \tilde{\bm p}_A d^3 \tilde{\bm p}_B d^3 \tilde{\bm p}_C (2\pi)^4 \delta(p_{\nu_\alpha} + p_{A} - p_{B} - p_{C}) S \sum \vert \mathcal{M} \vert^2 f_A(E_A) (1 \mp f_B(E_B) ) ( 1 \mp f_C(E_C) ) \mbox{,} \label{eq:scatteringintegral}
\end{align}
where the symbol $d^3 \tilde{\bm p}$ is shorthand for the Lorentz invariant phase space volume element $d^3 {\bm p}/[(2\pi)^3 2 E(\bm p)]$, the symbol $\sum \vert \mathcal{M} \vert^2$ is the absolute value of the matrix element squared and summed over all spin states, $S$ is a symmetry factor for identical particles in the initial and/or final states, and the $f(E)$s are appropriate Bose-Einstein/Fermi-Dirac phase space distributions depending on the statistics of the particles, with plus and minus signs for bosons and fermions respectively.

We follow the treatment in Ref.~\cite{Hannestad95} to reduce the nine-dimensional phase space integral of Eq.~\eqref{eq:scatteringintegral} to a numerically manageable three-dimensional integral over the variables $\vert \bm p_A \vert, \vert \bm p_B \vert$ and $\mu_B = \hat{\bm p}_B \cdot \hat{\bm p}_{\nu_\alpha}$. This procedure involves using the delta function to perform the integral over $\bm p_C$, and using the form of the matrix elements for tree level processes to analytically perform the integral over $\mu_A = \hat{\bm p}_A \cdot \hat{\bm p}_{\nu_\alpha}$. We refer the reader to Ref.~\cite{Hannestad95} for more details. The form of the matrix elements also lends itself to easy parameterization in terms of a small number of classes; along with the procedure described above, this enables a simple numerical implementation of the calculation of these reactions' contributions to the neutrino scattering rate. 

\subsection{Rates for neutrinos to go to one-particle final states}
\label{subsec:oneparticle}

We must also consider the contribution to the neutrino interaction rate, $\Gamma(E_{\nu_\alpha})$, from interactions with two-particle final states (``fusion'' or inverse decay). A four-fermion interaction such as the weak interaction (at $E\ll m_W,m_Z$) can produce such a final state in two ways. One, applicable at $T<T_{\rm QCD}$, is two-body fusion to produce a meson, e.g. $\nu_\mu + \mu^+ \rightarrow \pi^+$. The other is the `three-body fusion', e.g. $\nu_\mu + \bar\nu_e + e^- \rightarrow \mu^-$. By construction, these fusion processes are the inverse of a decay process. We describe our treatment of these processes in the rest of this section.

\subsubsection{Kinematics of two-body fusion}

A two-body fusion process must involve a meson in either the initial or the final state, and -- if it is to absorb a neutrino -- must then be semi-leptonic. The neutral current processes of this form (e.g. a neutral meson is created by the fusion of $\nu_\alpha\bar\nu_\alpha\rightarrow\pi^0$) are helicity-forbidden and have zero amplitude at tree level. The charged current processes can have either the meson in the initial state and the charged lepton in the final state (e.g. $K^-\nu_\tau\rightarrow\tau^-$) or the meson in the final state (e.g. $\nu_\mu\mu^+\rightarrow\pi^+$). The ``charged lepton in the final state'' case is possible only if the charged lepton is more massive than the meson, i.e. if that lepton is a $\tau$; at $T<T_{\rm QCD}$ this not energetically feasible for typical values of the incoming particles' momenta, since $m_\tau\gg T_{\rm QCD}$. Therefore, for the rest of this section, we focus on the problem of a charged meson in the final state. The reaction is
\begin{equation}
\nu_\alpha + \alpha^+ \rightarrow A^+
\label{eq:aplus}
\end{equation}
where $\alpha=e$ or $\mu$ and $A=\pi$ or $K$. We are interested in the thermal absorption rate $\Gamma_{\rm fusion}$ for the neutrinos as a function of temperature $T$ and neutrino energy $E_\nu$.

The simplest solution to this problem is to calculate the rate of the inverse reaction of Eq.~(\ref{eq:aplus}) and use detailed balance. In thermal equilibrium, there is a rate of decays given by
\begin{equation}
\frac{dN}{dV\,dt} = g_A \Gamma^{\rm vac}_{A^+\rightarrow \nu_\alpha \alpha^+}
\int_0^\infty \frac{4\pi p_A^2\,dp_A}{(2\pi)^3} \int_{-1}^1 \frac{d\mu'}{2}  \frac{m_A}{E_A} f_A(E_A) [1-f_{\nu_\alpha}(E_{\nu_\alpha})] [ 1 - f_\alpha(E_\alpha) ],
\label{eq:fusion.temp1}
\end{equation}
where $\mu'$ is the cosine of the angle of emission of the neutrino in the rest frame of the $A^+$, and the factor of $m_A/E_A$ is the inverse-Lorentz factor that accounts for the longer lab-frame lifetime of $A^+$ at high energies. 
The degeneracy factor is $g_A = 1$ for pions and kaons, but we include it for later use with heavy mesons.

With the help of relativistic kinematics, we see that the lab-frame neutrino energy is
\begin{equation}
E_{\nu_\alpha} = \frac12\left( 1 - \frac{m_\alpha^2}{m_A^2} \right) (E_A+ p_A \mu'),
\end{equation}
so that Eq.~(\ref{eq:fusion.temp1}) can be re-written in terms of a rate of decays per unit volume per unit neutrino energy:
\begin{equation}
\frac{dN}{dV\,dt\,dE_{\nu_\alpha}} = g_A \Gamma^{\rm vac}_{A^+ \rightarrow \nu_\alpha \alpha^+}
\int_{p_{A,\rm min}}^\infty \frac{4\pi p_A^2\,dp_A}{(2\pi)^3} \frac{1}{(1-m_\alpha^2/m_A^2)p_A} \frac{m_A}{E_A} f_A(E_A) [1-f_{\nu_\alpha}(E_{\nu_\alpha})] [ 1 - f_\alpha(E_\alpha) ].
\label{eq:fusion.temp2}
\end{equation}
The kinematically allowed range of $A^+$ momenta is given by
\begin{equation}
\frac12\left( 1 - \frac{m_\alpha^2}{m_A^2} \right) (E_A - p_A ) \le
E_{\nu_\alpha} \le \frac12\left( 1 - \frac{m_\alpha^2}{m_A^2} \right) (E_A + p_A )
\end{equation}
or -- using $p_A = \sqrt{E_A^2-m_A^2}$ and with some algebraic manipulation --
\begin{equation}
E_A \ge E_{A,\rm min} = \frac{(1-m_\alpha^2/m_A^2)^2m_A^2 + 4E_{\nu_\alpha}^2}{4E_{\nu_\alpha}(1-m_\alpha^2/m_A^2)}.
\label{eq:EMIN}
\end{equation}

Turning the integral into one over the energy $E_A$ of $A^+$ gives
\begin{equation}
\frac{dN}{dV\,dt\,dE_{\nu_\alpha}} = g_A \frac{\Gamma^{\rm vac}_{A^+ \rightarrow \nu_\alpha \alpha^+}}{1-m_\alpha^2/m_A^2}
\int_{E_{A,\rm min}}^\infty \frac{4\pi m_A \,dE_A}{(2\pi)^3}  f_A(E_A) [1-f_{\nu_\alpha}(E_{\nu_\alpha})] [ 1 - f_\alpha(E_\alpha) ].
\end{equation}
Now this should equal the fusion rate of neutrinos, which is
\begin{equation}
\frac{dN}{dV\,dt\,dE_{\nu_\alpha}} = \frac{4\pi E_{\nu_\alpha}^2}{(2\pi)^3} f_{\nu_\alpha}(E_{\nu_\alpha})\,\Gamma_{\rm fusion}(E_{\nu_\alpha}).
\end{equation}
We therefore conclude that
\begin{equation}
\Gamma_{\rm fusion}(E_{\nu_\alpha})
= \frac{g_A m_A \Gamma^{\rm vac}_{A^+ \rightarrow \nu_\alpha \alpha^+}}{(1-m_\alpha^2/m_A^2)E_{\nu_\alpha}^2}
\int_{E_{A,\rm min}}^\infty
f_A(E_A) \frac{1-f_{\nu_\alpha}(E_{\nu_\alpha})}{f_{\nu_\alpha}(E_{\nu_\alpha})} [ 1 - f_\alpha(E_\alpha) ]
\,dE_A.
\end{equation}
Next we substitute in the Bose-Einstein or Fermi-Dirac phase space distributions, and note that $E_\alpha=E_A-E_{\nu_\alpha}$, yielding
\begin{eqnarray}
\Gamma_{\rm fusion}(E_{\nu_\alpha})
&=& \frac{g_A m_A \Gamma^{\rm vac}_{A^+ \rightarrow \nu_\alpha \alpha^+}}{(1-m_\alpha^2/m_A^2)E_{\nu_\alpha}^2}
\int_{E_{A,\rm min}}^\infty
\frac{e^{E_{\nu_\alpha}/T}}{(e^{E_A/T}-1)(e^{-E_A/T}e^{E_{\nu_\alpha}/T}+1)}
\,dE_A
\nonumber \\
&=& \frac{g_A m_A \Gamma^{\rm vac}_{A^+ \rightarrow \nu_\alpha \alpha^+} T}{(1-m_\alpha^2/m_A^2)E_{\nu_\alpha}^2}
\Phi\left(
\frac{E_{A,\rm min}}T,
\frac{E_{A,\rm min}-E_{\nu_\alpha}}T
\right).
\end{eqnarray}
Here we have defined the dimensionless integral
\begin{eqnarray}
\Phi(a,b) &=& \int_0^\infty \frac{e^{a-b}\,dx}{(e^{x+a}-1)(e^{-x-b}+1)}
\nonumber \\
&=& \sum_{j=0}^\infty \sum_{k=0}^\infty \int_0^\infty e^{a-b} (-1)^k e^{-(1+j)(x+a)} e^{-k(x+b)}\,dx
\nonumber \\
&=& e^{a-b} \sum_{j=0}^\infty \sum_{k=0}^\infty \frac{(-1)^k e^{-(1+j)a-kb}}{1+j+k}
\nonumber \\
&=& e^{a-b} \sum_{m=1}^\infty \frac{e^{-ma}}m \sum_{k=0}^{m-1} (-e^{a-b})^k
\nonumber \\
&=& e^{a-b} \sum_{m=1}^\infty \frac{e^{-ma}}m \frac{1-(-e^{a-b})^m}{1+e^{a-b}}
\nonumber \\
&=& \frac1{e^{b-a}+1} \left[ \sum_{m=1}^\infty \frac{e^{-ma}}m + \sum_{m=1}^\infty (-1)^{m-1} \frac{e^{-mb}}m \right]
\nonumber \\
&=& \frac1{e^{b-a}+1} \left[ -\ln(1-e^{-a}) + \ln(1+e^{-b}) \right]
\nonumber \\
&=& \frac1{e^{b-a}+1} \ln\frac{1+e^{-b}}{1-e^{-a}}.
\end{eqnarray}
Recall that $E_{A,\rm min}$ is a function of $E_{\nu_\alpha}$ and is given by Eq.~(\ref{eq:EMIN}).
We then achieve the final simplification:
\begin{equation}
\Gamma_{\rm fusion}(E_{\nu_\alpha})
= \frac{g_A m_A \Gamma^{\rm vac}_{A^+ \rightarrow \nu_\alpha \alpha^+} T}{\upsilon(1+e^{-E_{\nu_\alpha}/T})E_{\nu_\alpha}^2}
 \ln\frac{1+e^{E_{\nu_\alpha}/T}e^{-(\upsilon^2m_A^2 + 4E_{\nu_\alpha}^2)/(4\upsilon E_{\nu_\alpha} T)}}{1-e^{-(\upsilon^2m_A^2 + 4E_{\nu_\alpha}^2)/(4\upsilon E_{\nu_\alpha} T)}},
\label{eq:Gfusion}
\end{equation}
where $\upsilon = 1-m_\alpha^2/m_A^2$. Note that the numerical calculation of the logarithm must be treated carefully since for $E_{A,\rm min}-E_{\nu_\alpha}\gg T$ we are taking the logarithm of a number that is very close to 1. For calculational purposes, we replace the logarithm in Eq.~\eqref{eq:Gfusion} by a truncation of its Taylor expansion at the fifth order wherever the argument deviates from unity by less than $\epsilon = 10^{-3}$. 

\subsubsection{Rates for two-body fusion processes}

The rate parameters for the key two-body fusion reactions are shown in Table~\ref{tab:2.body.rate}.

Some parameters were not available in the Review of Particle Properties. Key among these are the decay parameters for the weak decays of the $\rho$, $\omega(782)$, and $K^\ast(892)$ vector mesons. These mesons are actually extremely broad resonances, and their principal decay mode is into lighter mesons. Electromagnetic and especially weak decay modes are less common. Note that the helicity suppression arguments that forbid e.g. $\pi^0\rightarrow \nu_e \bar \nu_e$ do not apply to the vector mesons.

The relevant decays of the charged vector mesons $\rho^+$ and $K^\ast(892)^+$ can be obtained by noting that the virtual-$W$ diagram results in an effective vertex
\begin{equation}
{\cal L}_{\rm eff} \ni i\frac{e}{2\sqrt2\,\sin\theta_W} \bar \ell_X \gamma^\mu (1-\gamma^5) \nu_X \frac1{m_W^2}
\frac{e}{2\sqrt2\,\sin\theta_W} V_{ud} m_\rho f_\rho [\epsilon(\rho^-)]_\mu + {\rm h.c.},
\label{eq:rho.L1}
\end{equation}
where $f_\rho$ is the $\rho$ decay constant, and $[\epsilon(\rho^-)]_\mu$ is the polarization of the ``on-shell'' $\rho$ meson. The advantage of this Lagrangian is that the well-measured decay $\tau^+\rightarrow\rho^+\bar\nu_\tau$ is related to the decay of $\rho^+$ to a charged lepton and a neutrino (predicted branching fraction $\sim 2\times 10^{-11}$). From a tree-level calculation with this Lagrangian, we infer a ratio
\begin{equation}
\frac{\Gamma(\rho^+\rightarrow \mu^+\bar\nu_\mu)}{\Gamma(\tau^+\rightarrow\rho^+\bar\nu_\tau)} =
\frac{2m_\tau^3(m_\rho^2-m_\mu^2)^2(2m_\rho^2+m_\mu^2)}{3m_\rho^3(m_\tau^2-m_\rho^2)^2(2m_\rho^2+m_\tau^2)}.
\end{equation}
These rates are included in Table~\ref{tab:2.body.rate}.

\begin{table}[t]
\caption{\label{tab:2.body.rate}The parameters for reactions that go into Eq.~(\ref{eq:Gfusion}). Reactions relevant for the neutrino opacity are shown; antineutrinos are similar. Particle masses are obtained from the Particle Data Group. Decay partial widths are obtained from the sources indicated. All reactions in which a neutrino can produce a hadronic resonance below 1 GeV are included.}
\begin{tabular}{ccrccclcccl}
\hline\hline
Reaction & & $m_A$~ & & $g_A$ & & ~~~$\upsilon$ & & $\Gamma^{\rm vac}_{\rm reverse}$ & & Rate method \\
 & & MeV & & & & & & MeV & & \\
\hline
\multicolumn{11}c{Reactions involving the pseudoscalar meson octet}\\
$\nu_e + e^+ \rightarrow \pi^+$ & & 139.57 & & 1 & & 0.999987 & & $3.110\times 10^{-18}$ & & PDG \\
$\nu_\mu + \mu^+ \rightarrow \pi^+$ & & 139.57 & & 1 & & 0.4269 & & $2.528\times 10^{-14}$ & & PDG \\
$\nu_e + e^+ \rightarrow K^+$ & & 493.68 & & 1 & & 0.9999989 & & $8.41\phantom0\times 10^{-19}$ & & PDG \\
$\nu_\mu + \mu^+ \rightarrow K^+$ & & 493.68 & & 1 & & 0.95419 & & $3.38\phantom0\times 10^{-14}$ & & PDG \\
\hline
\multicolumn{11}c{Reactions involving vector mesons with nonzero isospin}\\
$\nu_X + \bar \nu_X \rightarrow \rho^0$ & & 775.26 & & 3 & & 1 & & $9.78\phantom{0}\times 10^{-12}$ & & Average of $\tau$ decay and $e^+e^-\rightarrow\rho^0$; assumed isospin $SU(2)$\footnotemark[1] \\
$\nu_e + e^+ \rightarrow \rho^+$ & & 775.26 & & 3 & & 0.9999996 & & $7.00\phantom0\times 10^{-11}$ & & $\tau$ decay \\
$\nu_\mu + \mu^+ \rightarrow \rho^+$ & & 775.26 & & 3 & & 0.98143 & & $6.80\phantom0\times 10^{-11}$ & & $\tau$ decay \\
$\nu_e + e^+ \rightarrow K^\ast(892)^+$ & & 891.66 & & 3 & & 0.9999997 & & $5.45\phantom0\times 10^{-12}$ & & $\tau$ decay \\
$\nu_\mu + \mu^+ \rightarrow K^\ast(892)^+$ & & 891.66 & & 3 & & 0.98596 & & $5.33\phantom0\times 10^{-12}$ & & $\tau$ decay \\
\hline
\multicolumn{11}c{Reactions involving vector mesons with zero isospin}\\
$\nu_X + \bar \nu_X \rightarrow \omega(782)$ & & 782.65 & & 3 & & 1 & & $7\phantom{.000}\times 10^{-13}$ & & $e^+e^-\rightarrow\omega(782)$; assumed quark content $(\bar uu+\bar dd)/\sqrt2$ \\
\hline\hline
\footnotetext[1]{The $\tau$ decay gives $1.01\times 10^{-11}$ and the $e^+e^-\rightarrow\rho^0$ computation gives $9.5\times 10^{-12}$.}
\end{tabular}
\end{table}

The rate for $\rho^0\rightarrow \nu_X\bar\nu_X$ can be obtained by replacing the terms in Eq.~(\ref{eq:rho.L1}) with the $Z$ couplings and propagator:
\begin{equation}
{\cal L}_{\rm eff} \ni i\frac{e}{4\sin\theta_W\cos\theta_W} \bar \nu_X \gamma^\mu (1-\gamma^5) \nu_X \frac1{m_Z^2}
\frac{e}{\sqrt2\,\sin\theta_W\cos\theta_W}\left( \frac12-\sin^2\theta_W \right) m_\rho f_\rho [\epsilon(\rho^0)]_\mu + {\rm h.c.}.
\label{eq:rho.L2}
\end{equation}
Here the coupling of the vector $Z$ current to the $\rho^0$ meson was related to the coupling of the vector $W$ current to the $\rho^+$ using isospin $SU(2)$ symmetry. The conclusion is that
\begin{equation}
\frac{\Gamma(\rho^0 \rightarrow \nu_e\bar\nu_e)}{\Gamma(\tau^+\rightarrow\rho^+\bar\nu_\tau)} =
\frac{8m_\tau^3m_\rho^3}{3(m_\tau^2-m_\rho^2)^2(2m_\rho^2+m_\tau^2)} \frac{(1/2-\sin^2\theta_W)^2}{|V_{ud}|^2},
\end{equation}
and similarly for the other two flavors. In this relation, we have used that $m_W=m_Z\cos\theta_W$.

The same decay rate can be obtained by taking the ratio
\begin{equation}
\frac{\Gamma(\rho^0 \rightarrow \nu_e \bar \nu_e)}{\Gamma(\rho^0\rightarrow e^+ e^-)} = \frac12\left( \frac1{2\sin\theta_W\cos\theta_W}\right)^2 \left( \frac{m_\rho^2}{m_Z^2} \right)^2 \left( \frac{1}{\sin\theta_W\cos\theta_W}
\frac{\langle 0 | (g_{u,V}\bar u \gamma^i u +g_{d,V} \bar d \gamma^i d + g_{s,V} \bar s \gamma^i s) | \alpha^0 \rangle}{\langle 0 | (\frac23\bar u \gamma^i u - \frac13 \bar d \gamma^i d - \frac13 \bar s \gamma^i s) | \rho^0 \rangle} \right)^2.
\label{eq:flavor}
\end{equation}
Note that the reaction $\rho^0\rightarrow e^++e^-$ is mediated principally by the photon instead of the $Z$. The $\frac12$ is due to the fact that the photon couples to right-handed as well as left-handed electrons. The factor of $m_\alpha/m_\rho$ is a combination of kinematic factors appropriate if the decay constant is the same for all octet members. The factor of $1/(2\sin\theta_W\cos\theta_W)$ is the ratio of the $Z$ coupling to $\nu_{e,L}$ to the photon coupling to $e_{L}$ (or $e_{R}$). The factor of $m_\rho^2/m_Z^2$ is the ratio of $Z$ to photon propagators. The last term is the ratio of $Z$ coupling to $\alpha^0$ to $\gamma$ coupling to $\rho^0$, with $g_{u,V} = \frac14-\frac23\sin^2\theta_W$ and $g_{d,V}=g_{s,V} = -\frac14+\frac13\sin^2\theta_W$. For the $\rho^0$, the last term can be computed using isospin symmetry.

The agreement between $\tau$ decay and $e^+e^-\rightarrow\rho^0$ is good: the former predicts a partial width for $\rho^0\rightarrow\nu_e\bar\nu_e$ of $1.01\times 10^{-11}$ MeV, and the latter predicts $9.5\times 10^{-12}$ MeV, a difference of only 6\%. The average is shown in the table.

No similar decay is allowed (i.e. it is not possible with a single intermediate vector boson propagator) for the $K^\ast(892)^0$ or $\bar K^\ast(892)^0$ mesons because the current that couples to the $Z$ cannot change strangeness.

It is less clear how this procedure should be applied to the $\omega(782)$ meson, which has no isospin. One might approximate it as a pure $(\bar u u + \bar d d)/\sqrt2$ state (i.e. with no strange quark), and repeat the argument used for the $\rho^0\rightarrow e^+e^-$ calculation with Eq.~(\ref{eq:flavor}). This result is shown in the table; it is much more uncertain than the calculation for the $\rho^0$ since mixing with $\bar ss$ is allowed. Nevertheless, the small rate for $\omega(782)$ production (as compared with $\rho^0$) suggests that it leads to an overall small correction to neutrino opacities.

\subsubsection{Three-body fusion processes}
\label{subsubsec:threebody}

The final set of reactions that contribute to the neutrino opacity are three-body fusions. As earlier, these reactions can be either leptonic or hadronic in nature. We adopt the prescription outlined in Sec.~\ref{subsubsec:matrixelts} for the hadronic reactions. Given the hadronic and leptonic currents coupling to the SM electroweak gauge bosons, we can enumerate all three-body reactions that contribute to the neutrino opacity in same manner as earlier. We must keep in mind the kinematic constraint that the rest mass of the product must be greater than that of the reactants. 

The matrix element for any three-body fusion reaction is related to one for a two-particle to two-particle reaction by crossing symmetry. Thus, we do not need to compute any new matrix elements for this section. However, we need to modify the treatment of the kinematics from the previous case. Consider a general three-body fusion reaction, $\nu_\alpha + A + B \rightarrow C$. The scattering rate for an input neutrino energy $E_{\nu_\alpha}$ is given by the collision integral:
\begin{align}
  \Gamma(E_{\nu_\alpha}) & = \frac{1}{2 E_{\nu_\alpha}} \int d^3 \tilde{\bm p}_A d^3 \tilde{\bm p}_B d^3 \tilde{\bm p}_C (2\pi)^4 \delta(p_{\nu_\alpha} + p_{A} + p_{B} - p_{C}) S \sum \vert \mathcal{M} \vert^2 f_A(E_A) f_B(E_B) ( 1 \mp f_C(E_C) ) \mbox{.} \label{eq:3bodyscatteringintegral}
\end{align}
All the symbols are defined identically to Eq.~\eqref{eq:scatteringintegral}. The procedure to reduce the dimensionality of this integral is exactly analogous to that in Sec.~\ref{subsubsec:twopartrates} and Ref.~\cite{Hannestad95}, with one important difference. The variables finally left to numerically integrate over are, as earlier, $\vert \bm p_A \vert, \vert \bm p_B \vert$ and $\mu_B = \hat{\bm p}_{\nu_\alpha} \cdot \hat{\bm p}_B$. If we consider the integration domain for the two-particle to two-particle case, for a given value of $\vert \bm p_A \vert$, energy constraints allow a maximum value of $\vert \bm p_B \vert$. For a three-body fusion, $\vert \bm p_B \vert$ has no upper bound, which greatly expands the allowed phase-space. With this caveat, the rest of the procedure proceeds as it did for the other case.

\bibliography{references,references_suscep,sterile_intro}

\end{document}